\documentclass[prx,twocolumn,showkeys,showpacs,superscriptaddress,nofootinbib,10pt,floatfix
]{revtex4-1}
\usepackage{graphicx}  
\usepackage{colortbl} 
\usepackage{amsthm}
\usepackage{epsfig}
\usepackage{amsmath,amssymb,amsfonts,mathtools,bbm,MnSymbol,mathrsfs,xfrac}   
\usepackage{longtable} 
\usepackage{tabularx}
\usepackage[breaklinks=true,colorlinks=true,linkcolor=blue,urlcolor=blue,citecolor=blue]{hyperref}
\usepackage{comment}
\usepackage{color}
\usepackage[makeroom]{cancel}
\usepackage{ifthen}

\usepackage{soul}

\renewcommand{\[}{\begin{equation}}
\renewcommand{\]}{\end{equation}}
\renewcommand{\Re}{\mathfrak{Re}}

\newcommand{\ket}[1]{|#1\rangle}
\newcommand{\bra}[1]{\langle#1|}
\newcommand{\braket}[2]{\langle#1|#2\rangle}
\newcommand{\pro}[2]{|#1\rangle\langle#2|}
\newcommand{\mean}[1]{\langle#1\rangle}

\newcommand{\ov}[1]{\overline{#1}}
\newcommand{\tr}{\mathrm{tr}}
\newcommand{\rank}{\mathrm{rank}}

\newcommand{\R}{{\hat{\rho}}}
\newcommand{\Rt}{{\hat{\rho}_{th}}}
\newcommand{\Rtt}{{\hat{\rho}_{th}^-}}

\newcommand{\C}{{\mathcal{C}}}
\renewcommand{\P}{\hat{P}}
\newcommand{\bi}{{\boldsymbol{i}}}

\newcommand{\Ha}{{\hat{H}}}
\newcommand{\Hi}{{\hat{H}^{(\mathrm{int})}}}
\newcommand{\HS}{\mathcal{H}}

\newboolean{showcomments}
\setboolean{showcomments}{false}

\newcommand{\ana}[1]{\ifthenelse{\boolean{showcomments}}{\textcolor{cyan}{[{\bf AA}: #1]}}{}}
\newcommand{\jd}[1]{\ifthenelse{\boolean{showcomments}}{\textcolor{green}{[{\bf JD}: #1]}}{}}
\newcommand{\ds}[1]{\ifthenelse{\boolean{showcomments}}{\textcolor{red}{[{\bf DS}: #1]}}{}}
\definecolor{mygray}{gray}{0.6}

\theoremstyle{definition}
\newtheorem{definition}{Definition}
\newtheorem{theorem}{Theorem}
\newtheorem{lemma}{Lemma}
\newtheorem{corollary}{Corollary}[theorem]

\begin{document}

\title{Quantum coarse-grained entropy\\
 and thermalization in closed systems
}

\author{Dominik \v{S}afr\'{a}nek}
\email{dsafrane@ucsc.edu}
\affiliation{SCIPP and Department of Physics, University of California, Santa Cruz, CA 95064, USA}
\author{J. M. Deutsch}
\affiliation{Department of Physics, University of California, Santa Cruz, CA 95064, USA}
\author{Anthony Aguirre}
\affiliation{SCIPP and Department of Physics, University of California, Santa Cruz, CA 95064, USA}

\date{\today}

\begin{abstract}
We investigate the detailed properties of {\em Observational entropy}, introduced by \v{S}afr\'{a}nek \emph{et al}. [\href{https://link.aps.org/doi/10.1103/PhysRevA.99.010101}{Phys. Rev. A
99, 010101 (2019)}] as a generalization of Boltzmann entropy to quantum mechanics. This quantity can involve multiple coarse-grainings, even those that do not commute with each other, without losing any of its properties. It is well-defined out of equilibrium, and for some coarse-grainings it generically rises to the correct thermodynamic value even in a genuinely isolated quantum system. The quantity contains several other entropy definitions as special cases, it has interesting information-theoretic interpretations, and mathematical properties -- such as extensivity and upper and lower bounds -- suitable for an entropy. Here we describe and provide proofs for many of its properties, discuss its interpretation and connection to other quantities, and provide numerous simulations and analytic arguments supporting the claims of its relationship to thermodynamic entropy. This quantity may thus provide a clear and well-defined foundation on which to build a satisfactory understanding of the second thermodynamical law in quantum mechanics.
\end{abstract}

\maketitle

\section{Introduction}

The second law of thermodynamics is widely regarded as one of the most fundamental in nature.  Yet there is a striking lack of consensus as to {\em exactly} what its content is.  Taking as its basic definition that ``entropy is non-decreasing in a closed system," there is disagreement as to how to define entropy, about to what systems the second law applies, and on what ``non-decreasing" means.  Let us consider these in turn.

``Entropy" has a host of thermodynamic, statistical-mechanical, and information-theoretic definitions.  As motivation for this paper consider four common entropies from physics:
\begin{itemize}
\item {\bf Thermodynamic entropy}, defined by $dS=dQ/T$ for a reversible change to a system receiving heat $dQ$ at absolute temperature $T$, within an axiomatic set of thermodynamics definitions.
\item {\bf Classical Gibbs entropy}, defined up to a multiplicative constant for a discrete state-space by $S_G(p_j) = -\sum_j p_j\ln p_j$, where $p_j$ is the probability of being in a state $j$; the sum is extendable to an integral over phase space given a phase-space state density $\rho$.
\item {\bf Classical Boltzmann entropy}, defined by $S_B(V_i) = \ln V_i$, where $V_i$ is the number (or phase space volume) of classical microstates belonging to the $i$th macrostate $A_i$.
\item {\bf von Neumann entropy}, defined by $S(\hat\rho) = -\tr[\R\ln\R]$, with $\R$ being the density matrix of a quantum system.
\end{itemize}
Equilibrium classical and quantum statistical mechanics draw a host of connections between these concepts. Gibbs and Boltzmann entropies are related if the microstates are given equal probability; thermodynamic entropy is expressible as Gibbs entropy~\cite{jaynes1965gibbs}; and
von Neumann entropy is a natural generalization of Gibbs entropy.\footnote{As $S(\hat\rho)$  can be written as $-\sum_i p_i\ln p_i$ if the $p_i$ are probabilities for elements of the spectral decomposition of $\hat\rho$.}

In terms of being ``closed," a system can be closed {\em in practice} if its interactions with its environment are sufficiently weak that the system is well-described for all practical purposes by a Hamiltonian operating on just the system's degrees of freedom.  A system might be {\em in principle} closed if it were, for example, the entire Universe, or if it were closed due to causality constraints.\footnote{The domain of dependence of a compact achronal surface would constitute such a system at the classical level and at the level of quantum fields in a fixed spacetime background. Such a system generally (but not in all cases) has a finite lifetime.  It is unclear whether fully closed subsystems exist at the quantum gravitational level.}  In addition, but often unmentioned, a closed system undergoing measurement or observation may have the observer {\em within} the system or {\em external to} the system.  Observers within a system are tricky to treat because much physics formalism implicitly assumes an external observer.  And an external observer performing a measurement on a closed system must necessarily interact with the system; by definition the system is then no longer closed. While this is not a problem in classical systems because properties of classical systems are considered to be robust against  such interaction in principle, and do not change when measured, in quantum systems any such interaction disturbs the system and changes its inherent properties.

A related issue is that in quantum theory different observers should not disagree about the quantum state of a system, which is regarded as un-improvable knowledge of the system.  Uncertainties in the state are described using the density matrix $\hat\rho$, upon which observers {\em can} disagree, for example due to different states of knowledge.  The meaning of the probabilities inherent in the density matrix is subject to the same interpretative issue as for probabilities in general: they may be interpreted as subjective credences, or relative frequencies in an ensemble, or otherwise.  In any case, $\hat\rho$ can be altered not just by the system's evolution, but also by {\em conditioning} and/or by {\em marginalizing}.  Conditioning would correspond to the observer performing some measurement or taking into account additional data.  Marginalizing would correspond to tracing over some degrees of freedom, for example when considering a quantum subsystem, or computing an expectation value of an observable depending on only a subset of the degrees of freedom.

Finally, let us turn to the ``non-decreasing" nature of entropy.  In classical thermodynamics, this is taken as axiomatic.  However, the thermodynamical quantity defined by $dS=dQ/T$ can fluctuate, and ``fluctuation theorems"~(e.g.,~\cite{sevick2008fluctuation,
aberg2016fully}) for open systems
quantify these fluctuations and indicate that over a time $\Delta t$, the relative probability of a downward versus upward change in entropy of magnitude $\Delta S$ is given by
$
{p(-\Delta S)/ p(+\Delta S)} = \exp(-\Delta S),
$
so that decreasing entropy in a given system is exponentially suppressed, rather than impossible.

In marked contrast, the Gibbs or von Neumann entropy for a closed system undergoing classical or quantum Hamiltonian evolution stays strictly constant; this expresses the preservation of information in the evolution of the density matrix $\hat\rho$ or density of states $\rho$, which in either case provide a full encoding of the statistics of the system.  Flows of information into or out of a system are represented by changes to $S_G(\rho)$ or $S(\hat\rho).$  This may occur as mentioned above, if the observer gains additional information (generally decreasing entropy) or marginalizes over more degrees of freedom (generally increasing entropy)~\cite{gharibyan2014sharpening}.\footnote{Or via similar re-definitions of the statistical ensemble underlying the density matrix or density of states.}
In terms of observer-independent time evolution, an {\em open} system (e.g.~a subsystem of a closed system), $S_G(\rho)$ or $S(\hat\rho)$ obtained by marginalizing (or tracing out) the rest of the closed system tends to increase toward a maximal value under some general conditions~\cite{cramer2008exact,reimann2008foundation,linden2009quantum,cramer2012thermalization,short2012quantum,reimann2012equilibration,brandao2012convergence,vznidarivc2012subsystem,masanes2013complexity,garcia2015equilibration,gluza2016equilibration}, which are reviewed in~\cite{eisert2015quantum,gogolin2016equilibration}, although some recent work has shown that this increase does not happen for every initial state~\cite{reimann2016typical}.

In the absence of external influences, however, we still expect an increase in some sort of entropy (or ``disorder"): a closed house tends to get messier even if nothing comes in or leaves! Dissatisfaction with the fact that von Neumann entropy does not increase\footnote{von Neumann wrote himself~\cite{von2010proof}: ``The expressions for entropy given by
the author [von Neumann] are not applicable here in the way they were intended, as they were
computed from the perspective of an observer who can carry out all measurements
that are possible in principle, i.e., regardless of whether they are macroscopic (for
example, there every pure state has entropy 0, only mixtures have entropies greater
than 0!)''} has led papers to propose different measures in order to better capture the essence of the second thermodynamical law.

One approach is to take the Shannon entropy of diagonal elements of the density matrix written in an eigenbasis of the instantaneous Hamiltonian~\cite{tolman1938principles,ter1954elements,jaynes1957information2}. This approach has been extensively studied recently and the entropy  named the ``diagonal entropy''~\cite{polkovnikov2011microscopic}. In its native form, this entropy remains constant in genuinely closed systems, which evolve through a time-independent Hamiltonian, and therefore suffers of the same problem as the von Neumann entropy. However it has been shown to increase for systems evolving through a time-dependent Hamiltonian due to non-zero transition probability between different instantaneous energy levels.

Another way of getting a physically relevant entropy is to trace out parts of the density matrix, leading to a reduced density matrix that looks locally thermal. At this point one can either take the von Neumann entropy (entanglement entropy), or the diagonal entropy of these reduced density matrices, and the results will be very similar. The resulting \emph{sum of local diagonal entropies} will generically increase even for genuinely closed systems, and it models local regions equilibrating with each other~\cite{polkovnikov2011microscopic}. We discuss later on in Section~\ref{sec-comparison} how this sum differs from the Factorized Observational entropy with energy coarse-graining (FOE) introduced in this paper. FOE keeps correlations between different regions.

Instead of defining entropy by instantaneous Hamiltonian, it is also possible to consider a more general case defined by any conceivable observable. The Shannon entropy of the diagonal elements of the density matrix written in an eigenbasis of an observable, which has been named
``entropy of an observable" has been introduced in the 1960s~\cite{ingarden1962quantum,grabowski1977continuity} and studied recently~\cite{anza2017information,anza2018new,anza2018eigenstate}. This entropy is similar to what we study in this paper, however it does not take into account different sizes of the respective macrostates, and weights each of them the same, making it much less like the Boltzmann entropy; rather, it describes the statistics of measurement outcomes.

Another path to understanding thermodynamics of a closed system leads through use of entropies that directly measure certain features the density matrix. One such entropy is the {\em entanglement entropy}.
The system is divided into two subsystems $A$ and $B$. Considering the reduced density matrix
$\R_A=\tr_B \R_{AB}$, we define the entanglement entropy to be the von Neumann entropy of
$\hat\rho_A$, $S_A = -\tr[ \R_A\ln\R_A]$. For pure states, this quantity measures mutual information between systems $A$ and $B$. Although the entanglement entropy $S_A$ was shown to be the same as the thermodynamic entropy of $A$ in the limit of large system size~\cite{deutsch2010thermodynamic,deutsch2013microscopic,santos2012weak}, it still primarily measures the information exchange rather than heat exchange. Another common entropy of a similar information-theoretic type is the \emph{quantum relative entropy}~\cite{sagawa2012nonequilibrium}, which measures how close two quantum states are to each other, and \emph{max-entropy}~\cite{konig2009operational}, which measures maximum fidelity of
$\R_{AB}$ with a product state that is completely mixed on $A$. However, although these entropies have been connected to certain parts of thermodynamics, such as entropy production~\cite{modi2010unified,deffner2010generalized} and the minimal thermodynamic cost of information erasure~\cite{del2011thermodynamic},
they are not directly related to macrostates and macro-observable quantities such as energy or angular momentum.

We will thus pursue a different approach, inspired by classical thermodynamics, where it is the {\em Boltzmann entropy} that most clearly addresses the spontaneous creation of disorder in a closed system. In classical thermodynamics, a microstate that evolves unitarily without regard to the definition of the macrostates $A_i$ will tend to enter progressively higher-entropy macrostates; it is then natural to expect evolution to higher-entropy macrostates to be exponentially more common than evolution to lower-entropy macrostates.

Somewhat  surprisingly,  until recently {\em a quantum generalization of Boltzmann entropy along these lines has not been well-developed}. In this paper, we extend results of the recently published paper~\cite{SafranekDeutschAguirreObservationalEntropyLetter} by the current authors, which developed a measure of entropy called the ``Observational entropy." We provide detailed definitions and proofs to theorems published there. Unlike previous studies involving a quantum coarse-grained entropy, we show how to generalize Observational entropy to include multiple coarse-grainings, even those that do not commute with each other. In addition,  we provide a comparison to other kinds of entropies, detailed interpretation, discussions of open systems, convergence, and extended simulations including the integrable systems. We show that Observational entropy has a number of desirable properties, is defined out of equilibrium, generically increases with time under unitary evolution (in given cases to the correct thermodynamic value), can be interpreted in terms of macroscopic measurements chosen by an observer, and has a compelling thermodynamic interpretation.

Insofar as Boltzmann entropy gives a measure of the spontaneous increase in disorder in a fully-closed classical system, this paper argues in detail that Observational entropy provides a closely analogous measure in the quantum case.

The main results of this paper are 1) set of definitions that introduces the framework of Observational entropy 2) theorems showing numerous properties of Observational entropy with general coarse-grainings (equivalency to Boltzmann entropy, monotonicity with finer coarse-graining, lower and upper bounds, extensivity, conditions for it to be constant in time, and conditions for it to rise); 3) finding two types of coarse-graining that leads to definition of entropy consistent with thermodynamics (they describe subsystems equilibrating with each other, they rise to the correct thermodynamic value, even in isolated quantum systems described by time-independent Hamiltonian);
4) simulations that support our analytical findings; and 5) discussion and comparison with other entropy measures.

The paper is structured as follows.  Sec.~\ref{sec-definitions} lays out the motivation and basic definitions of the proposed entropy measure, which takes as input both a density matrix and a coarse-graining. Sec.~\ref{sec-properties} enumerates a number of properties this of this entropy, including limiting behaviors, bounds, extensivity, and time-dependence in particular circumstances. Sec.~\ref{sec-entropy_increase} describes what  ``entropy increase'' means in terms of the Observational entropy, provides a simple example in terms of positional coarse-graining, and briefly discusses Observational entropy with more general coarse-grainings. Sec.~\ref{sec-multicoarse} treats the issue, particular to quantum rather than classical physics, of multiple coarse-grainings using variables that may or may not commute, and discusses properties of Observational entropy with multiple coarse-grainings. In Sec.~\ref{sec-thermodynamics} we introduce the two aforementioned thermodynamically-promising Observational entropies  (the factorized Observational entropy, and the Observational entropy of measuring position and energy). We show that these special cases of Observational entropies have number of desirable properties: mainly, they are perfectly defined out of equilibrium, they are bounded by the thermodynamic entropy, and they converge to thermodynamic entropy in the long-time limit, even in genuinely closed quantum systems. Sec.~\ref{sec-simulation} presents numerical simulations of evolving quantum systems that provide evidence for the analytical claims presented in the previous section.  In Section~\ref{sec-comparison} we bring all of this together, comparing with other entropy measures and providing several useful interpretations of the Observational entropy.  Finally, in Sec.~\ref{sec-conclusions} we summarize our results and point out interesting avenues of future research.
A number of mathematical proofs and technical results are left for appendices.

\section{Background and the definition}
\label{sec-definitions}

Let us first describe the partitioning of the state space of a a quantum system into a discrete set of ``macrostates."
Let $\{\P_i\}_i$ be a trace-preserving set of non-zero projectors acting on a Hilbert space $\HS$. This set represents a measurement that an observer can choose to perform on a quantum state. Each element $\P_i$ of the set corresponds to obtaining outcome $i$ from the measurement, and the set itself represent the measurement basis chosen by the observer. This measurement may not be complete: it does not necessarily project onto a pure state. (Equivalently,  there may be some projector $\P_i$ in the set with rank larger than 1.) This set of projectors splits the Hilbert space into smaller subspaces $\HS_i\equiv\P_i\HS \P_i$. The Hilbert space is then a direct sum of these subspaces, $\HS=\bigoplus_i\HS_i$. Any subspace $\HS_i$ may have bigger dimension than one, and this dimension is equal to the rank of the associated projector, which is itself equal to its trace,
\[\label{eq:dim_rank_tr_equivalence}
\mathrm{dim}\HS_i=\rank{\P_i}=\tr[\P_i].
\]
Each subspace $\HS_i$ represents a ``macrostate,''\footnote{These subspaces and the corresponding projectors are denoted ``properties" in the decoherence literature (see e.g. Ref.~\cite{griffiths2014consistent}), which is an approach to coarse-graining broadly similar to that adopted here.} and its dimension can be viewed as its volume. In quantum mechanics the state of the system is described by the density matrix $\R$, which is a positive semi-definite operator acting on the Hilbert space $\HS$. The probability of finding the state of the system in subspace $\HS_i$ is equal to $p_i=\tr[\P_i\R]$.

We now assume that observer has access to information about the system {\em only}  by making measurements of the type represented by the projectors $\P_i$. Thus the observer cannot distinguish between different states inside the subspace $\HS_i$ by learning the outcome of the measurement. We therefore assume that observer considers the quantum state having equal chances of being in any basis state of this subspace, which gives probabilities $\tilde{p}^{(k)}_i=\frac{p_i}{\tr{\P_i}}$, 
$k=1,\dots,\tr{\P_i}$.  (The corresponding assumption in the classical context is sometimes termed ``democracy of states.")

We define the Observational entropy as the Shannon entropy of these modified probabilities:
\begin{definition}\label{def:oe}
Let $\HS$ be a Hilbert space, let $\C=\{\P_i\}_{i}$, $\sum_i \P_i=\hat{I}$, be a trace-preserving set of non-zero projectors, which we call the coarse-graining, and $p_i=\tr[\P_i\R]$ the probability of measuring the density matrix to be in one of the subspaces $\HS_i\equiv\P_i\HS \P_i$. We define \emph{the Observational entropy with coarse-graining} $\C$ as
\[
S_{O(\C)}(\R)=-\sum_{i}p_i\ln \frac{p_i}{\tr\P_i}
\]
where the sum goes over all elements such that $p_i\neq 0$.
\end{definition}

The idea of coarse-grained projections is mentioned very early on by von Neumann~\cite{von2010proof} with an expression similar to this for the particular case of coarse-grained energies, that he attributes to Eugene Wigner. It is mentioned later in his book~\cite{von1955mathematical} for general coarse-grainings, and it has appeared several times after that, for example, in qualitative arguments supporting the emergence of macroscopic behavior~\cite{lebowitz1999microscopic}, in connection with developing a quantum mechanical master equation~\cite{wehrl1978general}, or in connection with fluctuation theorems~\cite{callens2004quantum}. However, this definition by itself does not partition phase space sufficiently to define the equivalent of a coarse-grained classical entropy that is defined for the system out of equilibrium, and corresponds to thermodynamic entropy in equilibrium. For that we need either multiple coarse-grainings, or partitioned coarse-graining, both of which will be introduced later.

The Observational entropy elegantly generalizes the Boltzmann entropy to quantum mechanics.\footnote{For an initial microstate belonging to macrostate of phase-space volume $V_i$, we associate entropy with the Boltzmann entropy $S_B=\ln V_i$. Theorem~\ref{thm:Boltzmann_equivalent} is then in a loose analogy to Boltzmann entropy. However, it is possible to imagine an exact classical analog of the Observational entropy, defined as $S_{BO}(m)=-\sum_{i}p_i\ln \frac{p_i}{\tr V_i}$, where $p_i$ denotes the probability of microstate $m$ being in $i$th macrostate of volume $V_i$. This kind of classical entropy has been previously considered in literature, see~\cite{alonso2017coarse}.} In our notation $\HS_i$ plays the role of a macrostate and $\R$ plays the role of a microstate. According to Boltzmann, if a physical system is in a certain microstate, then the entropy we associate with it is equal to the log of the phase-space volume of the macrostate the microstate belongs to. We obtain a similar statement for the Observational entropy.
\begin{theorem}\label{thm:Boltzmann_equivalent}
(Observational entropy is a quantum equivalent of the Boltzmann entropy)
If the density matrix is contained in one of the subspaces $\HS_i$, i.e., $\P_i\R\P_i=\R$, then $S_{O(\C)}(\R)=\ln {\tr\P_i}=\ln \dim\HS_i$.
\end{theorem}

A key difference between the Observational entropy and the classical Boltzmann entropy is that via superposition, the quantum system can be in a superposition of microstates belonging to different macrostates at the same time.

The above theorem also suggests an additional interpretation of the Observational entropy. Rewriting the Observational entropy as
\[\label{eq-splitup}
S_{O(\C)}(\R)=-\sum_{i}p_i\ln {p_i}+\sum_{i}p_i\ln {\tr\P_i},
\]
we can interpret it as the sum of two contributions.  Consider an observer who chooses to perform a measurement on the system in the basis given by the coarse-graining $\C$. Then the first term represents the expected amount of ``macrosopic" information -- i.e. regarding the macrostates -- that the measurement gives. In other words, this term measures the mean uncertainty as to which macrostate the system is in, and the mean reduction in uncertainty that would occur were the measurements performed. The second term represents the mean value of uncertainty an observer has about the system after the measurement outcome is learned, assuming he or she does not have an ability to distinguish between different microstates in a given macrostate. This can be also seen as follows.

First, note that {\em if} the observer had access to the density matrix $\R$, it would make sense to attribute von Neumann entropy $S(\R)$ to the system.  Then, after obtaining a measurement outcome $i$, the observer would update $\R$ to $\R_i\equiv\frac{\P_i\R\P_i}{\tr(\P_i\R\P_i)}$, and would attribute entropy $S(\R_i)$ to the system, which is lower than $S(\R)$ on average~\cite{gharibyan2014sharpening}.

However, an observer does {\em not} generally have access to the density matrix unless they have knowledge or control over the system's preparation, or access to an ensemble of identically prepared systems and the ability to measure its members in sufficiently many different bases.\footnote{We consider the density matrix to encode all information necessary to make predictions about the system, as well as all of the information {\em in principle} -- but generally not in practice -- extractable by an observer making a sequence of measurements.}
This being so, after measuring outcome $\R_i$ the observer should, following the logic of Boltzmann entropy, attribute to the system the maximally-uncertain density matrix compatible with the knowledge that the system is in macrostate ${\cal H_i}$, and thus attribute to it entropy $\ln\dim{\cal H}_i$.

Considering then the spectrum of possible measurement outcomes, the average uncertainty about the system  after the measurement is then equal to $\sum_{i}p_i\ln {\tr\P_i}$, which is the second contribution in Eq.~\eqref{eq-splitup}.\footnote{Note that the Observational entropy decreases on average when measuring in the basis given by the coarse-graining $\C$. However, that means that the measurement always decreases the entropy. The entropy can increase when the observer obtains a very unlikely outcome $i$, $p_i\ll 1$, which is connected to a large macrostate $\HS_i$. Then  $S_{O(\C)}(\R_i)=\ln \dim\HS_i>S_{O(\C)}(\R)$.}

There are additional ways of interpreting Observational entropy, which are collected in Sec.~\ref{sec-comparison}.

Note that for a single coarse-graining, it is possible to view the Observational entropy as the von Neumann entropy of a density matrix constructed out of the coarse-graining and the probabilities,
\[\label{eq:vonNeumannvsO}
S_{O(\C)}(\R)=S\left(\sum_i\frac{p_i}{\tr\P_i}\P_i\right).
\]
However, we have not found an analogous expression for Observational entropy with multiple coarse-grainings (defined in Section~\ref{sec-multicoarse} below).

\section{Properties}
\label{sec-properties}

We move on to studying the mathematical properties of the Observational entropy. First, we introduce two definitions that will be useful for the theorems to come.
\begin{definition}\label{def:finer_coarse_graining} (Relations between coarse-grainings)
We say that coarse-graining $\C_2$ is {\em finer} than coarse-graining $\C_1$ (and denote $\C_1\hookrightarrow \C_2$) when for every $\P_{i_1}\in\C_1$ there exists an index set $I^{(i_1)}$ such that $\hat{P}_{i_1}=\sum_{i_2\in I^{(i_1)}}\P_{i_2}$, $\hat{P}_{i_2}\in \C_2$. (That is, each element of $\C_1$ can be partitioned using elements of $\C_2$.) We correspondingly say that coarse-graining $\C_1$ is {\em rougher} than coarse-graining $\C_2$. Two coarse-grainings $\C_1$ and $\C_2$  are said to commute when for all $i_1$ and $i_2$, $[\P_{i_1},\P_{i_2}]=0$.
\end{definition}
Note that relation $\hookrightarrow$ represents a partial order on the set of all coarse-grainings. There is a common maximal element $\C_{\hat{I}}=\{\hat{I}\}$, i.e., $\C_{\hat{I}}\hookrightarrow \C$ for every coarse-graining $\C$.

\begin{definition}\label{def:c_g_by_observable} (Coarse-graining defined by an observable)
Let $\hat{A}=\sum_ia_i\P_{a_i}$, where $a_i\neq a_j$ for $i\neq j$, be the spectral decomposition of an observable\footnote{More generally, we can use any Hermitian matrix $\hat{A}$ to define a coarse-graining since every Hermitian matrix has a spectral decomposition. Later in Theorems~\ref{thm:bounded} and \ref{thm:bounded_multiple} we will use coarse-graining $\C_\R$ given by the density matrix $\R$ although the density matrix is not usually considered as to be an observable in a physical sense.} $\hat{A}$. We define coarse-graining given by the observable $\hat{A}$ as ${\C_{\hat{A}}=\{\P_{a_i}\}_{a_i}}$. We say that coarse-graining $\C$ commutes with an observable $\hat{A}$ if $[\P_k,\hat{A}]=0$ for all $\P_k\in\C$.
\end{definition}
The spectral decomposition is unique when written in terms of projectors associated with different eigenvalues, therefore also coarse-graining $\C_{\hat{A}}$ is uniquely defined. If coarse-graining $\C$ commutes with an observable $\hat{A}$ then the coarse-graining $\C$ also commutes with $\C_{\hat{A}}$. This is because two Hermitian operators commute if and only if projectors from their spectral decomposition commute~\cite{blank2008hilbert}.

Intuitively, we would expect that an observer with access to more information about a quantum system will attach a lower entropy to this system. This is described in the following theorem, which says that the observer with a higher resolution in measurement (a finer coarse-graining) attaches a lower entropy to the same density matrix.

\begin{theorem}\label{thm:monotonic} (Observational entropy is a monotonic function of the coarse-graining.)
If $\C_1\hookrightarrow \C_2$ then
\[
S_{O(\C_1)}(\R)\geq S_{O(\C_2)}(\R).
\]
\end{theorem}

It turns out that the Observational entropy is a bounded function: it is bounded below by the von Neumann entropy $S(\R)$ and above by the logarithm of the volume of the entire Hilbert space. This shows that the von Neumann entropy represents the ultimate knowledge that can be obtained about a quantum state. After achieving a certain resolution no better resolution can help to obtain better information.

\begin{theorem}\label{thm:bounded} (Observational entropy is bounded.)
\[
S(\R)\leq S_{O(\C)}(\R)\leq \ln \mathrm{dim}\HS
\]
for any coarse-graining $\C$ and any density matrix $\R$. $S(\R)= S_{O(\C)}(\R)$ if and only if $\C_{\R}\hookrightarrow \C$. $S_O(\R)=\ln \mathrm{dim}\HS$ if and only if $\forall i$, $p_i=\frac{\tr \P_i}{\mathrm{dim}\HS}$.
\end{theorem}
$C_{\R}$ is the coarse-graining given by the density matrix $\R$, as defined in Def.~\eqref{def:c_g_by_observable}. Let us explain the equality conditions. The Observational entropy is the same as the von Neumann entropy when the measurement represented by coarse-graining $\C$ is performed in the eigenbasis of the density matrix, and the resolution of the measurement is sufficient to distinguish between different eigenvalues of the density matrix. In other words, the two entropies are equal when a measurement performed on an eigenvector of the density matrix is enough to predict the eigenvalue associated with this eigenvector. On the other hand, the entropy is maximal when probabilities of obtaining measurement result $i$ are linearly proportional to the volumes of the respective subspaces $\HS_i$.

The previous theorem also suggests that obtaining the maximal Observational entropy is largely unrelated to the purity of the state.  The Observational entropy does not distinguish between pure and mixed states, but only between different probability distribution of measurement outcomes. In other words, it distinguishes only between different probability
 distributions over macrostates given by the coarse-graining. This is illustrated in the following corollary.
\begin{corollary}\label{thm:pure_state_max_entropy} (Pure states can achieve the maximal entropy.)
Both $\R=\pro{\psi}{\psi}$, $\ket{\psi}=\sum_i\sqrt{\frac{\tr \P_i}{\mathrm{dim}\HS}}\ket{\psi_i}$ where $\ket{\psi_i}\in \HS_i$, and $\R_{\mathrm{id}}=\frac{1}{\mathrm{dim}\HS}\hat{I}$ give the same maximal entropy,
\[
S_{O(\C)}(\R)=S_{O(\C)}(\R_{\mathrm{id}})=\ln \mathrm{dim}\; \!\HS.
\]
\end{corollary}

Any entropy describing a physical system should be extensive. This is because an entropy measures the amount of uncertainty about a physical system, and the total amount of this uncertainty should not change just because we consider two or more  systems at the same time. Thermodynamic entropy is extensive, and we would expect that if interactions between systems are sufficiently weak or local, the total uncertainty about them should not change by considering them combined into a single system. The following theorem says that the Observational entropy is extensive, i.e., it is additive on separable states.
\begin{theorem}\label{thm:extensivity} (Extensivity)
Let $\HS=\HS^{(1)}\otimes\cdots\otimes\HS^{(m)}$ be a Hilbert space of a composite system with the coarse-graining defined as $\C=\C^{(1)}\otimes\dots\otimes \C^{(m)}=\{\P_{i_1}\otimes\dots\otimes\P_{i_m}\}_{i_1,\dots,i_m}$. For a separable state $\R=\R^{(1)}\otimes\cdots\otimes\R^{(m)}$ we have
\[
S_{O(\C)}(\R)=\sum_{k=1}^mS_{O(\C^{(k)})}\left(\R^{(k)}\right).
\]
Thus $S_{O(C)}$ is an extensive quantity.
\end{theorem}

When an observer decides to track degrees of freedom that do not change in time, we would expect the observer to see no change in entropy. This is described by the following theorem.
\begin{theorem}\label{thm:constant_entropies} (Constant Observational entropies)
Let $\hat{H}$ be a Hamiltonian governing the evolution of the system, $\R_t=U(t)\R_0 U(t)^\dag$, $U(t)=e^{-i\hat{H}t}$. If coarse-graining $\C$ commutes with the Hamiltonian then
\[
S_{O(\C)}(\R_t)=S_{O(\C)}(\R_0)=\mathrm{const.}
\]
Specifically, if $[\hat{A},\hat{H}]=0$ then $S_{O(\C_{\hat{A}})}(\R_t)=S_{O(\C_{\hat{A}})}(\R_0)=\mathrm{const.}$
Thus if a coarse-graining is defined by an observable that is a conserved quantity, the Observational entropy remains constant in time for any choice of the initial state.
\end{theorem}

\begin{figure}[t!]
\centering
\begin{tabular}{cc}
a) &
 \\
b) &
\includegraphics[width=0.8\linewidth]{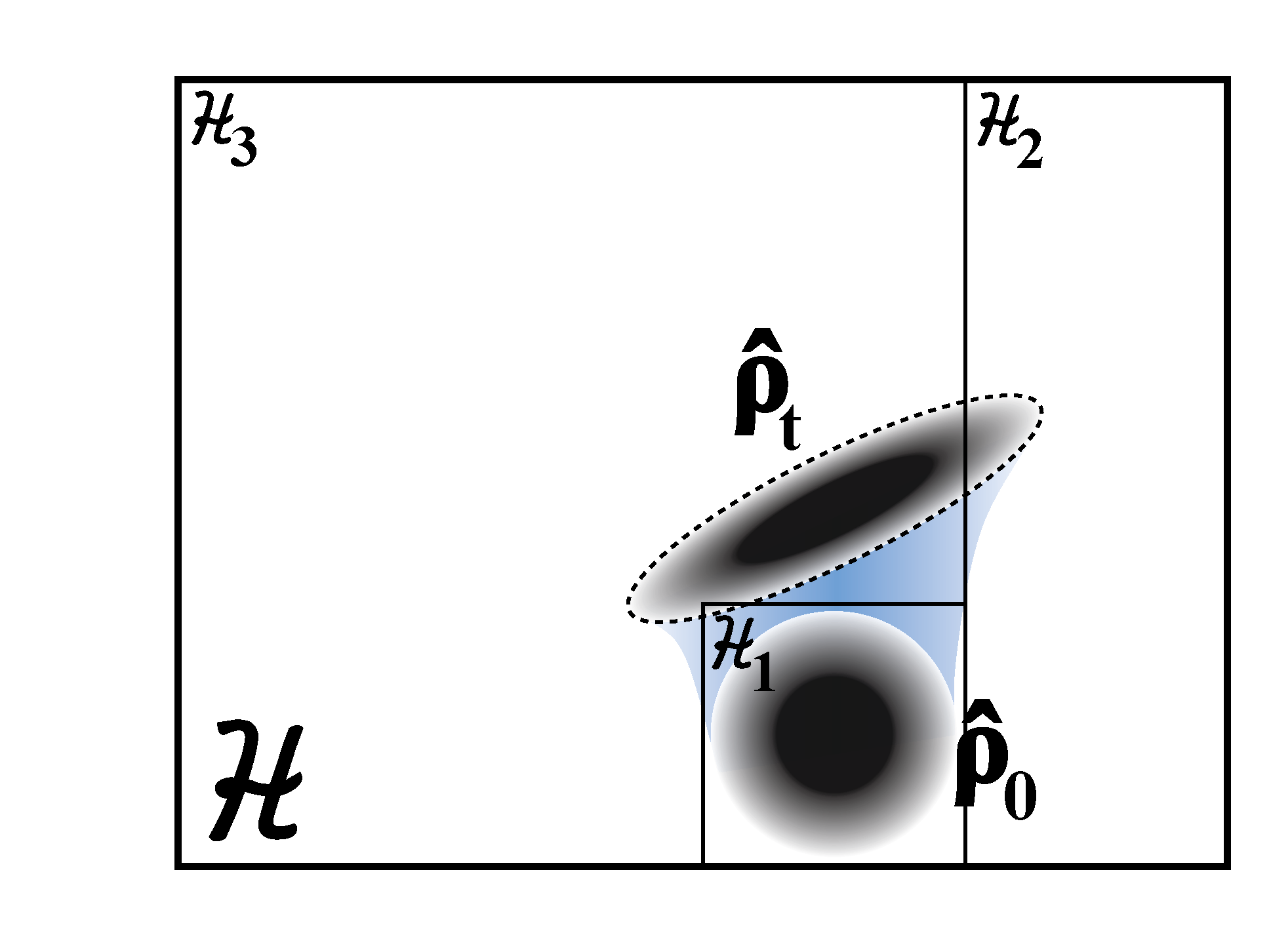} \\
c) & \includegraphics[width=0.5\linewidth]{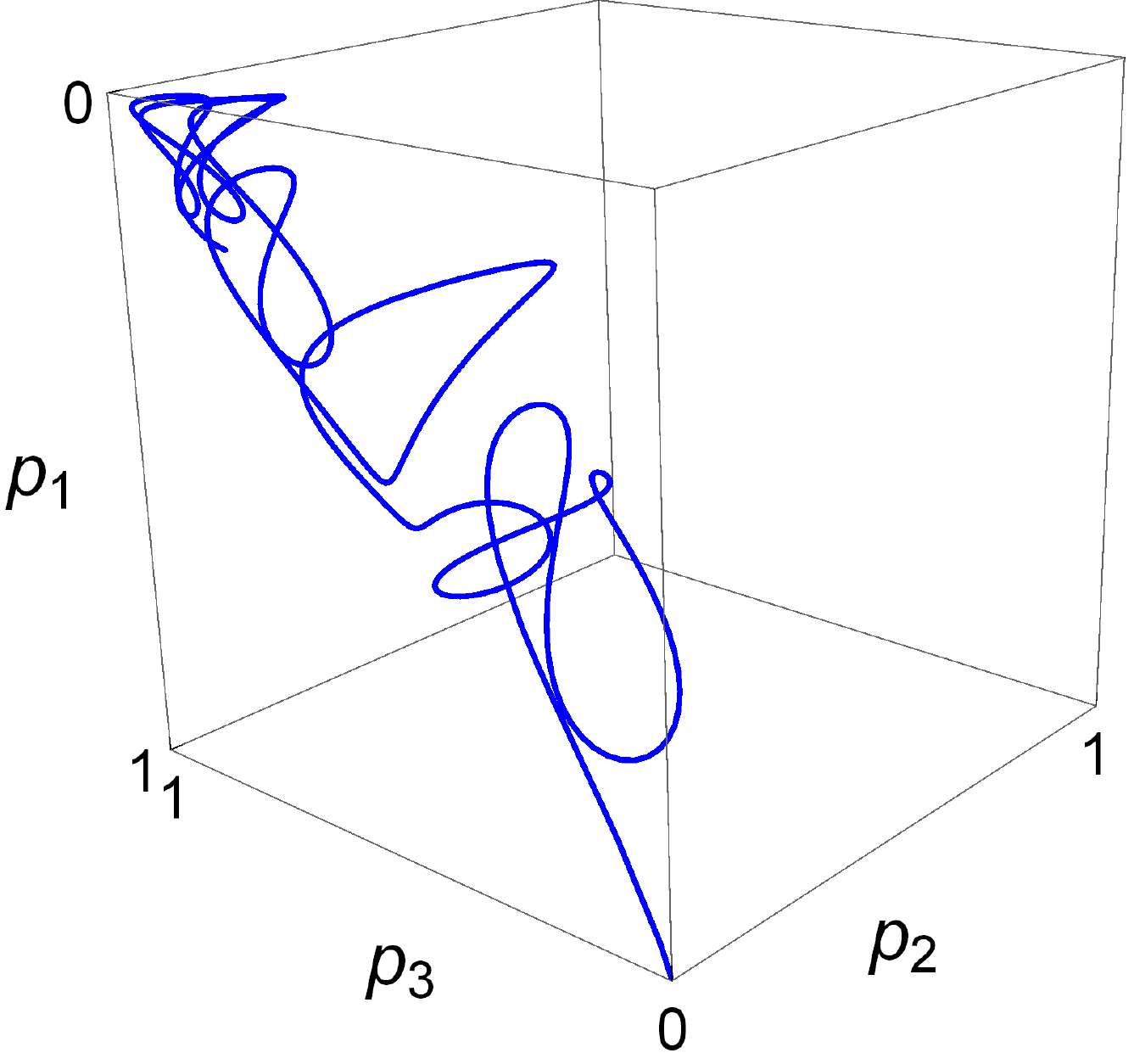}\\
 & \hspace{-2em}\includegraphics[width=1\linewidth]{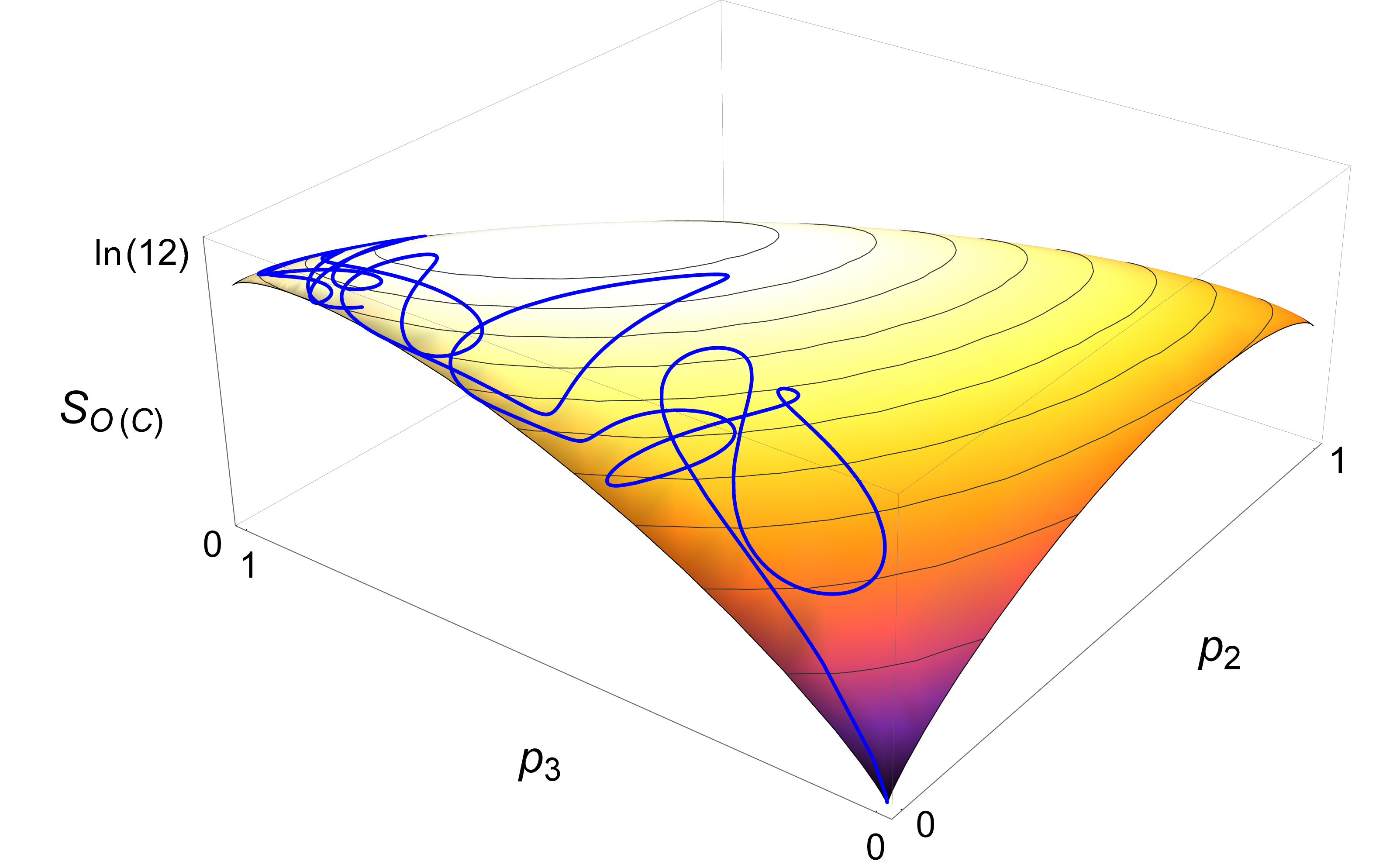}
 \end{tabular}\caption{a) Sketch of Hilbert space, b) evolution of probabilities, and c) graph of Observational entropy, in 12-dimensional Hilbert space with subspaces of dimensions 1, 3, and 8 respectively. Blobs in a) represent the amount of probability projected into each Hilbert subspace, but it should be kept in mind that the right picture is projecting a density matrix that lives in 12-d space into these lower-d subspaces; this cannot be depicted here. The blue curve in b) represents a possible evolution of probabilities $p_i(t)=\tr[\P_i\R_t]$, with density matrix starting in the 1-d subspace Hilbert subspace $\HS_1$. Panel c) depicts Observational entropy as function of probabilities $p_2$ and $p_3$ (where $p_1=1-p_2-p_3$), and the blue curve is the corresponding entropy $S_{O(C)}(\R_t)$ from evolution b). Observational entropy is a strictly concave function; since each corner of its graph represents one of the subspaces, the entropy must increase at least for a short time when starting in one of them. }\label{fig:scheme}
\end{figure}

In later sections we extensively investigate the tendency of Observation entropy to be non-decreasing under generic time evolution.  Here we give relevant analytic result that applies when starting in a macrostate.

\begin{theorem}\label{second_law_thermo} (Non-decreasing entropy for a macrostate over short times)
If the density matrix is initially contained in one of the subspaces $\HS_i$, i.e., $\P_i\R_0\P_i=\R_0$, then
\[
S_{O(\C)}(\R_{t})\geq S_{O(\C)}(\R_{0})
\]
for any $t$ that is small enough to satisfy
\[
t\lessapprox \left(\tr\big[(\hat{I}-\P_i)\hat{H}\R_0\hat{H}\big]\left(1+\frac{\tr[\P_i]}{\min_{j\neq i}\tr[\P_j]}\right)\right)^{-\frac{1}{2}}.
\]
\end{theorem}
The approximate relation $\lessapprox$ comes from the proof where we have neglected terms of order $O(t^4)$ and higher, in the big-O notation. However, an exact statement of a different form is obtained there. Why this increase happens becomes clear when looking at the general shape of the Observational entropy as depicted on Fig.~\ref{fig:scheme}.

Note that this property is not strictly true of classical Boltzmann entropy (defined as the log of the number/volume of microstates in a macrostate): starting in a given macrostate, there is a small probability of \emph{immediately} evolving into a macrostate of lower entropy. For Observational entropy, the corresponding potential decrease due to evolution into a macrostate of a lower volume, which would be in the second term of Eq.~\eqref{eq-splitup}, is more than compensated for by the increase in uncertainty in the macroscopic information (the first term of that equation), leading to a net entropy increase, at least for a short time.

\section{Entropy increase for general coarse-graining}
\label{sec-entropy_increase}

As an example, we consider Observational entropy with a position coarse-graining. Given a one-dimensional system with $N$ indistinguishable particles,
we can coarse-grain them into
$p$ bins, each of width $\delta$. We wish to make observations that will give us the bin that every particle is in. To do this, we denote the binned particle positions
by $\vec{x}=(x^{(1)},\dots,x^{(N)})$, where each element can take one of the equally spaced values $x_1,\dots,x_p$.\footnote{This example considers 1-dimensional lattice. We could, of course, consider a more general example, such as three dimensional lattice. Then for each particle we would attach a 3-dimensional vector, for example, for the second particle contained in the bin at the bottom corner of the lattice, we could write $x^{(2)}=(x_1,x_1,x_1)$.} For example, when the second particle is contained in the first bin, we write $x^{(2)}=x_1$. Vector $\vec{x}$ therefore contains information about to which bin every particle belongs. For indistinguishable particles, any permutation $\pi$ of elements of $\vec{x}$ constitutes the same vector, $\vec{x}\equiv\pi(x^{(1)},\dots,x^{(N)})$. We define a set of coarse-grained projectors indexed by $\vec x$,
\[\label{eq:x_projectors}
\C_{{\hat{X}}^{(\delta)}}=\{\P_{\vec x}^{(\delta)}\}_{\vec x}, ~ {\rm where} ~ \P_{\vec{x}}^{(\delta)}=\sum_{\vec{\tilde{x}} \in C_{\vec{x}}} \pro{\vec{\tilde{x}}}{\vec{\tilde{x}}}
\]
and $C_{\vec{x}}$ represents a hypercube of dimension $N$ and width $\delta=x_{j+1}-x_j$. Vector $\ket{\vec{\tilde{x}}}$ contains the exact position of each particle, and corresponds to a basis vector in the Hilbert space. Each hypercube defines one macrostate, which by the above definition is formed by vectors of position $\ket{\vec{\tilde{x}}}$ that correspond to the same vector of positional bins $\vec{x}$. Our coarse-graining $\C_{{\hat{X}}^{(\delta)}}$ then represents
measurements that can be done that would characterize the system positional macrostate at a scale of
$\delta$.

For indistinguishable particles, this coarse-graining can be also understood as follows. It coarse-grains the space into boxes, and counts the number of particles in each box. For example, on a one-dimensional lattice of length $L=9$ of with $N=4$ indistinguishable particles coarse-grained into $p=3$ boxes of size $\delta=3$, the first particle could be in the box $\{1-3\}$, the next two could be in box $\{4-6\}$, and the final one
could be in the box $\{7-9\}$. This represents one projector of this coarse-graining by the ``signature" $[1, 2, 1]$,
which represents the number of particles in each box. The set of coarse-grainings projectors $C_{{\hat{X}}^{(\delta)}}$ is isomorphic
to the set of allowed signatures, $[4,0,0]$, $[3,1,0]$, etc. In other words, these projectors represent a measurement that measures number of particles in each box, and we can write  $\C_{{\hat{X}}^{(\delta)}}=\C_{\hat{N}_1\otimes\cdots\otimes \hat{N}_p}\equiv \C_{\hat{N}_1}\otimes\cdots\otimes \C_{\hat{N}_p}$. When a projector $\P_{\vec x}\in\C_{{\hat{X}}^{(\delta)}}$ acts on a
wavefunction, it is projecting out the components of the wavefunction with $\P_{\vec x}$'s signature.

\begin{figure}[t]
\centering
\includegraphics[width=1\hsize]{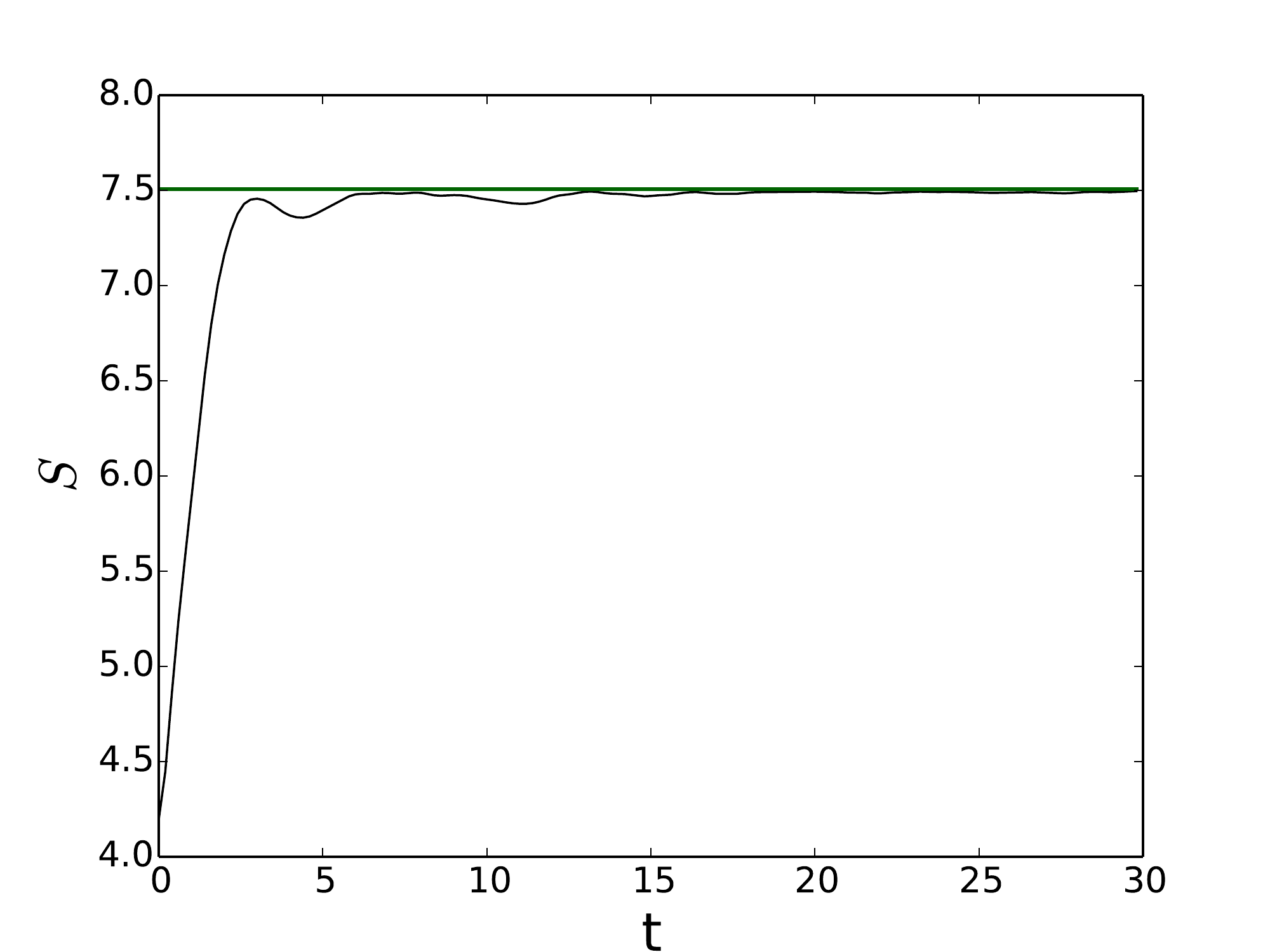}
\caption
{
Evolution of the Observational entropy with positional coarse-graining of the non-integrable system of size $L=16$, coarse-grained into $p=4$ parts of size $\delta=4$, starting at $t=0$ in a state of $N=4$ particles contained in the left side of the box (sites $\{1-8\}$). As time passes, particles expand through the entire box and the Observational entropy quickly increases, reaching value not far from the maximal value $S_{\max}=\ln \dim \HS$, where $\dim \HS={{L}\choose{N}}=1820$, depicted by the straight green line.
}
\label{fig:position_coarse_graining}
\end{figure}

We illustrate evolution of the Observational entropy with this positional coarse-graining in Fig.~\ref{fig:position_coarse_graining}, starting in a state that is confined to the first half of the lattice, subject to the Hamiltonian describing a non-integrable system, Eq.~\eqref{FermionHam}, that will be discussed later in detail. As we can see, particles quickly spread over the entire Hilbert space, filling it almost uniformly, and approximating the maximal value $S_{\max}=\ln \dim \HS$.\footnote{However, it is important to note that closed quantum systems are not in general ergodic in classical sense, and the Observational entropy does not usually reach the maximal value $S_{\max}=\ln \dim \HS$.  For example, starting from an energy eigenstate, this state never evolves and therefore the Observational entropy remains constant for any choice of coarse-graining. More precisely, the wave function of a closed system is contained on the surface of the hyper-sphere that is given by the decomposition of the wave function into eigenvectors of the Hamiltonian, and the system is then ergodic on this hypersphere~\cite{von2010proof,goldstein2010long}. The details of quantum ergodicity are still a topic of ongoing research~\cite{klein1952ergodic,bocchieri1958ergodic,goldstein2010normal,brody2007unitarity,asadi2015quantum,zhang2016ergodicity,dymarsky2018subsystem}.} This Observational entropy measures how uniformly distributed the particles are over the macrostates. The fact that it almost reaches the maximal value means that in the long-time limit, probability of each particle being in a given macrostate is linearly proportional to the macrostate's volume, as shown by Theorem~\ref{thm:bounded}.

The question of {\em to what value} the Observational entropy increases is interesting and not fully resolved for general coarse-grainings of closed quantum systems such as the system considered in the above example.

For open systems, which we model as systems interacting weakly with a thermal bath, the convergence can be readily calculated. We allow the system to exchange (for example) energy, number of particles, momentum, and angular momentum with the thermal bath. Considering the system and the thermal bath being a closed system as a whole, according to strong eigenstate thermalization hypothesis~\cite{garrison2015does}, at some long time in the future, the system of interest is very likely to be a in state that for all practical purposes closely resembles the generalized thermal density matrix  $\R_{\mathrm{th}}$ introduced in Ref.~\cite{jaynes1957information2}. We can therefore write
\[\label{eq:convergence_in_open_systems}
\R_t\overset{t\rightarrow \infty}{\leadsto}\R_{\mathrm{th}}\equiv \frac{1}{Z}\exp\Big(-\sum_j\lambda_j\hat{A}_j\Big), 
\]
where $\hat{A}_j$ describes the observable of a conserved quantity (such as energy or a particle number),  $\lambda_j$ its respective Lagrange multiplier (such as inverse temperature or a chemical potential), and $Z$ the partition function. The relation symbolized by $\overset{t\rightarrow \infty}{\leadsto}$ should be read as ``approximately approaches'' and might not be strictly a convergence in the mathematical sense. For example, if we consider the system and the thermal bath together as being a closed system, there is a timescale over which the closed system returns arbitrarily closely to the initial state~\cite{bocchieri1957quantum} --  the Poincar\'e recurrence time. Hence, this convergence should be rather understood in a physical sense that the system spends exponentially more time at this entropy than at lower entropies.

For such an open system, the probabilities of finding the density matrix in the Hilbert space $\HS_i$ for a general coarse-graining $\C$ will then approach a value given by this limiting density matrix,
\[\label{eq:hypothesis3}
p_{i}(t)\equiv\tr[\P_i\R_t]\overset{t\rightarrow \infty}{\leadsto}p_{i}^{(th)}=\tr[\P_i \R_{th}].
\]
The time-dependent Observational entropy
\[\label{eq:observational_entropy_evolution}
S_{O(\C)}(\R_t)=-\sum_{i}p_i(t)\ln \frac{p_i(t)}{\tr\P_i}
\]
then converges (in the same sense) to the value
\[\label{eq:limiting_behavior_entropy}
S_{O(\C)}(\R_t)\overset{t\rightarrow \infty}{\leadsto}S_{O(\C)}^{(th)}=-\sum_{i}p_{i}^{(th)}\ln \frac{p_{i}^{(th)}}{\tr\P_i}.
\]

The situation in closed quantum systems, which is the main topic of this paper, is more difficult to analyze, since the density matrix does not approach the generalized canonical density matrix $\R_{th}$. In a closed quantum system the amplitudes of the wave-function written in an eigenbasis of the Hamiltonian stay fixed, and only the respective phases change. The relevant figure of merit is then the micro-canonical ensemble rather than canonical, and the situation is described by arguments similar to the eigenstate thermalization hypothesis, as we will discuss in detail below in Section~\ref{sec-thermodynamics} and Appendices~\ref{sec-FOE_for_eigenstates} and \ref{sec-S_xE_for_convergence}. Generally, in closed quantum systems the Observational entropy will tend to a limiting value that depends on the initial state, although entropy for many initial states will converge to a similar value.

Although not much can be said about the convergence of entropy in closed systems for general coarse-grainings, we have observed that even in closed quantum systems the Observational entropy tends to increase for most initial states.  And as shown by Theorem~\ref{second_law_thermo}, it always increases or stays the same, at least for a short time, for states that start their evolution contained in one of the macrostates. One instance of where the entropy increases is shown in Figure~\ref{fig:CoarseGrainedEigenstate}, and such an increase can be observed for almost any initial state. It is true that one can find states where the entropy decreases, but such states are rare and have to be either carefully designed (and fitted to the particular coarse-graining) or found as rare cases in a large number of random trials.\footnote{The statistical characterization of downward fluctuations in Observational entropy is a topic of current investigation by the authors.}

We can now further discuss the meaning of the Observational entropy. An Observer chooses a coarse-graining $\C=\{\P_i\}_i$ that defines macrostates of interest. This choice may be given for example by a coarse-grained measurement that can be in principle performed, or by coarse-grained (macroscopic) degrees of freedom that the observer wants to track. The time-dependent Observational entropy, Eq.~\eqref{eq:observational_entropy_evolution}, then describes the increasing amount of disorder in the system with respect to these chosen macrostates.

Low Observational entropy means that the state of the system is localized in a few small macrostates. The observer conceives of such a situation as a highly ordered state in his or her subjective point of view. This can be seen as an ability to say a lot about the system without the actual knowledge of the density matrix.

High entropy means that the state of the system is contained within a large macrostate, or spans across many small macrostates. Even though the system might be in a pure state, the fact that this pure state cannot be localized in a few small macrostates means that the such a system is considered as disordered. With such a disordered system, only a little can be said about the system from the observer's perspective. Given the arbitrary choice of macrostates, this entropy may or may not be connected to any particular {\em thermodynamic} quantity; this depends entirely on the choice of coarse-graining.

When the Observational entropy achieves (approximately) its maximum, we say that the system has thermalized with respect to coarse-graining $\C$. Growth of this entropy describes the loss of perceived order due to the time evolution. The state of the system spreads into more and larger macrostates, and the observer loses the ability to say much about the system as time passes.

The exception to this is if the observer chooses to track the degrees of freedom that do not change in time, i.e., $[\C,\hat{H}]=0$. According to Theorem~\ref{thm:constant_entropies}, the Observational entropy then remains constant. However, choosing macrostates that lead to rising entropy is often unavoidable and/or desirable, for example when coarse-grained quantities represent collective degrees of freedom in which the system can be understood well, or
when coarse-grained quantities determine the amount of work that can be extracted, or when we want to describe thermalization between initially separated systems. We will detail the latter in Section~\ref{sec-thermodynamics}.

There is also one important point to make. In infinite-dimensional Hilbert spaces, such as that of the quantum harmonic oscillator, there can be subspaces with infinite dimension. Clearly, if there is a non-zero probability that the density matrix belongs in such a subspace, then Observational entropy is formally infinite, indicating an infinite amount of uncertainty about the particular state of the system if {\em only} probabilities of macrostates are known. An infinite-dimensional Hilbert space is very unlikely to be relevant in real physical systems,\footnote{The Beckenstein bound,\cite{bekenstein1981universal} for example, indicates that for a system to have an infinite entropy it must be infinite in either extent or energy, and this entropy is often interpreted to refer to the ($\log$ of the) number of accessible microstates.} but it is desirable to apply our formalism to an in-principle infinite Hilbert space nonetheless. This can be achieved by choosing a coarse-graining in such a way that that none of the subspaces is infinite dimensional. In many physical situations, the high-dimensional subspaces will be then exponentially suppressed by low probabilities $p_i$. This is for example the case of the physically relevant entropies that we are going to introduce in Section~\ref{sec-thermodynamics}, which are provably bounded from above. There, the space is coarse-grained  in energy, and even though the Hilbert space can be in principle infinite dimensional, the high energy subspaces are exponentially suppressed due to the constraints on energy of the initial state. In such situations, it can be desirable to truncate the infinite-dimensional Hilbert space, and stop considering subspaces with low enough probabilities $p_i$. Then the Observational entropy on this truncated Hilbert space will approximate the Observational entropy on the full space. (It is straightforward to show that when $p_i=0$ for subspaces that have been taken away, both Observational entropies coincide.)

But before we take a closer look at the use of Observational entropy in thermodynamics, we first have to introduce Observational entropy with several coarse-grainings.

\section{Observational entropy with multiple coarse-grainings}
\label{sec-multicoarse}

Imagine an observer with the ability to perform two distinct measurements on the system. In an ideal case the second measurement would provide additional information about the system that the observer was unable to obtain from the first measurement. Since each measurement is represented by a coarse-graining, two measurements are represented by two coarse-grainings. Considering these two coarse-grainings together should give rise to a new definition of the Observational entropy that describes this additional ability to perform the second measurement. We first discuss the simple case when two coarse-grainings nicely fit together, i.e., they commute. We then generalize the definition of Observational entropy to non-commuting coarse-grainings.
\begin{definition}
Let $\C_{1}$ and $\C_{2}$ commute. We define the joint coarse-graining $\C_{1,2}$ as the roughest coarse-graining that is finer than both $\C_{1}$ and $\C_{2}$, i.e., $\C_{1,2}$ is such that
\[\label{def:joint_coarsegraining}
\C_{1}\hookrightarrow \C_{1,2},\  \C_{2}\hookrightarrow \C_{1,2},\ \mathrm{and}\
(\forall \C| \C_{1}\hookrightarrow \C, \C_{2}\hookrightarrow \C)(\C_{1,2}\hookrightarrow \C).
\]
\end{definition}

The joint coarse-graining always exists for any set of commuting coarse-grainings, and is uniquely defined by them, per the following lemma.

\begin{lemma}\label{thm:joint_coarsegraining}
Let coarse-grainings $\C_1=\{\P_{i_1}\}_{i_1}$ and $C_2=\{\P_{i_2}\}_{i_2}$ commute. Then the joint coarse-graining is uniquely defined and it is given by $\C_{1,2}=\{\P_{i_1}\P_{i_2}\}_{{i_1},{i_2}}\!\setminus\!\{0\}$, where $\setminus\{0\}$ means that the zero element has been taken out.
\end{lemma}

This lemma says that that elements of the joint coarse-graining $\C_{1,2}$ are given by products of elements of coarse-grainings $\C_1$ and $\C_2$. Although the joint coarse-graining $\C_{1,2}$ has been constructed from two different coarse-grainings, it is still a single coarse-graining. Therefore, we can use the above lemma and Definition~\ref{def:oe} to compute the Observational entropy with the joint coarse-graining as
\[\label{eq:joint_coarse_entropy}
S_{O(\C_{1,2})}(\R)=-\sum_{i_1,i_2}p_{i_1i_2}\ln \frac{p_{i_1i_2}}{\tr[{\P_{i_1}\P_{i_2}}]}.
\]
The sum goes over all elements such that $p_{i_1i_2}=\tr[{\P_{i_1}\P_{i_2}}\R]\neq 0$.

When the coarse-grainings do not commute,\footnote{The decoherence/consistent histories approach explicitly eschews non-commuting coarse-grainings~\cite{griffiths2014consistent}. However this seems overly limiting, as our usual macroscopic description of the world is quite comfortable with coarse-grained variables that do not technically commute but for which the non-commutation is a tiny effect in the large-mass or many-particle limit.  Thus where possible we attempt to extend our results to the non-commuting case.} the issue becomes somewhat more complicated. It can be easily shown that non-commuting coarse-grainings do not have a joint coarse-graining,\footnote{We assume that $\P_{i_1}\in \C_1$, $\P_{i_2}\in \C_2$ do not commute and that $\C_{1,2}=\{\P^{1,2}_k\}_k$ is the joint coarse-graining. Then $\P_{i_1}=\sum_{k\in I^{(i_1)}}\P^{1,2}_k$ and $\P_{i_2}=\sum_{k\in I^{(i_2)}}\P^{1,2}_k$. Because $\P_{i_1}$ and $\P_{i_2}$ do not commute, then there must exist projectors $\P^{1,2}_k$ and $\P^{1,2}_{\tilde{k}}$ that also do not commute, which contradicts with the fact that they are orthogonal projectors.} and do not correspond to a single direct sum of subspaces, hence the original Definition~\ref{def:oe} of the Observational entropy cannot be directly applied. However, Eq.~\eqref{eq:joint_coarse_entropy} motivates a more general definition that applies even for coarse-grainings that do not commute. The most compelling way\footnote{Considering Eq.~\eqref{eq:dim_rank_tr_equivalence}, other possible ways that could be considered are: (1) taking the generalization of the dimension of the corresponding subspace; this fails, because projectors that do not commute do not correspond to a subspace. (2) taking the rank of projector $\P_{i_1}\P_{i_2}$ as the volume; this is not desirable definition, since these ranks do not add the volume of the entire Hilbert space in non-commuting case. (3) An alternative way is to make use of von Neumann entropy of the reduced state $\frac{{\P_{i_2}\P_{i_1}\R\P_{i_1}\P_{i_2}}}{\tr[\P_{i_2}\P_{i_1}\R\P_{i_1}\P_{i_2}]}$, but this does not connect well with the interpretation of the volume and does not yield desirable properties. Generalizing the volume using trace $\tr[{\P_{i_2}\P_{i_1}\P_{i_2}}]$ is then the most meaningful, and it yields many desirable properties.} is to simply take $\tr[{\P_{i_2}\P_{i_1}\P_{i_2}}]$ and the corresponding generalization of probabilities $p_{i_1i_2}=\tr[\P_{i_2}\P_{i_1}\R\P_{i_1}\P_{i_2}]$. This is because $V_{i_1,i_2}\equiv\tr[{\P_{i_2}\P_{i_1}\P_{i_2}}]$ has a clear interpretation as a volume of a small part of the Hilbert space --- a multi-macrostate $(i_1,i_2)$. It is always positive and it sums up to the volume of the entire Hilbert space, $\sum_{i_1,i_2}V_{i_1,i_2}=\dim\HS$. $p_{i_1i_2}$ represents the probability of obtaining result $i_1$ in the first measurement while obtaining result $i_2$ in the second measurement when two consequent measurements in bases $\C_1$ and $\C_2$ are performed on the state described by the density matrix $\R$. Moreover, this definition gives an intuitive answer to the thermodynamical behavior that we describe in the next section. This discussion leads to the following definition.
\begin{definition}\label{def:o_entropy_general}
Let $(\C_1,\dots,\C_n)$ be an ordered set of coarse-grainings, and let $V_{i_1,\dots,i_n}\equiv \tr[\P_{i_n}\cdots\P_{i_1}\cdots\P_{i_n}]$ denote volume of macrostate $(i_1,\dots,i_n)$. We define \emph{Observational entropy with coarse-grainings $(\C_1,\dots,\C_n)$} as
\[
S_{O(\C_1,\dots,\C_n)}(\R)=-\!\!\sum_{i_1,\dots,i_n}\!\!p_{i_1,\dots,i_n}\ln \frac{p_{i_1,\dots,i_n}}{V_{i_1,\dots,i_n}},
\]
where the sum goes over elements such that $p_{i_1,\dots,i_n}=\tr\big[\P_{i_n}\cdots\P_{i_1}\R\P_{i_1}\cdots\P_{i_n}\big]\neq 0$.
\end{definition}
When the coarse-grainings commute, this definition coincides with commuting case, i.e., $S_{O(\C_1,\dots,\C_n)}(\R)=S_{O(\C_{1,\dots,n})}(\R)$. (Note the admittedly subtle notational difference.) However, it is important that the order of the coarse-grainings matters when they do not commute: in general $S_{O(\C_1,\C_2)}(\R)\neq S_{O(\C_2,\C_1)}(\R)$. We will see an example of Observational entropy implementing two different coarse-grainings in the next section.

As is the case for the Observational entropy with a single coarse-graining, the more general version is bounded by the same values. To show this, we first need to generalize Def.~\ref{def:finer_coarse_graining} and define a finer set of coarse-grainings.
\begin{definition}\label{def:finer_set_coarse_graining}~(Finer set of coarse-grainings)
We say that an ordered set of coarse-grainings $(\C_1,\dots,\C_n)$ is finer than coarse-grainings $\C$ (and denote $\C\hookrightarrow (\C_1,\dots,\C_n)$) when for every multi-index $\bi=(i_1,\dots,i_n)$ exists $\P_j\in\C$ such that\footnote{Looking at Eq.~\eqref{eq:finer_set_condition}, it is also suggestive to say that $\P_j$ ``dissolves'' in $\P_{i_n}\cdots\P_{i_1}$, or that coarse-graining $\C$ ``dissolves'' in the set $(\C_1,\dots,\C_n)$.}
\[\label{eq:finer_set_condition}
\P_{i_n}\cdots\P_{i_1}\P_j=\P_{i_n}\cdots\P_{i_1},
\]
where $\P_{i_k}\in \C_k$, $k=1,\dots,n$.
\end{definition}
Intuitively, the set of coarse-grainings $(\C_1,\dots,\C_n)$ is finer than coarse-graining $\C$, if each element $\P_{i_n}\cdots\P_{i_1}$ from the set projects into one of the subspaces $\HS_j$ given by the projectors from coarse-graining $\C$, before possibly projecting somewhere else. In other words, measuring the system first in the basis given by coarse-graining $\C$ is redundant, if the sequence of measurements given by the set $(\C_1,\dots,\C_n)$ is performed afterwards. This is because all information the first measurement could provide, will be also obtained just by performing the sequence of measurements.

For $\P_{i_n}\cdots\P_{i_1}\neq 0$ in the Def.~\ref{def:finer_set_coarse_graining}, the index $j$ that is mapped to each multi-index $\bi$ is unique, i.e., $\bi\rightarrow j$ forms a map. We also note that if $\C\hookrightarrow (\C_1,\dots,\C_n)$, then for any additional coarse-graining $\C_{n+1}$, also $\C\hookrightarrow (\C_1,\dots,\C_n,\C_{n+1})$. The reader may have noticed that Def.~\ref{def:finer_set_coarse_graining} does not look very similar to Def.~\ref{def:finer_coarse_graining}. Despite this, these two definitions coincide for $n=1$. We show all of these properties in Appendix~\ref{app:finer_definition}.

The theorem follows:
\begin{theorem}\label{thm:bounded_multiple} (Observational entropy with multiple coarse-grainings is bounded)
\[
S(\R)\leq S_{O(\C_1,\dots,\C_n)}(\R)\leq \ln \mathrm{dim}\HS
\]
for any ordered set of coarse-grainings $(\C_1,\dots,\C_n)$ and any density matrix $\R$. $S(\R)= S_{O(\C_1,\dots,\C_n)}(\R)$ if and only if $\C_{\R}\hookrightarrow (\C_1,\dots,\C_n)$. $S_O(\R)=\ln \mathrm{dim}\HS$ if and only if $\forall i_1,\dots,i_n$, $p_{i_1,\dots,i_n}=\frac{V_{i_1,\dots,i_n}}{\mathrm{dim}\HS}$.
\end{theorem}

The Observational entropy is therefore equal to the von Neumann entropy when the set of coarse-grainings is fine enough to distinguish between eigenvectors of the density matrix associated with different eigenvalues. To understand this, consider an observer that is given an eigenvector of the density matrix. The Observational entropy is equal to the von Neumann entropy if performing $n$ consequent measurements in measurement bases $(\C_1,\dots,\C_n)$ on this eigenvector is enough to determine the eigenvalue associated with this eigenvector \emph{with certainty}, no matter what eigenvector it is. The Observational entropy is equal to the maximal value when probabilities of obtaining measurement outcomes $i_1,\dots,i_n$ are linearly proportional to the volumes of the respective multi-macrostates $V_{i_1,\dots,i_n}$.

One could expect that performing more measurements of the system should give the observer better knowledge about the system, at least on average, corresponding to a decrease in entropy representing this knowledge. Observational entropy has this property, as follows.
\begin{theorem}\label{thm:non-increase} (Observational entropy is non-increasing with each added coarse-graining.)
\[
S_{O(\C_1,\dots,\C_{n})}(\R)\geq S_{O(\C_1,\dots,\C_n,\C_{n+1})}(\R)
\]
for any ordered set of coarse-grainings $(\C_1,\dots,\C_n,\C_{n+1})$ and any density matrix $\R$. The inequality becomes an equality if and only if $\forall i_1,\dots,i_{n+1}$, $p_{i_1,\dots,i_{n+1}}=\frac{V_{i_1,\dots,i_{n+1}}}{V_{i_1,\dots,i_n}}p_{i_1,\dots,i_n}$.
\end{theorem}
This has the interesting interpretation that entropy is unaffected by additional coarse-graining $\C_{n+1}$ if the corresponding measurement is ``uninformative" in the sense that the conditional probability of the outcome $i_{n+1}$ is given by the ratio of the volumes of macrostates.

There are two notable cases in which this is the case. The first is when measurements corresponding to the set of coarse-grainings $(\C_1,\dots,\C_{n})$ project onto a pure state, meaning that performing an additional measurement on the system about which the observer already has the perfect knowledge, does not provide any new information about the prior system. For example, if the first coarse-graining is given by operator $\hat{A}$ that has a non-degenerate spectrum, then for any coarse-graining $\C_2$, $S_{O(\C_{\hat{A}})}(\R)= S_{O(\C_{\hat{A}},\C_{2})}(\R)$. The inequality becomes an equality also when $\C_{n+1}\hookrightarrow (\C_n,\dots,\C_1)$ (note the reverse order in the set), meaning that the last measurement is redundant in a sense that all the information it could provide has been already provided by its preceding measurements.

Note that we could also define a relationship $\hookrightarrow$ between two sets of coarse-grainings and derive other theorems analogous to those stated in the previous chapter, such as monotonicity of the Observational entropy as a function of sets of coarse-grainings, or extensivity with multiple coarse-grainings. Another possible task would be finding sets of coarse-grainings that lead to a constant Observational entropy. This will be left for future work. Instead, we use this theory to introduce Observational entropies with multiple-coarse-grainings that have compelling interpretation in quantum thermodynamics.

\section{Observational entropy as thermodynamic entropy}
\label{sec-thermodynamics}

\begin{figure}[t]
\centering
\includegraphics[width=1\hsize]{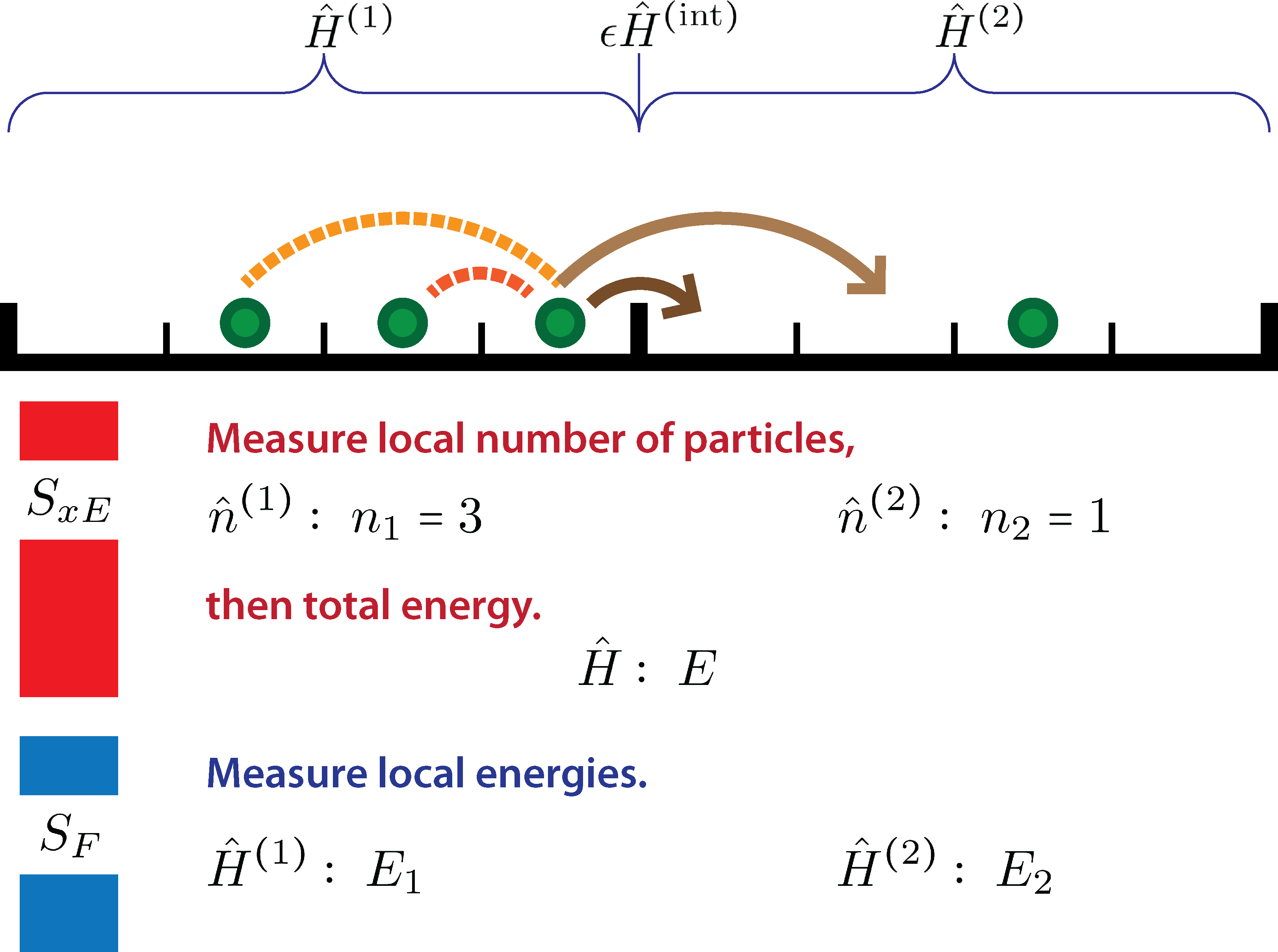}
\caption
{Schematic depiction of two relevant Observational entropies for thermodynamics, illustrated on a model of fermionic chain used in our simulations. We consider situation of $p=m=2$ regions (subsystems/partitions), evolving through Hamiltonian $\hat{H}=\hat{H}^{(1)}\otimes \hat{I}+\hat{I}\otimes\hat{H}^{(2)}+\epsilon\hat{H}^{(\mathrm{int})}$.  $S_{xE}$ uses positional coarse-graining, which for indistinguishable particles corresponds to measuring local particle densities, and then coarse-graining in total energy. Factorized Observational entropy $S_F$ uses local energy coarse-graining, corresponding to measuring energy of the first and the second region respectively. Dashed curves represent interparticle forces, solid curves represent particle hopping.
}
\label{fig:simulationscheme}
\end{figure}

In this section we introduce two entropies that are well-defined out of equilibrium, and correspond to thermodynamic entropy in equilibrium, even in closed quantum systems.  This provides a compelling answer to the question of {\em what kind of entropy} increases in closed quantum systems. These two entropies are similar in spirit, but they employ different coarse-graining and are  mathematically distinct.

First we introduce an entropy that corresponds to measuring the coarse-grained position of particles, and then the total energy of the system. This entropy measures whether the system is in a positional configuration that corresponds to many different energies; it thus describes how energy is distributed over many possible configurations of the system.

The second entropy employs coarse-graining in local quantities.  We primarily treat ``local" energy, but the technique could additionally treat the number of particles, angular momentum, etc. This entropy describes whether the total quantities are evenly distributed over the local systems. Both entropies are schematically summarized in Fig.~\ref{fig:simulationscheme}.

Interpretation of these two entropies are very similar. In Appendix~\ref{sec-S_xE_FOE_correspondence}  we show that these entropies are closely related analytically for small interaction strengths between partitions, or equivalently between positional coarse-grained bins. This similarity will be illustrated in Section~\ref{sec-simulation}, where we numerically evolve a system of fermions on a one-dimensional lattice.

\subsection{Observational entropy of position and energy}

Consider an observer who wishes to measure coarse-grained position of the particles (or equivalently, local particle number measurements) and energy of the system. The two relevant coarse-grainings are
\begin{subequations}\label{eq:S_Ex_entropy_projectors}
\begin{align}
\C_{{\hat{X}}^{(\delta)}}&=\{\P_{\vec x}^{(\delta)}\}_{\vec x},\quad \P_{\vec{x}}^{(\delta)}=\sum_{\vec{\tilde{x}} \in C_{\vec{x}}} \pro{\vec{\tilde{x}}}{\vec{\tilde{x}}},\\
\C_{\hat{H}}&=\{\P_E\}_E,\quad \P_E=\pro{E}{E},
\end{align}
\end{subequations}
where $\C_{{\hat{X}}^{(\delta)}}$, which corresponds to coarse-graining in position space with $p$ number of bins of size $\delta$, has been already explained in detail in Eq.~\eqref{eq:x_projectors}. For indistinguishable particles, this coarse-graining is equivalent to measuring number of particles in each box, and can be written as $\C_{{\hat{X}}^{(\delta)}}=\C_{\hat{N}_1\otimes\cdots\otimes \hat{N}_p}\equiv \C_{\hat{N}_1}\otimes\cdots\otimes \C_{\hat{N}_p}$. $\C_{\hat{H}}$ consists of projectors from the spectral decomposition of the total Hamiltonian $\hat{H}=\sum_E E\P_E$. There are two different types of Observational entropies that can be considered.

The first, $S_{Ex}(\R) \equiv S_{O(\C_{\hat{H}},\C_{{\hat{X}}^{(\delta)}})}(\R)$,  corresponds to measuring the energy and then coarse-grained position. It remains constant for any initial state and therefore does not have a meaningful interpretation describing the dynamics of a closed quantum system. One could also imagine a coarse-grained measurement of energy, but as we explain in Appendix~\ref{sec-S_Ex}, even this more general choice does not lead to something thermodynamically meaningful.

On the other hand,  we can switch the order of the non-commuting coarse-grainings, instead considering
\[
S_{xE}(\R) \equiv S_{O(\C_{{\hat{X}}^{(\delta)}},\C_{\hat{H}})}(\R),
\]
which corresponds to first measuring the coarse-grained position and then energy of the system.
This yields an entropy that rises in a closed system and reaches the correct thermodynamic value given by the microcanonical ensemble (microcanonical entropy) for initial pure states that are superpositions of energy eigenstates strongly peaked around a given value of energy  (denoted PS states; peaked superposition states), and reaches a value that is between the canonical entropy and the mean value of the microcanonical entropies (the mean given by the initial state) for other initial states. This is shown in simulations in Section~\ref{sec-simulation}, and analytically in Appendix~\ref{sec-S_xE_for_convergence}.

Denoting by $p_{xE}$ the probability of observing a given position macrostate $x$ then energy state $E$, and denoting the corresponding Hilbert space volume $V_{xE} = \tr[\P_E\P_x]$, Theorem~\ref{thm:bounded_multiple} implies that $S_{xE}$ is maximized when $p_{xE} \propto V_{xE}$.  In general, $S_{xE}$ is large if $p_{xE}$ is high for large volumes $V_{xE}$ while low for small volume; $S_{xE}$ is small if $p_{xE}$ is low for large volumes $V_{xE}$, and large for small volumes $V_{xE}$.  That is,  $S_{xE}$ is low to the extent the state is localized in a small region of space with a well-defined energy, and high otherwise.

As an example, imagine the positional coarse-graining on a one-dimensional lattice of length $L=9$ of with $N=3$ indistinguishable particles coarse-grained into $p=3$ boxes of size $\delta=3$ (for detailed explanation, see example below Eq.~\eqref{eq:x_projectors}). A state of all three particles contained in the first box of sites $\{1-3\}$ corresponds to signature $[3,0,0]$, which is a very small subspace of dimension $1$. The number of energy states (of the full Hamiltonian) corresponding to this positional configuration is also small, in fact proportional to the interaction strength between the first and the second box.  Therefore, the state is localized in a small region of space and has a relatively well-defined energy, resulting in low Observational entropy $S_{xE}$.

We will see in simulations (Sec.~\ref{sec-simulation}) that Observational entropy $S_{xE}$ is maximized by the evolution of the system, i.e., evolution of the system leads to positional configurations that have the highest uncertainty in energy. As mentioned before, and described in Appendix~\ref{sec-S_xE_FOE_correspondence}, this entropy is also closely connected to the factorized Observational entropy that will be introduced shortly -- in fact, they are identical when the interaction strength between different positional bins is zero (i.e., different bins-partitions do not interact), and they are closely connected for very small interaction strengths.\footnote{For this equivalence, we assume that the coarse-grained positional bins in $S_{xE}$ match the partitioning of the Hilbert space for the factorized Observational entropy. Trivially, this assumption implies that number of positional bins $p$ matches the number of partitions in FOE $m$, $p=m$.} Therefore many of properties of factorized Observational entropy we show below, such as its convergence to thermodynamic entropy, are also expected to hold for $S_{xE}$.

\subsection{Factorized Observational entropy}

Now we introduce the second relevant entropy. Compared with $S_{xE}$, while more complex in its definition, this entropy is more theoretically tractable, and it can be easily generalized to include several conserved quantities in addition to energy -- for example particle number, momentum, or angular momentum. This then leads to definitions of non-equilibrium entropy converging to thermodynamic entropy of generalized ensembles such as grand-canonical ensemble.

We start by considering Hilbert space divided into two parts $\HS^{(1)}$ and $\HS^{(2)}$, the joint system being $\HS=\HS^{(1)}\otimes\HS^{(2)}$. The Hamiltonian $\hat{H}$ can then be separated into three terms
\[
\label{eq:H=H1+H2+Hint}
\hat{H}=\hat{H}^{(1)}\otimes \hat{I}+\hat{I}\otimes\hat{H}^{(2)}+\epsilon\hat{H}^{(\mathrm{int})},
\]
where $\hat{H}^{(1)}$ and $\hat{H}^{(2)}$ are the Hamiltonians that describe internal interactions in the
first and second systems respectively, and $\hat{H}^{(\mathrm{int})}$ is an interaction term.
For large subsystems and local interactions, contribution of this term to the total energy is expected to be small and hence we have introduced a parameter (interaction strength) $\epsilon$ to indicate this. Consider a coarse-graining that projects to the eigenstates of the local Hamiltonians $\hat{H}^{(1)}$ and $\hat{H}^{(2)}$; this corresponds to simultaneous measurements of local energies. The Observational entropy built up from this coarse-graining, which we call the \emph{factorized Observational entropy} (FOE), can be formally written as
\[\label{def:non-equilibrium_entropy}
S_{F}(\R)\equiv S_{O{\displaystyle (}\C_{\hat{H}^{(1)}}\otimes \C_{\hat{H}^{(2)}}{\displaystyle )}}(\R).
\]
Explicitly, this factorized coarse-graining is given by $\{\P_{E_1}\otimes \P_{E_2}\}_{E_1E_2}$, and the projectors are given by spectral decompositions of local Hamiltonians, $\hat{H}^{(1)}=\sum_{E_1}E_1\P_{E_1}$, $\hat{H}^{(2)}=\sum_{E_1}E_2\P_{E_2}$. Later we will generalize this definition to an arbitrary number $m$ of local Hamiltonians, and additional observables representing other conserved quantities.

\subsubsection*{Properties}

This entropy has many interesting properties. In this section we will show that:
\begin{enumerate}
\item FOE is extensive on separable states. In other words, if the
total density matrix is separable, $\R=\R^{(1)}\otimes\R^{(2)}$, then the FOE is the sum of entropies of the subsystems.
\item FOE is upper-bounded by the von Neumann entropy of the diagonal density matrix, defined as the diagonal part of the density matrix written in an energy basis. This entropy is in turn  upper-bounded by the canonical entropy. This upper bound is achieved by a thermal (canonical) density matrix.
\item In the long-time limit for closed non-integrable systems, the FOE of a superposition of states peaked around a given value of energy (``PS states") converges to the microcanonical entropy. The FOE of other states (that span across many energy eigenstates) converges to the von Neumann entropy of the diagonal density matrix.
\item FOE converges to the canonical entropy in systems that weakly interact with a thermal bath.
\end{enumerate}

{\em Property 1.}
The first property follows immediately from Theorem~\ref{thm:extensivity}, which says that Observational entropy with a factorized coarse-graining is extensive on separable states, hence
\[
\begin{split}
&S_{O{\displaystyle (}\C_{\hat{H}^{(1)}}\otimes \C_{\hat{H}^{(2)}}{\displaystyle )}}\Big(\R^{(1)}\otimes\R^{(2)}\Big)=\\
&S_{O{\displaystyle (}\C_{\hat{H}^{(1)}}{\displaystyle )}}\Big(\R^{(1)}\Big)+S_{O{\displaystyle (}\C_{\hat{H}^{(2)}}{\displaystyle )}}\Big(\R^{(2)}\Big).
\end{split}
\]
Physically, this equation means that if an observer is able two describe and study the two systems separately (i.e., the coarse-graining is factorized), and the two systems are non-interacting (and do not contain any correlation), then it does not make a difference whether these two systems are described separately from the observer's point of view, or together. This property also ensures that one can indefinitely extend the Hilbert space by adding more particles, or more sites to the system, while the corresponding factorized Observational entropy of the entire system will be always a continuous function of time, even during sudden (discontinuous) changes of the Hamiltonian representing turning on the interactions between the old system and the new added subsystems.

\vskip0.1in

{\em Property 2.}
We start with some motivation. In a closed quantum system, the amplitudes of the initial state written in the eigenbasis of the full Hamiltonian do not change. This is a consequence of the identity
\[
\begin{split}
p_E(\R_t)&=\tr[\P_E\R_t]=\tr[\P_EU(t)\R_0U(t)^\dag]\\
&=\tr[U(t)\P_E\R_0U(t)^\dag]=\tr[\P_E\R_0]=p_E(\R_0),
\end{split}
\]
where $U(t)=\exp(-i\hat{H}t)$ represents the unitary evolution operator of the system. The above equation shows that all such probabilities are conserved. For closed quantum systems, it therefore makes sense to introduce the diagonal density matrix $\R_d$ that contains the information about these conserved quantities. This density matrix is defined to be diagonal in the energy basis, with its diagonal elements to be equal to the diagonal elements of the original density matrix written in the energy basis.
 However, we will average the diagonal values of $\R_d$ over the degenerate subspaces of the Hamiltonian, in order to reflect the fact that coarse-graining $\C_{\hat{H}}$ cannot distinguish between different eigenvectors with the same eigenvalue $E$.

Mathematically, assuming the system is described by density matrix $\R_t$, we define $\R_d$ by its elements in the energy basis as $\bra{E}\R_d\ket{E'}\equiv \frac{p_E(\R_t)}{\tr[\P_E]}\delta_{EE'}$, equivalent to
\[\label{eq:diagonal_density_matrix}
\R_d=\sum_E \frac{p_E(\R_t)}{\tr[\P_E]}\P_E,
\]
which, per the above, is a constant density matrix for closed quantum systems, defined fully by $\R_0$.

If the diagonal coefficients are strongly peaked around one set of invariants, such as energy and number of particles, then this density matrix is microcanonical. If the diagonal coefficients are spread out so that there is significant weight in many states with macroscopically different invariants, then this density matrix represents a macroscopic superposition of different thermal states~\cite{OPenrose1979FoundStatMech}. Such superpositions are real features of quantum mechanics that have been observed experimentally~\cite{girvin2009circuit}. If we concentrate on the former case, then this microcanonical density matrix is known to be equivalent to the canonical density matrix, in the limit of large system sizes and large local observables, meaning that relative differences between the von Neumann entropy of the microcanonical density matrix (which approximates the microcanonical entropy computed from the density of states) and von Neumann entropy of canonical density matrix (thermodynamic entropy) vanish as the system size grows to infinity~\cite{ruelle1999statistical}. The von Neumann entropy of the diagonal density matrix, $S(\R_d)$, is one of the two entropies that figure into the bound we define below. As per Eq.~\eqref{eq:vonNeumannvsO}, we can see that this quantity can be also simply written as Observational entropy with coarse-graining given by the total Hamiltonian,
\[\label{eq:diagonalEntropy}
S(\R_d)=S_{O{\displaystyle (}\C_{\hat{H}}{\displaystyle )}}(\R).
\]
For non-degenerate and time-independent Hamiltonians, this entropy corresponds to diagonal entropy~\cite{polkovnikov2011microscopic}.
For the second, assume the mean energy of the system is given by $\overline{E}\equiv\tr[\hat{H}\R_t]=\tr[\hat{H}\R_d]$. We can then define a corresponding canonical density matrix as $\R_{th}=\frac{1}{Z}\exp(-\beta\hat{H})$, where partition function is defined as $Z=\tr[\exp(-\beta\hat{H})]$, and inverse temperature $\beta$ is defined as solution to equation  $\overline{E}=-\frac{\partial\ln Z}{\beta}$. The canonical (thermodynamic) entropy is then defined as von Neumann entropy of the canonical state, $S_{th}\equiv S(\R_{th})$.

As we show in Appendix~\ref{app:bounds_on_FOE}, FOE is bounded by the von Neumann entropy of diagonal density matrix up to a correction term of order $\epsilon$ representing the interaction strength between the subsystems; this entropy is in turn bounded by the corresponding canonical entropy.  Thus:
\[\label{eq:bounds_on_FOE}
S_{F}(\R)+O(\epsilon)\leq S(\R_d) \leq S(\R_{th}).
\]
Having Eqs.~\eqref{def:non-equilibrium_entropy} and \eqref{eq:diagonalEntropy} in mind, this inequality shows that Observational entropy given by coarse-graining defined by the local Hamiltonians, is always lower than the Observational entropy given by the global Hamiltonian. The first inequality is derived using the particular form of the factorized Observational entropy and the properties of finer coarse-graining expressed by Theorem~\ref{thm:monotonic} and equality conditions in Theorem~\ref{thm:bounded}, while the second equality comes from the usual maximization procedure. At the same time, we have also derived
\[\label{eq:FOE_of_diagonal}
S_{F}(\R_d)+O(\epsilon)= S(\R_d).
\]
The canonical density matrix is diagonal in the energy basis, therefore it is a special case of the diagonal density matrix. Thus also
\[\label{eq:FOE_of_thermal}
S_{F}(\R_{th})+O(\epsilon)= S(\R_{th}).
\]

The correction term $O(\epsilon)$ is a finite size correction, and generically is expected to be negligible for systems which are sufficiently coarse-grained, i.e., if the energies of the subsystems are large enough in comparison to the energy of interaction between the subsystems. However, there are few exceptions to this: if the density matrix $\R$ in Eqs.~\eqref{eq:bounds_on_FOE} or \eqref{eq:FOE_of_diagonal} consists of a single energy eigenstate $\R=\pro{E}{E}$ (or is a mixture of very few energy eigenstates), then $S(\R_d)=0$ (assuming the Hamiltonian is non-degenerate), but as shown in Appendix~\ref{sec-FOE_for_eigenstates}, FOE is proportional to the microcanonical entropy, $S_{F}(\R)\approx S_{\rm micro}(E)$ (defined in Eq.~\eqref{eq:microcanonical_entropy}), resulting in $O(\epsilon)\approx S_{\rm micro}(E)$ and $S_{F}(\R)>S(\R_d)$. Similarly, if the density matrix $\R=\pro{E_-}{E_-}$ is an eigenstate of the Hamiltonian without the interaction terms $\Ha_-\equiv\hat{H}-\epsilon\hat{H}^{(\mathrm{int})}$, then $S(\R_d)\approx S_{\rm micro}(E)$ and $S_{F}(\R)=0$. In these cases, amplitudes of the perturbative expansion diverge, so such expansion is not valid, and $O(\epsilon)$ cannot be considered a first-order correction. In all other cases, the perturbative expansion $O(\epsilon)$ is well-defined, is expected to be small, and represent a finite-size correction.

We managed to find the exact form of the first-order correction in Eq.~\eqref{eq:FOE_of_thermal}, which turns out to be
\[\label{eq:correctionToThermal}
O(\epsilon)=-2\epsilon\beta^2\mean{\hat{H} \hat{H}^{(\mathrm{int})}}_C
\]
plus a quantum term (which we neglected) that comes from non-commutativity of $\hat{H}$ and $\hat{H}^{(\mathrm{int})}$. The covariance $\mean{\hat{H} \hat{H}^{(\mathrm{int})}}_C \equiv \mean{\hat{H}\hat{H}^{(\mathrm{int})}}_{\R_{th}}-\mean{\hat{H}}_{\R_{th}}\mean{\hat{H}^{(\mathrm{int})}}_{\R_{th}}$, where $\mean{\hat{A}}_{\R_{th}}\equiv \tr[\hat{A}\R_{th}]$, is proportional to the correlation length. The $O(\epsilon)$ term represents a finite-size correction, since it scales as $~N$ with the particle number in the thermodynamic limit, but it goes to zero when the coarse-grained regions are sufficiently large. See Appendix~\ref{app:bounds_on_FOE} for more details.

\vskip0.1in
{\em Property 3.}
The third property is that for an initial PS state, FOE converges to the microcanonical entropy for closed quantum non-integrable systems.  This emerges from similar arguments as the eigenstate thermalization hypothesis. Let us first mention the essential difference between integrable and non-integrable systems. In integrable systems the form of interaction does not sufficiently mix particles, resulting in a large number of constants of motion that prevent full thermalization. An example of such a system is a fermionic chain with only nearest-neighbor interactions. Non-integrable systems have a form of interaction that is sufficient to result in full thermalization, an example being a fermionic chain with both nearest-neighbor and next-nearest-neighbor interaction. Both systems have been studied~\cite{santos2010onset,deutsch2013microscopic,santos2012weak}, and we will examine them both in the next section in relation to Observational entropies.

We define microcanonical entropy as~\cite{reif2009fundamentals}
\[\label{eq:microcanonical_entropy}
S_{\rm micro}(E)=\ln({\rho(E)\Delta E}),
\]
where $\rho(E)$ denotes the (energy) density of states, and $\Delta E$ is the typical energy.\footnote{This typical energy is there to give it the right units, and is usually taken to be $\Delta E = \sigma(E)/\sqrt{N}$, where $\sigma$ computes the standard deviation and $N$ is the number of particles. This choice of $\Delta E$ is rather arbitrary as it is unimportant in the thermodynamic limit.}'\footnote{For initial states that have a spread in energy of order $\Delta E$, such as a uniform superposition of states $\ket{\psi_E}=\frac{1}{\mathcal{N}}\sum_{\tilde{E}\in[E,E+\Delta E]}\ket{\tilde{E}}$, where $\mathcal{N}$ denotes the normalization constant, we have $S_{\rm micro}(E)\approx S(\R_d)$, where $\R_d=\R\equiv\pro{\psi_E}{\psi_E}$.} In appendix~\ref{sec-FOE_for_eigenstates} we show, using the connection between non-integrable systems and random matrix theory, that for both energy eigenstates $\ket{E}$, and PS states with random phases $\R_E$, the FOE of such states gives the same value as the microcanonical entropy,
\[\label{eq:approx_micro}
S_F(\ket{E})\approx S_F(\R_E)\approx S_{\rm micro}(E).
\]
This is also illustrated on Fig.~\ref{fig:SXEvsE}, which shows FOE for such states in comparison with the microcanonical entropy.

Because of the evolution, after some time all phases of the state (written in the energy basis) will become random, and thus states with random phases are typical states of the system some time in future. As a consequence of Eq.~\eqref{eq:approx_micro}, considering a PS state $\R_E$ as the initial state, the FOE of an evolved state $\R_t=U(t)\R_EU(t)^\dag$ converges to the microcanonical entropy,
\[\label{eq:convergnce_to_microcanonical}
S_F(\R_t)\overset{t\rightarrow \infty}{\leadsto} S_{\rm micro}(E).
\]

In an alternative scenario, when the initial state of the system is a superposition of many energy eigenstates (not a PS state), the FOE converges to the von Neumann entropy of the diagonal density matrix,
\[
S_F(\R_t)\overset{t\rightarrow \infty}{\leadsto} S(\R_d).
\]

The above convergences can be combined, and we can write
\[\label{eq:convergence_combined}
S_F(\R_t)\overset{t\rightarrow \infty}{\leadsto}\max\Big\{\sum_{E}p_E(\R_0)S_{\rm micro}(E),\ \!S(\R_d)\Big\}.
\]
That is, in closed non-integrable systems and for any initial state, FOE will converge to the mean value of the corresponding microcanonical entropies, or the von Neumann entropy of the diagonal state, whichever is bigger, up to some order-unity corrections that become irrelevant in the thermodynamic limit  (see Appendix~\ref{sec-FOE_for_eigenstates} for details). In addition to previous cases, the above equation also applies to the macroscopic superpositions of different microcanonical states. We also remind that this maximum is still smaller than the canonical entropy (up to order $O(\epsilon)$), as shown by Eq.~\eqref{eq:bounds_on_FOE}.

The differences between microcanonical and canonical entropy disappear in the ``thermodynamic'' limit of large system sizes~\cite{ruelle1999statistical}. We can therefore conclude that for both typical (microcanonical) and atypical (macroscopic superpositions) states, and in non-integrable systems, the FOE of any state converges to a value that closely approximates  the thermodynamic entropy.

\vskip0.1in
{\em Property 4.}
We now turn to systems interacting with a thermal bath. As described in Eq.~\eqref{eq:convergence_in_open_systems}, in such systems the density matrix resembles the thermal density matrix at most times in future. According to Eq.~\eqref{eq:FOE_of_diagonal}, the FOE of the thermal density matrix is the thermodynamic (canonical) entropy. Combining these two equations we conclude that for systems interacting with the thermal bath, the FOE converges to the thermodynamic entropy,
\[\label{eq:convergence_to_cano}
S_{F}(\R_t)\overset{t\rightarrow \infty}{\leadsto} S(\R_{th})+O(\epsilon),
\]
up to order $\epsilon$ (which denoted the strength of the interaction between partitions inside of the system of interest.)

\subsubsection*{Generalization of FOE}
The idea of FOE can be generalized to multiple observables (beyond energy), and multiple partitions. To keep the notation compact, we identify
\[
\hat{O}^{(k)}\equiv \hat{I}\otimes\cdots\otimes \hat{I}\otimes
\hat{O}^{(k, \mathrm{local})}
\otimes\hat{I}\otimes\cdots\otimes \hat{I}.
\]
for any operator $\hat{O}^{(k)}$ with the upper number index, where $\hat{O}^{(k, \mathrm{local})}$ is on the $k$'th position. Operator $\hat{O}^{(k)}$ acts only on the $k$'th system via operator $\hat{O}^{(k, \mathrm{local})}$, and it leaves other subsystems intact. We assume that the evolution of the joint system is governed by a Hamiltonian with local terms and the interaction between subsystems,
\[
\hat{H}=\hat{H}^{(1)}+\cdots+\hat{H}^{(m)}+\epsilon\hat{H}^{(\mathrm{int})}.
\]
We will also assume that there are thermodynamic quantities
\[
\hat{A_j}=\hat{A}_j^{(1)}+\cdots+\hat{A}_j^{(m)},\quad j=1,\dots,n
\]
that can be measured locally, and that are conserved globally. We introduce the following definition.
\begin{definition}\label{def:factorized_observational_entropy}
We define the \emph{generalized factorized Observational entropy} (GFOE) on Hilbert space $\HS=\HS^{(1)}\otimes\cdots\otimes\HS^{(m)}$ with conserved quantities $\hat{A}_1,\dots,\hat{A}_n$  as
\[
S_{O\left(\C_1,\dots,\C_n\right)}(\R_t),
\]
where
\[
\C_j=\C_{\hat{A}_j^{(1,\mathrm{local})}}\otimes\cdots\otimes\C_{\hat{A}_j^{(m,\mathrm{local})}},\quad j=1,\dots,n.
\]
\end{definition}

The above definition inherits many of the properties of the original definition (in which $n=1$, $m=2$, and $\hat{A}_1 = \hat H$) --  in fact, all of those properties when the local conserved quantities commute, $\big[\hat{A}_j^{(k,\mathrm{local})},\hat{A}_{\tilde{j}}^{(k,\mathrm{local})}\big]=0$ for all $k$ and $j\neq \tilde{j}$. This is a consequence of the fact that in such a case, a common eigenbasis exists. (Also, in such a case the canonical density matrix will be replaced by the generalized thermal density matrix.)

The situation becomes more difficult when these observables (coarse-grainings) do not commute.\footnote{Some aspects of non-commutativity in quantum thermodynamics have been recently studied in Refs.~\cite{lostaglio2015thermodynamic,halpern2016microcanonical,guryanova2016thermodynamics}.} For example, it can be shown that the generalized thermal density matrix does not necessarily maximize the GFOE, and therefore an equivalent of Eq.~\eqref{eq:FOE_of_thermal} might not hold in general.

\subsection{Interpretation}
Finally let us turn to the interpretation of FOE in the light of above properties, and compare it to $S_{xE}$.

FOE, as any Observational entropy, is generally small when the state described by the density matrix is localized in a small subspace in the Hilbert space, corresponding to a projector of a small rank (trace), or when the state is localized over a few of such small subspaces (see Theorem~\ref{thm:Boltzmann_equivalent} or Sec.~\ref{sec-entropy_increase}). Therefore, FOE is small for example when for a projector $\P_{E_1E_2}\equiv \P_{E_1}\otimes \P_{E_2}$ of a small trace, $\P_{E_1E_2}\R\P_{E_1E_2}=\R$ holds. But this means that the state has a well-defined energy (low uncertainty/low variance) in the basis of local Hamiltonians, i.e. has a well defined energy in each subsystem. When the system starts to evolve from such a state, due to the interaction between the subsystems the local energies stop being well-defined, and the factorized Observational entropy starts to rise, until it reaches the thermodynamic entropy of the entire system.

Thus the FOE measures how local subsystems exchange heat with each other (and in case of GFOE corresponding flows connected to other observables), until they become thermalized. The FOE, and GFOE respectively, measures thermalization of the entire, possibly closed, system, in terms of local subsystems becoming equilibrated with each other.

FOE and $S_{xE}$ are rather similar: they can be both interpreted as measures of how close are subsystems to thermal equilibrium. These subsystems, or we could say physical regions or partitions, are defined by the positional coarse-graining (Eq.~\eqref{eq:S_Ex_entropy_projectors}) in case of $S_{xE}$, and by the separation of the total Hamiltonian into local Hamiltonians (Eq.~\eqref{def:non-equilibrium_entropy} and Def.~\ref{def:factorized_observational_entropy}) in the case of FOE. In practice, both entropies will be maximal when the total energy is uniformly distributed over these regions. This is because macroscopic state where all the local energies are equal, $\langle E_1\rangle=\langle E_2\rangle=\cdots$, has the highest number of possible configurations (microstates) that correspond to it of all macrostates that are allowed by the conservation of total energy. This will be generally true up to some pathological cases: for example, state consisting of eigenstates of local Hamiltonians $\ket{\psi}=\ket{E_1}\ket{E_2}\cdots$ will have FOE equal to zero, even when all the energies are equal.

Therefore up to these pathological cases, we can say that entropy is maximal when each region contains roughly the same energy; and this is exactly what we could call thermalization. It is then not a surprise that both $S_{xE}$ and FOE then correspond to the thermodynamic entropy. The dynamical process of rising entropy then describes thermalization of different regions (in our simplest case, two regions) with respect to each other. In other words, it demonstrates heat exchanges between the two regions until the flow between partitions is zero on average, which we can intuitively describe as a macroscopic state where the temperature of the first and temperature of the second partition became identical.

What is special about  these two entropies? And why do we consider entropies that need coarse-graining in position and energy, or coarse-graining in local energies? Since thermodynamic entropy is fundamentally connected with energy, a coarse-grained entropy definition that matches thermodynamic entropy must have coarse-graining in energy.  However, if the coarse-graining is {\em only} in total energy then the entropy is preserved, reflecting the conservation of information under unitary dynamical evolution. This is not the kind of entropy that models dynamics of thermalization, but rather has to do with information.  Coarse-graining in some other observable that does not commute with energy yields a kind of entropy that has second-law-like behavior and is some measure of “disorder,” but by itself cannot be quantitatively connected to the thermodynamic entropy. Position is special in that interactions in physics tend to be local in position rather than in momentum or other variables, and therefore a positional coarse-graining creates a locally-conserved quantity and associated entropy that both evolve on the dynamical timescale of the system.

This is why we need to define entropy using either a combination of two non-commutative coarse-grainings, first in positional configuration, and then in energy (which gives $S_{xE}$), or by a single coarse-graining that is, however, constructed from local energy coarse-grainings (which gives FOE). Intuitively, both such entropies contain both locality, and energy, which seems to be the two crucial ingredients for finding a meaningful definition of a non-equilibrium thermodynamic entropy.

\section{Simulations}
\label{sec-simulation}

\begin{figure}[htp]
\centering
\includegraphics[width=1\hsize]{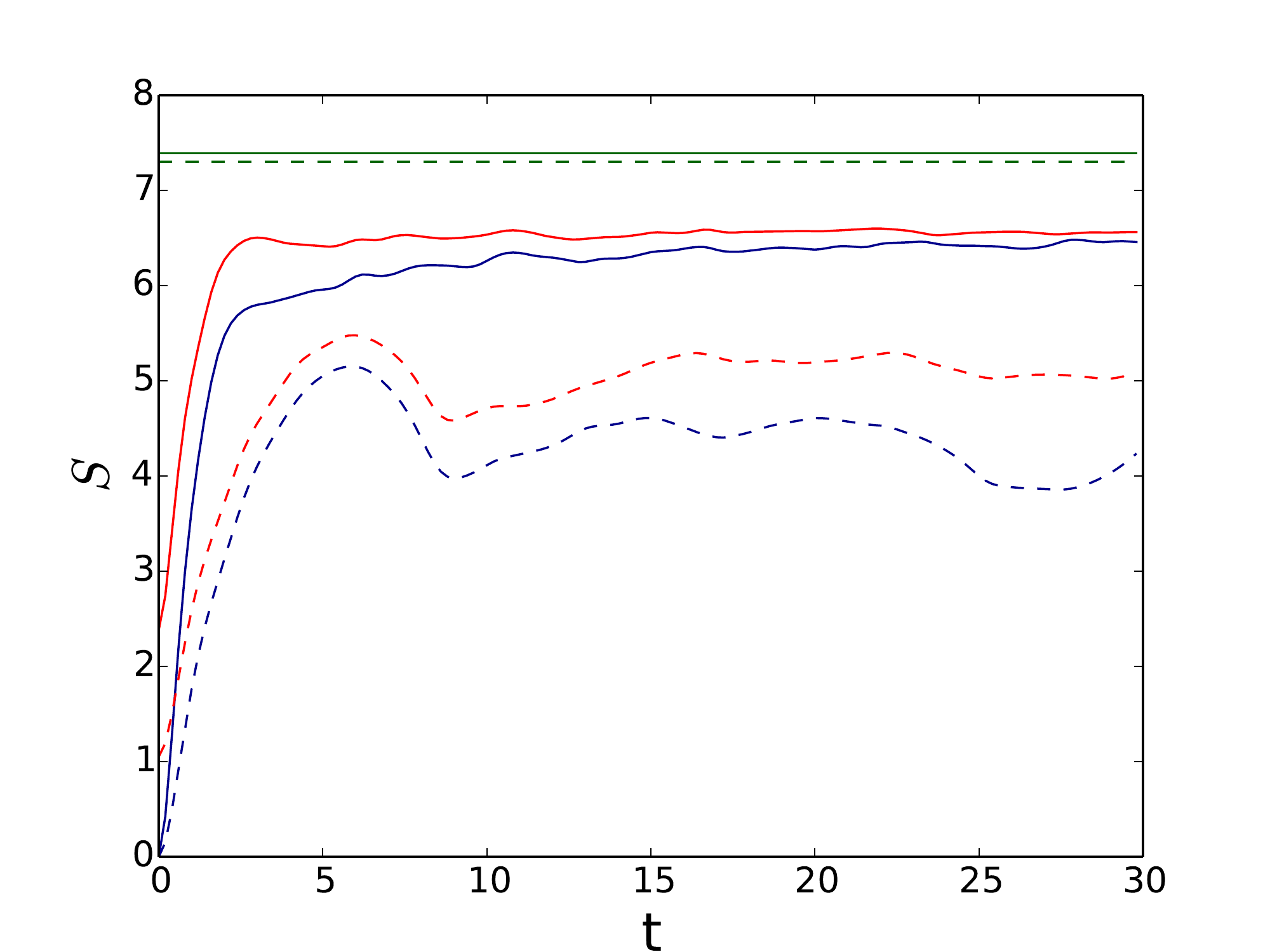}\\
~~~(a)\\
\includegraphics[width=1\hsize]{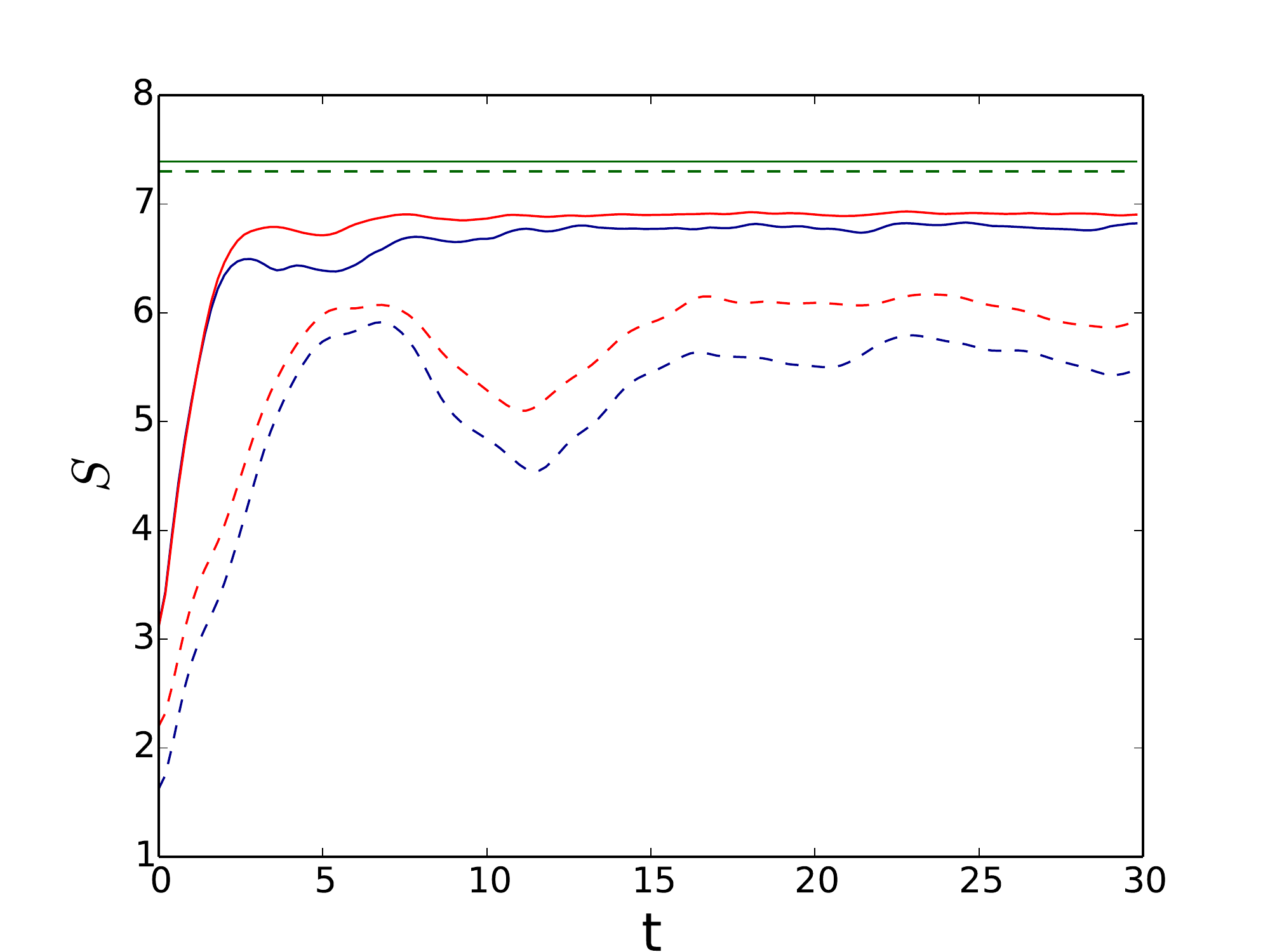}\\
~~~(b)\\
\caption
{(a) The factorized Observational entropy $S_F$ (dark blue lines) and Observational entropy $S_{xE}$ (light red lines) with non-integrable dynamics ($t'=V'=0.96$; full lines) and integrable dynamics ($t'=V'=0.0$; dashed lines). The system of length $L=8$ with hardwall boundary conditions starts in the 11th energy eigenstate of the reduced Hamiltonian $\hat{H}^{(1-8)}$. At $t=0$ the right wall is expanded so that $L=16$ and the system evolve. Coarse-graining is given by local Hamiltonians for $S_F$, and local position operators for $S_{xE}$ respectively. We coarse-grain into $p=m=2$ partitions, corresponding in FOE to $\C=\C_{\hat{H}^{(1-8)}}\otimes\C_{\hat{H}^{(8-16)}}$. The thermodynamic entropies are $7.389708$ in the non-integrable case (straight green line), and $7.301491$ in the integrable case (straight green dashed line). Because the system is initially in an energy eigenstate of the reduced Hamiltonian, which corresponds to the coarse-graining in FOE, the FOE only has one probability value that is nonzero. Hence this entropy is initially zero. (b) The same quantities as in (a) but with the coarse-graining to 4 partitions, which for FOE corresponds to $\C=\C_{\hat{H}^{(1-4)}}\otimes\C_{\hat{H}^{(4-8)}}\otimes\C_{\hat{H}^{(8-12)}}\otimes\C_{\hat{H}^{(12-16)}}$.}
\label{fig:CoarseGrainedEigenstate}
\end{figure}

\begin{figure}[htp]
\centering
\includegraphics[width=1\hsize]{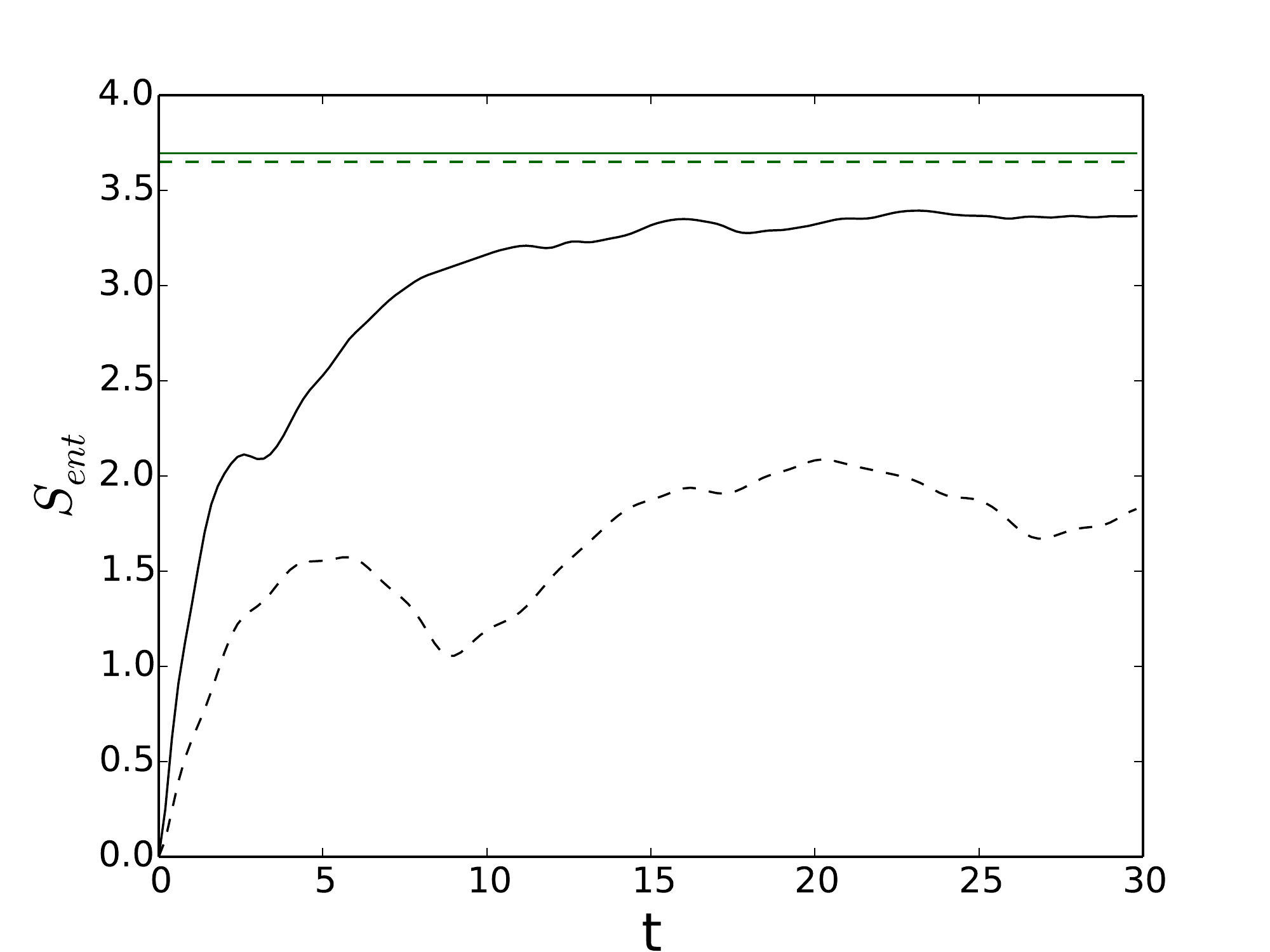}
\caption
{The entanglement entropy between the first and last 8 sites, is measured as a function of time for the same systems as in Fig.~\ref{fig:CoarseGrainedEigenstate}.
In the non-integrable case ($t'=V'=0.96$), the entanglement entropy is expected to asymptote to $1/2$ of the thermodynamic entropy of the complete system (shown as green straight lines), in the limit of large system sizes. The curve for non-integrable system (full line) is above the integrable one (dashed line).}
\label{fig:EntanglementEntropy}
\end{figure}

\begin{figure}[htp]
\begin{center}
\includegraphics[width=1\hsize]{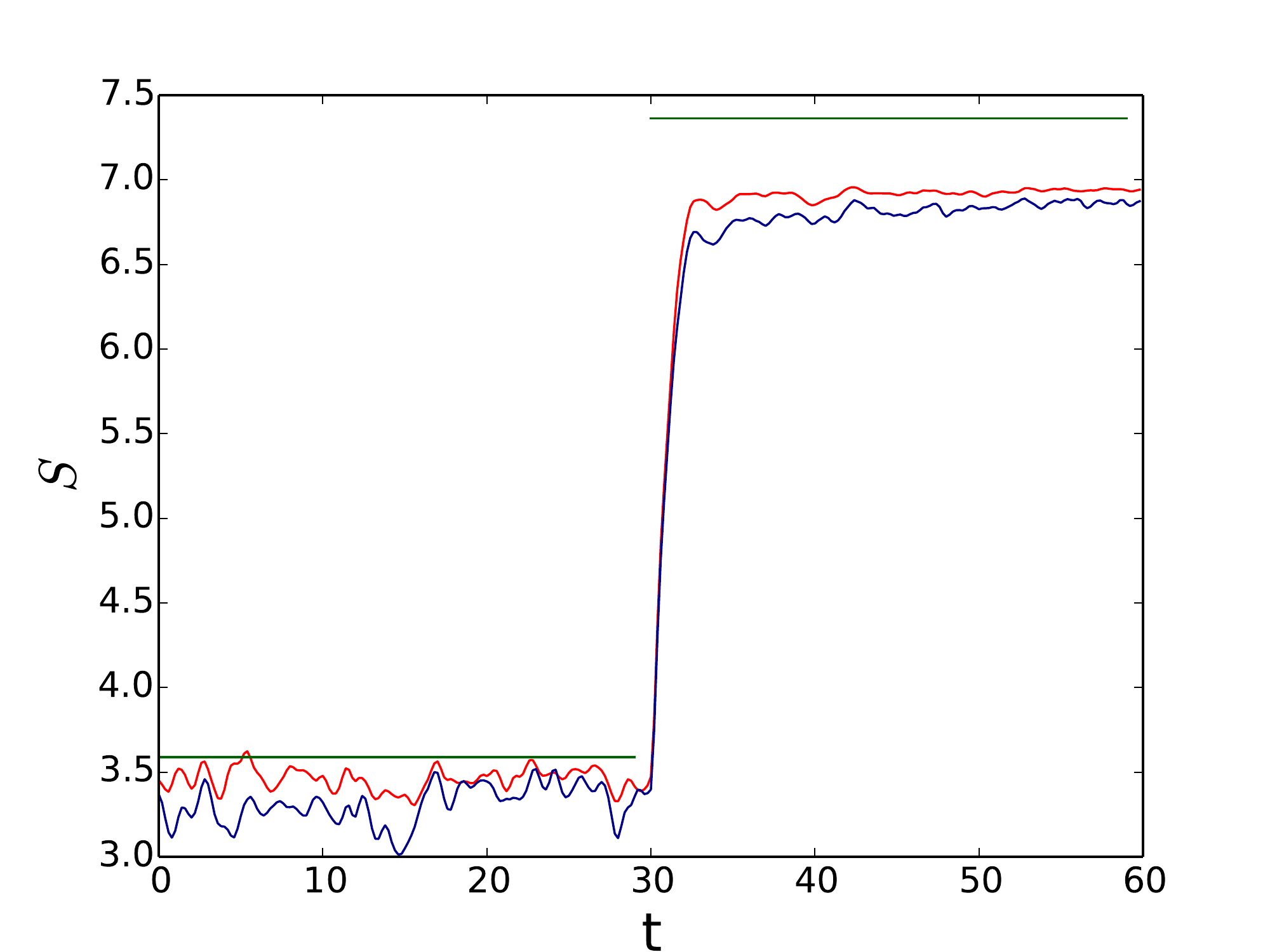}\\
~~~(a)\\
\includegraphics[width=1\hsize]{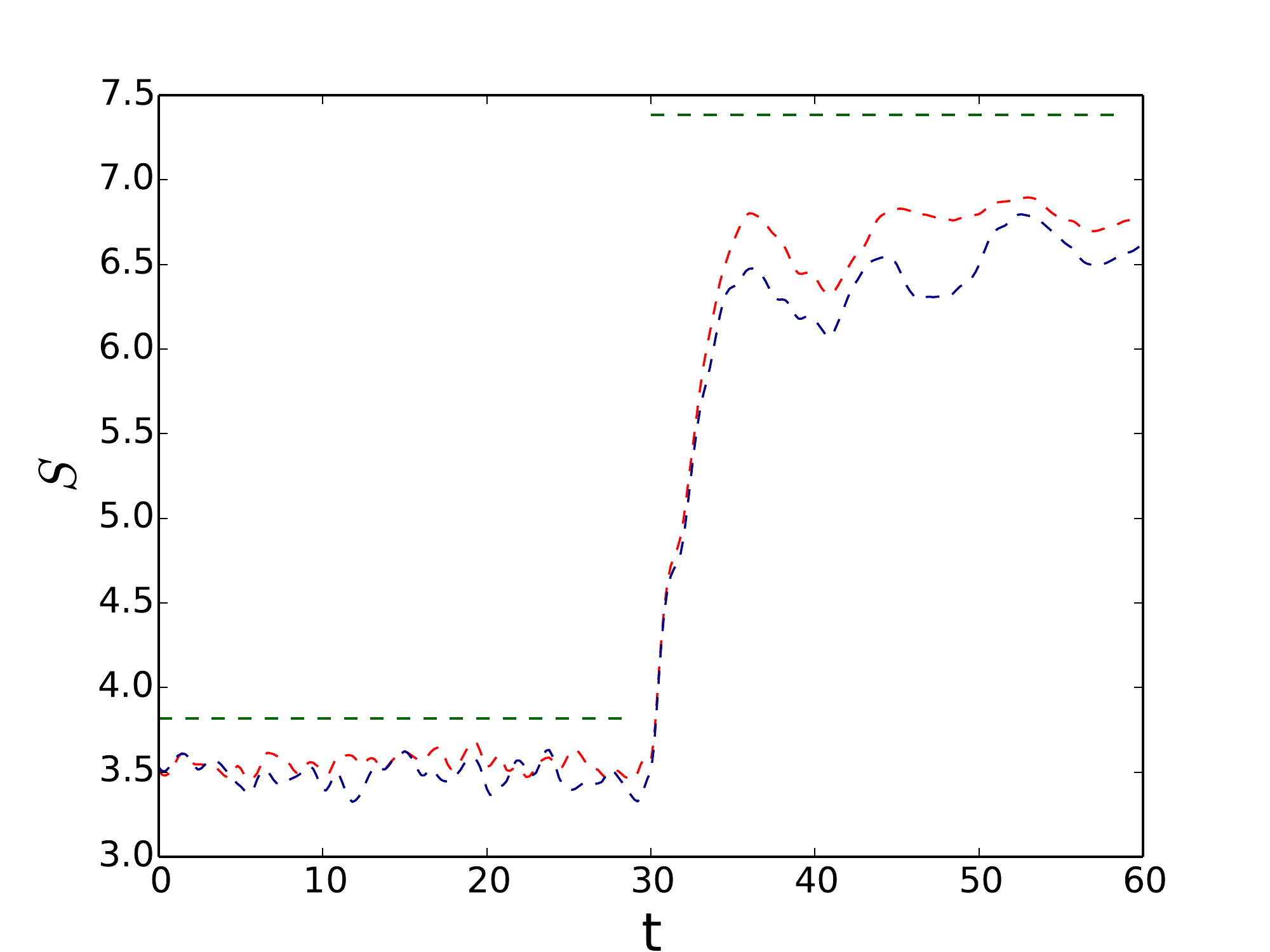}\\
~~~(b)
\caption
{(a) The factorized Observational entropy $S_F$ (dark blue) and Observational entropy $S_{xE}$ (light red)
that start in a ``pure thermal state'' with inverse temperature $\beta=1$, in the system of size $L=8$. At $t=30$, the right wall is expanded to double the system size so that
$L=16$ and the system continues to evolve. The straight green lines represent the thermodynamic (canonical) entropy $S(\R_{th})$ before and after the expansion. We coarse-grain into $p=m=4$ partitions that for FOE corresponds to $\C=\C_{\hat{H}^{(1-4)}}\otimes\C_{\hat{H}^{(4-8)}}\otimes\C_{\hat{H}^{(8-12)}}\otimes\C_{\hat{H}^{(12-16)}}$. This graph shows non-integrable dynamics ($t'=V'=0.96$).
(b)
The same as (a) but for integrable dynamics ($t'=V'=0$).}
\label{fig:FOE2}
\end{center}
\end{figure}

\begin{figure}[htp]
\centering
\includegraphics[width=1\hsize]{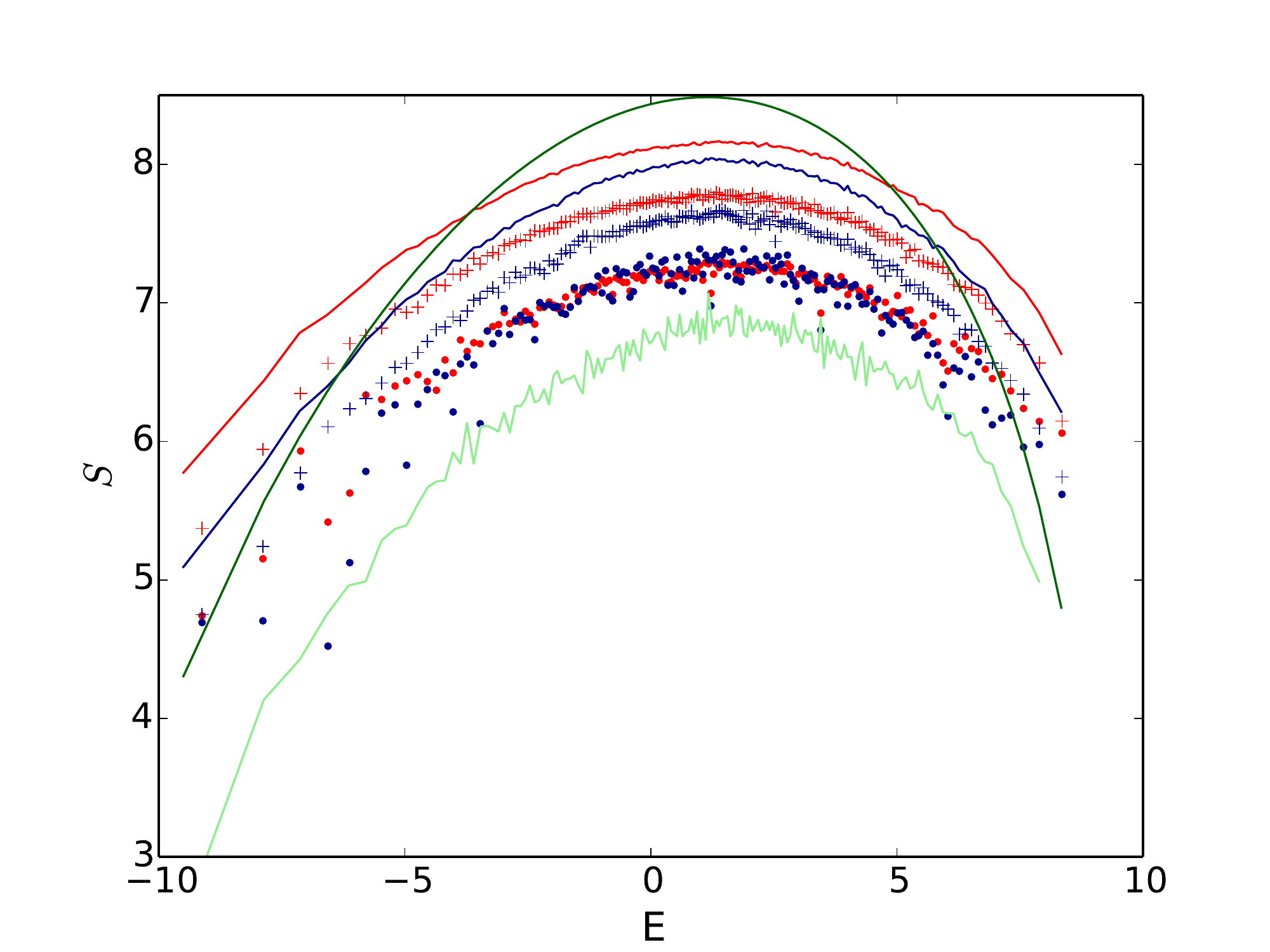}\\
~~~~~(a)\\
\includegraphics[width=1\hsize]{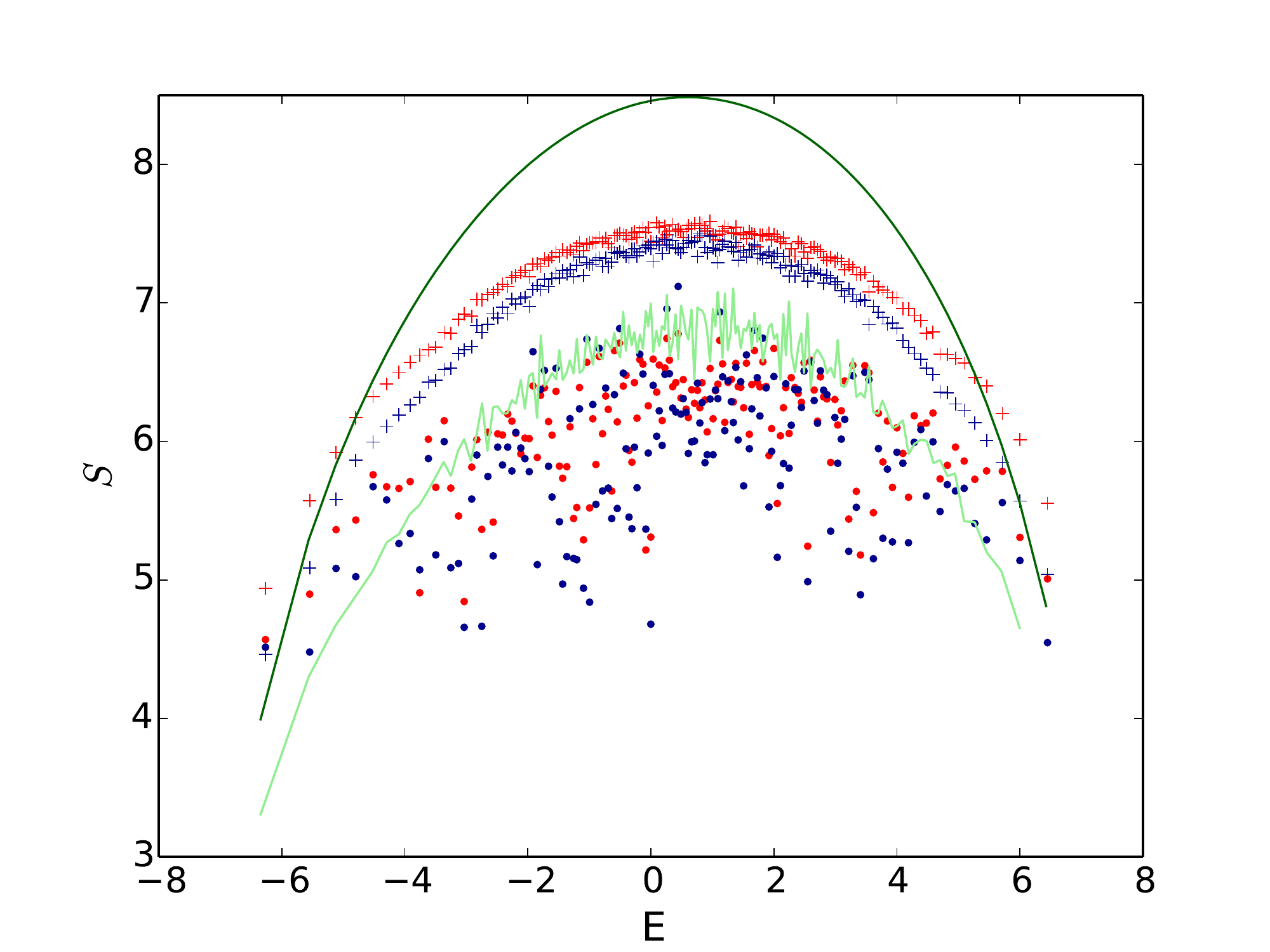}\\
~~~~~(b)
\caption
{(a) The red and blue curves in the middle show Observational entropies $S_{xE}$ (light red) and $S_F$ (dark blue) for microcanonical states (line), random superpositions of neighboring energy eigenstates (crosses), and energy eigenstates (dots), from top to bottom. The lowest (full light green) curve is the microcanonical entropy $S_{\rm micro}(E)$ given by logarithm of the density of states. The top (full dark green) curve that has a slightly different shape than the other curves is the thermodynamic (canonical) entropy $S(\R_{th})$, with the inverse temperature calculated such that the mean value of energy of the thermal state $\R_{th}$ corresponds to the energy $E$ depicted the horizontal axis. This graph shows non-integrable system ($t'=V'=0.96$). (b) depicts the same but for integrable system ($t'=V'=0$). (For increased visibility, we decided not to plot Observational entropies for microcanonical states.)}
\label{fig:SXEvsE}
\end{figure}

In this section we examine observation entropy using explicit numerical simulations of simple $N$-qubit quantum systems.  We examine primarily the factorized Observational entropy with local energy coarse-grainings $S_F$, and Observational entropy of measuring the coarse-grained position and then energy $S_{xE}$, but we also add some results for entanglement entropy $S_{\rm ent}$ for comparison.

We consider a one-dimensional lattice model of spinless fermions, with both nearest-neighbor (NN) and next-nearest-neighbor (NNN) hopping and interactions. This model is illustrated in Fig.~\ref{fig:simulationscheme}. Following the notation of Santos and Rigol~\cite{santos2010onset}, the Hamiltonian $\hat{H}^{(k-l)}$ that describes fermions moving between sites number $k$ and $l$ is
\[\label{FermionHam}
\begin{split}
\hat{H}^{(k-l)} = \sum_{i=k}^{l} \Big[ &-t \left( f_i^{\dagger} f_{i+1} + h.c. \right) +V n_i^f   n_{i+1}^f \\
&- t' \left( f_i^{\dagger} f_{i+2} + h.c. \right) +V' n_{i}^f  n_{i+2}^f\Big].
\end{split}
\]
$f_i$ and $f_i^{\dagger}$ are the fermionic annihilation and creation operators for site $i$.  $n_i^f= f_i^{\dagger} f_i$ is the local density operator. Operators anti-commute on different sites. We employ hard wall boundary conditions for our numerical experiments so that we can study the expansion of a gas from a smaller to a larger box. The Hamiltonian of the full system is $\hat{H}\equiv \hat{H}^{(1-L)}$, where $L$ is the length of the chain; however we will also require ``local" forms of the Hamiltonian using smaller ranges $(k-l)$.  We compute the eigenvalues and eigenvectors of relevant Hamiltonians using exact diagonalization.

The NN and NNN hopping strengths are respectively $t$  and $t'$. The interaction strengths are $V$ and $V'$ respectively. We always take $\hbar=V=t=1$. We choose $V'=t'=0$ to investigate the integrable system, and $V'=t'= 0.96$ to study the non-integrable (generic) system. We have chosen these parameters because they have been studied extensively in previous work~\cite{santos2010onset,deutsch2013microscopic,santos2012weak} relevant to our paper. Evolution of the integrable system is solvable by ansatz~\cite{bethe1931theorie,karabach1997introduction}. The  non-integrable system displays level spacing statistics in good numerical agreement with the Wigner-Dyson distribution~\cite{mehta2004random,wigner1951class}; these parameters were also used to study thermalization~\cite{santos2010onset,rigol2008thermalization,alba2015eigenstate,beugeling2014finite}, and found to obey the Eigenstate Thermalization Hypothesis~\cite{deutsch1991quantum,srednicki1994chaos}, which has been successfully tested in experiments (e.g., \cite{kinoshita2006quantum,hofferberth2007non,trotzky2012probing}), reviewed in~\cite{polkovnikov2011nonequilibrium}.

We first test this model where initially the system of $N = 4$ particles  is confined by hard walls to $L=8$ sites and evolves through Hamiltonian $\hat{H}^{(1-8)}$. At time $t=0$ we change the
position of the right hard wall to $L=16$, and allow the fermionic gas to expand through evolution of the full Hamiltonian $\hat{H}^{(1-16)}$. We investigate the FOE and $S_{xE}$ for two different coarse-grainings, first when the full system has been coarse-grained into $p=m=4$ sites, and then coarse-graining into $p=m=2$ sites. The initial state is always taken to be the $11$th energy eigenstate of the Hamiltonian $\hat{H}^{(1-8)}$ reduced system. We also investigate also both non-integrable and integrable dynamics. Because the full Hamiltonian has an interaction term between the first and last $8$ sites, the local energy representation quickly populates many other basis vectors. As a result the entropy rapidly increases. Evolution of these Observational entropies is shown as a function of time in Fig.~\ref{fig:CoarseGrainedEigenstate} for both the non-integrable (generic) and integrable cases, and for the two different types of coarse-graining. We also plot the thermodynamic entropy $S(\R_{\mathrm{th}})$. This entropy is always above the curves, which corresponds with the theory, Eq.~\eqref{eq:bounds_on_FOE}.

As a comparison, we also calculate the entanglement entropy for both the non-integrable and integrable systems above. We measure this as a function time, starting at $t=0$, where the right hand wall is is moved from position $8$ to $16$. We measure the entanglement between the first $8$ and last $8$ sites. This is shown in Fig.~\ref{fig:EntanglementEntropy}. The entanglement entropy starts at zero because initially the wavefunction is zero on the right hand side of the box. The entanglement entropy grows and is expected for generic systems to go to $1/2$ of the total thermodynamic entropy of the system~\cite{deutsch2013microscopic}.

We now consider a different initial condition, and a slightly different scenario. We start the system in a pure state that models the
canonical ensemble. We consider our initial wave function as the sum over all energy eigenstates
\[
\ket{\psi} = \sum_E d_E \ket{E}
\]
with coefficients $d_E$ that are complex random values so that $|d_E|^2 \propto \exp(-\beta E)$. This state correspond to what we could call a ``pure thermal state,'' since the amplitudes have been taken randomly from the ensemble that imitates the canonical ensemble. In this case, we set the inverse temperature $\beta=1$. At $t=0$ the system in the smaller box with $L=8$ sites and let it evolve. At $t=30$, we expand the box to size $L=16$. Both entropies increase rapidly but smoothly, and in the non-integrable case they quickly reach equilibrium. Fig.~\ref{fig:FOE2}(a) shows Observational entropies $S_{xE}$ and $S_{F}$ as functions of time. Fig.~\ref{fig:FOE2}(b) shows the same situation for the integrable case. The horizontal lines show the thermodynamic entropy $S(\R_{\mathrm{th}})$. This differs from the computed limit values of the $S_{xE}$ and $S_{F}$ by approximately $10\%$, which we attribute to finite-size effects. In both Fig.~\ref{fig:CoarseGrainedEigenstate} and Fig.~\ref{fig:FOE2} the
fluctuations in $S_{xE}$ and $S_{F}$ are substantially larger  for integrable system dynamics, as expected~\cite{deutsch1991quantum,deutsch2013microscopic}.

To investigate behavior of these two entropies in more detail, we also plot $S_{xE}$ and $S_{F}$ as functions of energy for various equilibrium states as shown in Fig.~\ref{fig:SXEvsE} for both integrable and non-integrable system; this is particularly relevant for studying the long-time limit. Both entropies are coarse-grained into 4 subsystems ($p=m=4$) of the full system of size $L=20$, and computed for energy eigenstates, random PS states peaked around energy $E$, and microcanonical mixed states peaked around energy $E$. The random PS states were obtained by superposing $k=30$ neighboring energy eigenstates with complex amplitudes drawn uniformly from the unit disk, then normalizing. The microcanonical states were obtained by adding together the density matrices of $k=30$ neighboring energy eigenstates with equal weights. In the non-integrable case, all shapes fit very well the microcanonical entropy $S_{\rm micro}(E)$, defined in Eq.~\eqref{eq:microcanonical_entropy}. This is expected, because all states considered represent a microcanonical ensemble, since they are all peaked around a given value of energy, and confirms our analytical results presented in the previous section. One can also notice the order-unity differences between energy eigenstates, random superpositions, and microcanonical mixed states. These differences come from the randomness of phases and amplitudes and are predicted by the theory using Central Limit Theorem (see Appendix~\ref{sec-FOE_for_eigenstates}, Eq.~\eqref{eq:differences_in_FOE}). A graph that adjusts for these theoretically-estimated differences is depicted in Fig.~\ref{fig:SXEvsEsubtract}. The integrable case shows much larger fluctuations,  signifying that such systems do not thermalize well.

We also plot the thermodynamic (canonical) entropy $S(\R_{\mathrm{th}})$ for comparison. This entropy has a different shape: this is because for middle-range energies, the thermodynamic density matrix $\R_{th}$ contains many more non-negligible energy states, while towards the ends of energy spectrum, the number of non-negligible energy states dwindles and the thermodynamic density matrix practically becomes the microcanonical state. Contrary to the usual rule, $S_{F}(\R)\lessapprox S(\R_{th})$ (see Eq.~\eqref{eq:bounds_on_FOE}), thermodynamic entropy also drops below the FOE and $S_{xE}$ towards the ends. This happens because as the number of non-negligible energy eigenstates in thermal density matrix $\R_{th}$ goes to zero (which happens when we try to push the mean value of energy towards the end of the spectrum), the von Neumann entropy $S(\R_{th})$ approaches zero, and attains that value when $\R_{th}$ becomes a single energy eigenstate. But in such a situation, FOE is still non-zero, and is approximated by the microcanonical entropy. Similar arguments for exceptions to the rule $S_{F}(\R)\lessapprox S(\R_{th})$ have been already presented below Eq.~\eqref{eq:bounds_on_FOE}. This unusual effect becomes less evident for larger system sizes, and coarser coarse-grainings. Then $S_{F}(\R)\lessapprox S(\R_{th})$ holds for almost any energy $E$. For example, in Fig.~\ref{fig:SXEvsE} presented here, we used coarse-graining into four parts, $p=m=4$, to explicitly show this pathological but easily understandable behavior. But when we focused on coarse-graining only into two parts, $p=m=2$, this effect was much less obvious, and $S_{F}(\R)\lessapprox S(\R_{th})$ was violated only at the very ends of the energy spectrum.

Because the microcanonical entropy and thermodynamic (canonical) entropy are equivalent in the thermodynamic limit~\cite{ruelle1999statistical}, and since FOE and $S_{xE}$ of random PS states (meaning that both amplitudes and phases are random, which represent typical states of the system in the long-time limit) approximate well the microcanonical and thermodynamic entropy, this graph further supports the claim that has been presented in the previous section, that in the long-time limit, both FOE and $S_{xE}$ converge to the thermodynamic entropy in closed quantum systems.

\section{Connection with experiments}\label{sec:experiments}

Experimentally, it would be interesting to probe both the FOE and $S_{xE}$,
particularly in systems out of equilibrium. In cold atom experiments, it is possible to measure density, both of individual atoms and at a coarse grained scale~\cite{kaufman2016quantum,trotzky2012probing,hofferberth2007non,levin2012ultracold}. There have also been proposals for how to measure a system's total energy~\cite{Villa2018cavityassisted}.

For the FOE, we must perform a measurement in the local energy basis. Experimentally, this can be accomplished by increasing the height of the barrier separating the wells between the two regions. In a one-dimensional model,
this requires the creation of a secondary light field that can act as a potential barrier. Since non-periodic light fields are used in cold atom experiments~\cite{schreiber2015observation}, this appears to be feasible. Then the energy is measured for each region separately. Because of the potential barrier, these two regions are now noninteracting, so such measurements would constitute an extension of the total-energy method~\cite{Villa2018cavityassisted}.
Alternatively, other approximate estimates, based on
local quantities~\cite{kaufman2016quantum} could
be employed. Measurements would be performed, multiple times, each giving an energy for each region.  These measurements would give us the probability distribution over energies, from which we can compute FOE of this system.

To determine $S_{xE}$ experimentally, two measurements are perfomed: first of the coarse-grained density of the system,
and then its energy. The denominator in the observational entropy requires also knowing the volume $V_{xE}=|\braket{x}{E}|^2$. This is
the probability of observing a coarse-grained density for an energy eigenfunction. If we call the
measured coarse-grained density as a function of position $n(x)$, this implies we need to estimate
the probability density of obtaining a particular $n(x)$. If we confine ourselves
to coarse graining over length scales longer than the correlation length, density fluctuations for different $x$
should be independent. Therefore, the part of the observational entropy involving the
denominator can be estimated theoretically from equilibrium statistical mechanics.
The numerator (the probability distribution $p_{xE}$) can be measured experimentally by repeatedly measuring coarse-grained density and then energy. This gives data points in a space containing density bins for each coarse grained region, and energy bins.
After many repeated measurements, we would obtain the probability distribution $p_{xE}$ and compute $S_{xE}$. With small enough system sizes, comparable to ones currently
employed~\cite{kaufman2016quantum}, it might be within the bounds of current technology to
perform such measurements.  Even if it turns out that the resolution of
the apparatus is not fine-grained enough to get individual eigenstates,
an Observational entropy with finite energy coarse-graining can
still be calculated theoretically, and compared with experimental
data.

\section{Comparison with other measures and interpretation}
\label{sec-comparison}

\begin{table*}
\centering
\caption{\label{tab:relations}  Relation and connection of the Observational entropy to other measures}
\begin{ruledtabular}
\begin{tabular}{ll}
{\bf Information-theoretic or thermodynamic quantity} &  {\bf Relation(s)} \\
\hline
Boltzmann entropy $S_B(V_i)=\ln V_i$ for a macrostate $i$. & In analogy,
$S_{O(\C)}(\R)=\ln\dim\HS_i$ for $\R\in\HS_i$.\\
von Neumann entropy $S(\R)=-\tr[\R\ln\R]$  &  $S_{O(\C_1,\dots,\C_n)}(\R)\geq S(\R)$,\ \ \  $S(\R) = S_{O(\C_{\R})}(\R)$\\
Maximal entropy $S_{\max}\equiv S(\R_{\mathrm{id}})=\ln\dim\HS$ & $S_{O(\C_1,\dots,\C_n)}(\R)\leq S_{\max}$ \\

Microcanonical entropy $S_{\mathrm{micro}}(E)=\ln({\rho(E)\Delta E})$ &
$S_F(\R_t)\overset{t\rightarrow \infty}{\leadsto} S_{\mathrm{micro}}(E)$,  $S_{xE}(\R_t)\overset{t\rightarrow \infty}{\leadsto} S_{\mathrm{micro}}(E)$\\
 & in closed quantum non-integrable systems, for a \\
 & superposition of states peaked around energy $E$,\\
 & and $S_F(\ket{E})\approx S_{xE}(\ket{E})\approx S_{\mathrm{micro}}(E)$ for energy\\
 &  eigenstates $\ket{E}$.\\

Thermodynamic entropy $S_{th}\equiv S(\R_{th})=\ln Z +\sum_j \lambda_j \overline{A}_j$ & $S_F(\R_t)\lessapprox S_{th}$, $S_F(\R_{th})\approx S_{th}$, and \\
$\ \ \ \ \ \ \ \ \ \ $(Thermodynamic entropy is equal to microcanonical & $S_F(\R_t)\overset{t\rightarrow \infty}{\leadsto} S_{th}$ in quantum non-integrable systems\\
$\ \ \ \ \ \ \ \ \ \ \ $entropy in the thermodynamic limit.)&  interacting with a thermal bath.\\
Diagonal entropy $S_{\mathrm{diag}}(\R)$ with ``instantaneous'' Hamiltonian $\hat{H}$ & $S_{\mathrm{diag}}(\R)=S_{O{\displaystyle (}\C_{\hat{H}}{\displaystyle )}}(\R)$,  $S_F(\R_t)\lessapprox S_{\mathrm{diag}}(\R)$, and \\
$\ \ \ \ \ \ \ \ \ \ $(Assuming that the Hamiltonian is non-degenerate & $S_F(\R_t)\overset{t\rightarrow \infty}{\leadsto} S_{\mathrm{diag}}(\R)$ in closed quantum non-integrable\\
$\ \ \ \ \ \ \ \ \ \ $ and time-independent.) & systems for states with a high variance in energy.\\
Sum of local diagonal entropies $\sum_{i=1}^mS_{\mathrm{diag}}(\R_{i})$
&
$\sum_{i=1}^m S_{\mathrm{diag}}(\R_{i})=S_F(\R)+C(E_1,\dots,E_m)$\\
Kullback-Leibler divergence 
$D_{KL}$ & $S_{O(\C_1,\dots,\C_n)}(\R)=\ln\dim\HS - D_{KL}\big(P(\R)\big|\big|P\left(\R_{\mathrm{id}}\right)\!\!\big)$\\
Entropy of an observable $S_{\hat{A}}(\R)$ &
$S_{\hat{A}}(\R)=S_{O{\displaystyle (}\C_{\hat{A}}{\displaystyle )}}(\R)-\sum_ap_a\ln\tr\P_a$\\
\end{tabular}
\end{ruledtabular}
\end{table*}

In this section we compare the Observational entropy to other information-theoretic and thermodynamic quantities. Then we highlight the most prominent interpretations of the Observational entropy that we have encountered in this paper.

The relations between the Observational entropy and other quantities are collected in Table~\ref{tab:relations}. We have already discussed several such relations: we have shown that the Observational entropy is a quantum analog of the Boltzmann entropy (Theorem~\ref{thm:Boltzmann_equivalent}), that it is bounded below by the von Neumann entropy, and that it is bounded above by the maximal entropy (Theorems~\ref{thm:bounded} and~\ref{thm:bounded_multiple}). We have also shown that both  $S_{xE}$ and the factorized Observational entropy in non-integrable closed quantum systems converge (in a physical sense) to the microcanonical entropy for initial states that are a superposition of close energy eigenstates (Eq.~\eqref{eq:convergnce_to_microcanonical}), and that $S_{xE}$ and FOE of energy eigenstates are approximately equal to the the microcanonical entropy  (Eq.~\eqref{eq:approx_micro}, Appendices \ref{sec-S_xE_for_convergence} and \ref{sec-FOE_for_eigenstates}, and Fig.~\ref{fig:SXEvsE}). FOE is approximately bounded by the thermodynamic (canonical) entropy (Eq.~\ref{eq:bounds_on_FOE}), up to an order $\epsilon$ representing the interaction strength between partitions. In non-integrable quantum systems weakly interacting with a thermal bath, the FOE of the system converges to the canonical entropy (Eq.~\eqref{eq:convergence_to_cano}).

Now we turn to connection with entropy-related measures not detailed in previous sections.
First, there is an important connection with the \emph{Kullback-Leibler divergence}, which measures a distance between two probability distributions $P=\{p_i\}_i$ and $Q=\{q_i\}_i$ and is defined as $D_{KL}(P||Q)=\sum_ip_i\ln\frac{p_i}{q_i}$. Assuming that the dimension of the Hilbert space is finite, from the definitions it directly follows that
\[\label{eq:KL_divergence_entropy}
S_{O(\C_1,\dots,\C_n)}(\R)=\ln \dim \HS - D_{KL}\big(P(\R)\big|\big|P\left(\R_{\mathrm{id}}\right)\!\!\big),
\]
where $\R_{\mathrm{id}}=\frac{\hat{I}}{\dim \HS}$ and the probability distributions are defined as
\begin{subequations}
\begin{align}
P_{i_1,\dots,i_n}(\R)&=p_{i_1,\dots,i_n}=\tr\big[\P_{i_n}\cdots\P_{i_1}\R\P_{i_1}\cdots\P_{i_n}\big],\\
P_{i_1,\dots,i_n}\left(\R_{\mathrm{id}}\right)&=\frac{V_{i_1,\dots,i_n}}{\dim \HS}=\tr\big[\P_{i_n}\cdots\P_{i_1}\R_{\mathrm{id}}\P_{i_1}\cdots\P_{i_n}\big].
\end{align}
\end{subequations}

This identity shows that maximizing the Observational entropy is equivalent to finding the density matrix that produces statistics of measurement outcomes that is the closest to the statistics produced by the uniform (maximally uncertain) state $\R_{\mathrm{id}}$. Various tasks and uses of the Observational entropy in Sections~\ref{sec-entropy_increase}, \ref{sec-thermodynamics}, and \ref{sec-simulation} have shown that evolution of the system maximizes the Observational entropy subject to constraints. This provides the following prescription that is an interesting general statement about physical systems. Consider a set of density matrices that are in correspondence with the mean values of conserved quantities. Then {\em the density matrix evolves towards a density matrix from this set that has a probability distribution of measurement outcomes that most closely resembles the probability distribution produced by the uniform state.}

Observational entropy is also connected to  \emph{diagonal entropy}, which has been introduced in~\cite{tolman1938principles,ter1954elements}, mentioned in \cite{jaynes1957information2}, and developing in depth in~\cite{polkovnikov2011microscopic}. Diagonal entropy is the Shannon entropy of diagonal elements of the density matrix written in what is referred to as the ``instantaneous energy basis.'' By instantaneous it is meant that in the ideal case when the system is genuinely closed, the system is evolving according to the instantaneous Hamiltonian $\hat{H}$. The diagonal entropy can be then defined as $S_{\mathrm{diag}}(\R)=-\sum_E\bra{E}\R\ket{E}\ln \bra{E}\R\ket{E}$, where $\ket{E}$ are eigenvectors of the instantaneous Hamiltonian $\hat{H}$. Assuming that the Hamiltonian is non-degenerate, which is a typical assumption for non-integrable systems, the diagonal entropy can be written as the Observational entropy with the coarse-graining given by the instantaneous Hamiltonian, as listed in the table. Assuming that the system is genuinely closed and the system evolves according to the time-independent Hamiltonian $\hat{H}$, the diagonal entropy is identical to entropy $S(\R_d)$ (see Eqs.~\eqref{eq:diagonalEntropy} and \eqref{eq:bounds_on_FOE}), and according to Theorem~\ref{thm:constant_entropies}, it must stay constant. This behavior is mentioned in the pioneering paper~\cite{polkovnikov2011microscopic}, however, it is argued that it is impossible avoid transitions between different energy levels in the thermodynamic system of many particles. Therefore, a more general case of a time-dependent Hamiltonian is considered that may lead to such transitions, and the diagonal entropy defined by the instantaneous Hamiltonian (by which is meant $\hat{H}(t=0)$) increases. This is not in contradiction with Theorem~\ref{thm:constant_entropies}, because in such scenario the instantaneous Hamiltonian $\hat{H}$ that defines the coarse-graining does not commute with the actual Hamiltonian governing the evolution. The diagonal entropy has been also found to increase in other scenarios, for example when external operations are performed on the system~\cite{ikeda2015second}. In comparison to the diagonal entropy, both $S_{xE}$ and FOE rise even in a genuinely closed system described by a time-independent Hamiltonian, without the need to introduce transitions between different energy levels, or external operations.

A different situation occurs when we look at the sum of the local diagonal entropies, studied in the same paper~\cite{polkovnikov2011microscopic}. This sum is time-dependent even for genuinely closed systems evolving through time-independent Hamiltonian. We define diagonal entropy of region $i$ as $S_{\mathrm{diag}}(\R_{i})=S_{O{\displaystyle (}\C_{\hat{H}^{(i)}}{\displaystyle )}}(\R)$, with local Hamiltonian $\hat{H}^{(i)}$ defined by Eq.~\eqref{eq:H=H1+H2+Hint}, and the local density matrix defined as $\R_{i}=\tr_{\neg i}[\R]$, where the partial trace goes over all subsystems but $i$. For $m$ subsystems it turns out that
\[\label{eq:diagFOEcomparison}
\sum_{i=1}^m S_{\mathrm{diag}}(\R_{i})=S_F(\R)+C(E_1,\dots,E_m),
\]
where $C(E_1,\dots,E_m)=D_{KL}(p_{E_1\dots E_m}||p_{E_1}\cdots p_{E_m})$ is the {\em total correlation.} In case of $m=2$ regions, this quantity reduces to mutual information $I(E_1;E_2)$. This shows that while FOE takes into account correlations in energy of different subsystems, the sum of local diagonal entropies ignores them, and therefore overshoots the total entropy. This is also explains why sum of local diagonal entropies is larger than the total entropy in simulation performed in~\cite{polkovnikov2011microscopic}, while FOE is lower (Eq.~\eqref{eq:bounds_on_FOE}, Figs.~\ref{fig:CoarseGrainedEigenstate} and \ref{fig:FOE2}). Let us take a look at what this means in practice. Consider a situation where interaction between the regions is severed. In such a situation, both $\sum_{i=1}^m S_{\mathrm{diag}}(\R_{i})$ and $S_F(\R)$ become constant. Since the diagonal entropy is a good measure of thermodynamic entropy, elements $S_{\mathrm{diag}}(\R_{i})$ model thermodynamic entropy of each region, as if they were treated separately. Therefore, the sum corresponds to the total entropy of the system, treating the regions as independent. Qualitatively, such entropy then describes possible extractable work from the system, if one extracted work from each region one at the time, while ignoring correlations between them. $S_F$, on the other hand, corresponds also to the total entropy of the system, but without neglecting the correlations between the regions, leading to a lower total entropy, and therefore possibly larger extractable work. Thus we can speculate that FOE corresponds to the amount of extractable from from the entire system as a whole, where the protocols for the extraction from each region may be interdependent. For example if one extracts some amount of work from the first region, value of this amount may affect the protocol in which the work is extracted from the second region.

A mild generalization of the diagonal entropy is the \emph{entropy of an observable}~\cite{ingarden1962quantum,grabowski1977continuity,anza2017information}, which is the Shannon entropy of probability outcomes obtained by measuring an observable $\hat{A}$. Assuming the observable has spectral decoposition $\hat{A}=\sum_a a \P_a$, the probability of measuring outcome $a$ is given by $p_a=\tr[\P_a\R]$, and the entropy of an observable is defined as $S_{\hat{A}}(\R)=-\sum_a p_a\ln p_a$. Unlike the Observational entropy, entropy of an observable does not take into account uncertainty within a macrostate $\HS_a=\P_a\HS\P_a$, which is why these two entropies do not coincide in general. The relation can be easily derived to be $S_{\hat{A}}(\R)=S_{O{\displaystyle (}\C_{\hat{A}}{\displaystyle )}}(\R)-\sum_ap_a\ln\tr\P_a$.

It is also worthwhile briefly comparing the above approach of Observational entropy with a well-known entropy used for closed quantum systems, {\em the entanglement entropy} in a system divided into subsystems $A$ and $B$. Entanglement entropy can be interpreted various ways. For pure states, this entropy measures the mutual information between $A$ and $B$, and is defined as the von Neumann entropy of the reduced state, $S_A = -\tr[ \R_A\ln\R_A]$, where $\R_A=\tr_B[\R_{AB}]$.  In the context of qubits, the entanglement entropy is the number of entangled bits between $A$ and $B$. For a generic (i.e. non-integrable) system at nonzero temperature, we expect  this entanglement to be very close to its maximum, and theoretical and numerical results indicate that for a homogeneous system in equilibrium, this is the case. Taking system $A$ to be of lower dimension than $B$, one can think of $B$ as a bath for $A$, with $A$ corresponding to a system at some temperature chosen to match the system's total energy.
The entanglement entropy $S_A$ was then shown by~\cite{deutsch2010thermodynamic,deutsch2013microscopic,santos2012weak} to be the same as the thermodynamic entropy of $A$ in the limit of large system sizes. But it is a distinct quantity that is fundamentally different from $S_{xE}$ or FOE. For example, if the state is a product state of $A$ and $B$, then the entanglement entropy is zero, but $S_{xE}$ is not. We expect that even in such product states, the thermodynamic entropy of the complete system should still be large, and thus the entanglement entropy cannot give us a sensible measure, at least in this case, for the thermodynamic entropy. $S_{xE}$ is largely unaffected by this lack of entanglement for short ranged systems. Entanglement entropy also bears only indirect connection to macrostates and macro-observables that can be tracked and measured by an observer of the system, and its relation to classical entropy is somewhat less clear.

To complete the picture, we collect the most prominent interpretations of the Observational entropy encountered in this paper, as follows.
\begin{enumerate}
  \item (information-theoretic) Given a set of measurements, the Observational entropy represents the mean uncertainty in the measurement outcomes (in the sense of the mean information that would be gained by performing them) plus the mean remaining uncertainty about the system after these measurements. (Eq.~\eqref{eq-splitup})
  \item (statistical) The Observational entropy measures how closely the probability distribution of outcomes (of measurements given by the set of coarse-grainings) resembles the probability distribution of outcomes produced by the maximally uncertain state. (Eq.~\eqref{eq:KL_divergence_entropy})
  \item (physical-subjective) The Observational entropy measures how much the state of the system differs from what the observer thinks of as an ordered state, where the perceived (subjective) order is given by the choice of the coarse-graining. (The system is ordered when it is contained within a small macrostate.) Growth of the Observational entropy then describes the loss of perceived order of the system due to the time evolution. (Sec.~\ref{sec-entropy_increase} and
Eq.~\eqref{eq:observational_entropy_evolution})
\item (information theoretic/physical-subjective) Observational entropy measures how much information an observer would obtain about the system, if he or she would measure the system in the bases given by the macrostates. The von Neumann entropy then describes the lowest uncertainty observer can have about the system. (Theorem~\ref{thm:bounded_multiple} and Theorem~\ref{thm:non-increase}).
\item (thermodynamic)
Considering a system consisted of smaller subsystems, where the coarse-graining of the system is given by a tensor product of thermodynamical observables of subsystems, the Factorized Observational entropy measures how close these subsystems are to being in thermal equilibrium with each other. (Sec.~\ref{sec-thermodynamics}, Def.~\ref{def:factorized_observational_entropy}, and Eq.~\eqref{eq:convergence_combined})
\end{enumerate}

\section{Conclusions and prospects}
\label{sec-conclusions}

In this paper we have developed the theory behind Observational entropy, introduced earlier by the present authors in~\cite{SafranekDeutschAguirreObservationalEntropyLetter}. Although similar ideas have occasionally been mentioned since von Neumann in 1927~\cite{von2010proof}, this approach was until now essentially unexplored.  The quantity is crisply defined in terms of a Hilbert space partitioned by one or more ordered sets of operators corresponding to sequences of potential measurements, and the probabilities of outcomes of those measurements.  The partitioning provides an operational definition of macrostates in terms of an observer's potential measurements; the Observational entropy is related to the uncertainty inherent in those un-made measurements as given by their outcome probabilities, combined with the uncertainty that would remain after making them.

We have argued that this captures the real physical effect that macroscopically-defined ``disorder" tends to increase in physical systems, even under unitary evolution describing the closed-system dynamics. Underlying this argument are a large number of formal mathematical results revealing desirable and appropriate properties of the definition for describing entropy, as well as a suite of numerical investigations of simulated quantum systems that connect Observational entropy to standard quantities such as thermodynamic entropy.

While analogous to the classical Boltzmann entropy, Observational entropy has crucial differences stemming from its quantum context.  Boltzmann entropy is generally defined on a phase space (with an appeal to the quantum effect of non-commuting position and momentum to regularizing the minimal size of phase-space bins.)  It is thus not obvious how to generalize Boltzmann entropy to quantum systems defined using Hilbert space. Observational entropy does so in a general and rigorous way. But rather than ``glossing over'' fundamentally quantum effects, we find that they are crucial in the particular forms of Observational entropy that we have found to correspond to thermodynamic entropy.

For example, we have found that Observational entropy corresponding to measuring  coarse-grained position, and then measuring energy, defines a non-equilibrium entropy (denoted $S_{xE}$) that converges to the thermodynamic entropy. On the other hand, switching the order of operations immediately gives the total entropy of the system (which is constant) in case of non-degenerate Hamiltonian, but is difficult to interpret in case of degenerate Hamiltonian. Another special case of Observational entropy, which we called the Factorized Observational entropy (FOE), is based on a factorization of operators corresponding to conserved quantities (energy in particular) and generally describes situations where local systems are  equilibrating with each other. Similarly to $S_{xE}$, this entropy is well-defined out of equilibrium, and converges to the thermodynamic entropy even for genuinely closed quantum systems. Moreover, this entropy naturally incorporates micro-canonical, canonical, grand-canonical, and other ensembles, based on the specific physical situations.

Both $S_{xE}$ and FOE ``work" very well in the sense of giving a close approximation to thermodynamic quantities even in quite small quantum systems. Even for as few as $4$ particles contained on $16$ sites, the difference between the relevant Observational entropies and thermodynamic entropies fell within $10\%$, and the relative change of such entropies from one equilibrium situation to another was under $5\%$ as compared to change in equilibrium entropies. From general arguments, these differences are expected to become unimportant as a system is scaled up in number of constituents. Thus, for thermodynamic systems where the number of particles cannot be counted on one hand, these entropies should give an extremely accurate measure of thermodynamic entropy (indeed we might argue that these {\em are} the quantities thermodynamically measured) while also being well-defined and applicable in small systems and out of equilibrium.

An open question is the precise connection between the entropies we have discussed and work extraction. Considering a system consisted of smaller subsystems, we have speculated that the Factorized Observational entropy measures the amount of extractable work from the system as a whole, including correlations between the subsystems.

While this paper has focused on developing the mathematical framework and basic properties of observational entropy, the theory merits further development and there could be a great number of applications for it.

Experimentally, the definition could be quite directly applied to simple ``closed" systems resembling those we have simulated. In cold atoms, experiments on isolated quantum systems are now becoming feasible~\cite{kaufman2016quantum,trotzky2012probing}, and as we explain in Section~\ref{sec:experiments}, measuring thermodynamically relevant Observational entropies in such systems could be within experimental reach.

Theoretically, there are many important  results -- including fluctuations theorems, limits on work extraction, computation, etc. -- that are formulated in classical statistical mechanics and are lacking a convincing quantum generalization.  Observational entropy and its related formalism could supply a framework for creating such generalizations. This could eventually have practical applications in thermodynamic systems using few quantum particles (such as nano-engines) or in refrigeration at extremely low temperatures.
Observational entropy may also elucidate situations in which ``the observer" plays a major role, such as in Maxwell's demon and information engines in general, or in the difference between thermodynamic entropies ascribed to the same physical system by two observers with different knowledge (the ``Gibbs paradox.'')

Finally, there is a great amount of work in fundamental physics, including gravitational physics and cosmology, concerning entropy of black holes, general horizons, the Universe as a whole, etc.  Most of these works take ``entropy" to correspond to either the size of the full Hilbert space, or entanglement entropy.  In some subset of these investigations, however, we suspect Observational entropy may be the more appropriate notion. It will therefore be very interesting to see if black-hole thermodynamics, the cosmological arrow of time, and other vexing issues might be elucidated by this new framework.

\acknowledgements{We are grateful to Tom Banks, Onuttom Narayan, Benjamin Lev, Joseph C. Schindler, and Dana Faiez for helpful discussions. This research was supported by the Foundational Questions Institute (FQXi.org), of which AA is Associate Director, and by the Faggin Presidential Chair Fund.}

\appendix
\section{Proofs}\label{app:proofs}

In this appendix we are going to provide proofs for all theorems in the main text. To prove theorems~\ref{thm:monotonic}, \ref{thm:bounded},~\ref{thm:bounded_multiple}, and \ref{thm:non-increase} we use the well known Jensen's inequality, which we state as follows:
\begin{theorem}(Jensen)\label{thm:Jensen}
Let $f$ be a strictly concave function, $0\leq a_i \leq 1$, $\sum_i a_i=1$. Then for any $b_i\in\mathbb{R}$,
\[
f\big(\sum_i a_i b_i\big)\geq \sum_i a_i f(b_i).
\]
$f(\sum_i a_i b_i)= \sum_i a_i f(b_i)$ if and only if
$
(\forall i,j|a_i\neq 0, a_j\neq 0)(b_i=b_j).
$
\end{theorem}

\subsection{Proof of Theorem~\ref{thm:Boltzmann_equivalent}}
\begin{proof}
If $\P_i\R\P_i=\R$, then $p_i=\tr[\P_i\R\P_i]=\tr[\R]=1$. Using Eq.~\eqref{def:oe} and $\tr{\P_i}=\mathrm{dim}\HS_i$ we immediately obtain $S_O(\R)=\ln \dim\HS_i$.
\end{proof}

\subsection{Proof of Theorem~\ref{thm:monotonic}}
\begin{proof}
Let $\C_1\hookrightarrow \C_2$. Then $\P_{i_1}=\sum_{i_2\in I^{(i_1)}}\P_{i_2}$ for all $\hat{P}_{i_1}\in \C_1$. Inequality follows
\[
\begin{split}
&S_{O(\C_1)}(\R)=-\sum_{i_1}\tr[\R\hat{P}_{i_1}]\ln \frac{\tr[\R\hat{P}_{i_1}]}{\tr\hat{P}_{i_1}}\\
&=-\sum_{i_1}\tr[\R{\textstyle \sum_{i_2\in I^{(i_1)}}\P_{i_2}}]\ln \frac{\tr[\R\sum_{i_2\in I^{(i_1)}}\P_{i_2}]}{\tr\hat{P}_{i_1}}\\
&=\sum_{i_1}\tr\hat{P}_{i_1}\bigg(-\sum_{i_2\in I^{(i_1)}}\frac{p_{i_2}}{\tr\hat{P}_{i_1}}\ln \sum_{i_2\in I^{(i_1)}}\frac{p_{i_2}}{\tr\hat{P}_{i_1}}\bigg)\\
&=\sum_{i_1}\tr\hat{P}_{i_1}\bigg(-\sum_{i_2\in I^{(i_1)}}\frac{\tr\hat{P}_{i_2}}{\tr\hat{P}_{i_1}}\frac{p_{i_2}}{\tr\hat{P}_{i_2}}\ln \sum_{i_2\in I^{(i_1)}} \frac{\tr\hat{P}_{i_2}}{\tr\hat{P}_{i_1}}\frac{p_{i_2}}{\tr\hat{P}_{i_2}}\bigg)\\
&\geq \sum_{i_1}\tr\hat{P}_{i_1}\bigg(-\sum_{i_2\in I^{(i_1)}}\frac{\tr\hat{P}_{i_2}}{\tr\hat{P}_{i_1}}\frac{p_{i_2}}{\tr\hat{P}_{i_2}}\ln \frac{p_{i_2}}{\tr\hat{P}_{i_2}}\bigg)\\
&=-\sum_{i_1}\sum_{i_2\in I^{(i_1)}}p_{i_2}\ln \frac{p_{i_2}}{\tr\hat{P}_{i_2}}=S_{O(\C_2)}(\R),
\end{split}
\]
where we have chosen a strictly concave function $f(x)=-x\ln x$, $a_{i_2}=\frac{\tr\hat{P}_{i_2}}{\tr\hat{P}_{i_1}}$ and $b_{i_2}=\frac{p_{i_2}}{\tr\hat{P}_{i_2}}$ for $i_2\in I^{(i_1)}$ for the Jensen's inequality, which proves the theorem.

The equality conditions from the Jensen's inequality show that $S_{O(\C_1)}(\R)=S_{O(\C_2)}(\R)$ if and only if
\[\label{eq:equality_monotonic}
(\forall i_1)(\forall i_2,\tilde{i}_2\in I^{(i_1)})\left(\frac{p_{i_2}}{\tr \P_{p_{i_2}}}=\frac{p_{\tilde{i}_2}}{\tr \P_{p_{\tilde{i}_2}}}=c^{(i_1)}\right).
\]
To determine the constant $c^{(i_1)}$ we multiply the equation by $\tr \P_{p_{\tilde{i}_2}}$ and sum over all $\forall i_2\in I^{(i_1)}$, which gives
\[
c^{(i_1)}=\frac{p_{i_1}}{\tr\P_{i_1}}.
\]
Therefore, $S_{O(\C_1)}(\R)=S_{O(\C_2)}(\R)$ if and only if
\[
(\forall i_1)(\forall i_2\in I^{(i_1)})\left(p_{i_2}=\frac{\tr\P_{p_{i_2}}}{\tr\P_{i_1}}p_{i_1}\right).
\]
\end{proof}

\subsection{Proof of Theorem~\ref{thm:bounded} and Theorem~\ref{thm:bounded_multiple}}
\begin{proof}
Since Theorem~\ref{thm:bounded} is a special case of Theorem~\ref{thm:bounded_multiple}, we are going to prove only Theorem~\ref{thm:bounded_multiple}.  First we prove $S(\R)\leq S_{O(\C_1,\dots,\C_n)}(\R)$ plus the equality condition and then $S_{O(\C_1,\dots,\C_n)}(\R)\leq \ln \dim \HS$ plus the equality condition. Before we start we define necessary notation. We define the spectral decomposition of the density matrix in terms of its eigenvectors as $\R=\sum_x\rho_x\pro{x}{x}$ where eigenvalues $\rho_x$ do not have to be necessarily different for different $x$, and therefore this decomposition is not unique. We also define of the density matrix in terms of its projectors $\R=\sum_{\rho}\rho\P_{\rho}$, where eigenvalues $\rho$ are now different from each other. This decomposition is unique. It follows that for each $x$ there exists $\lambda$ such that $\rho_x=\lambda$. We define a multi-index $\bi=(i_1,\dots,i_n)$, probability of the state being in multi-macrostate $\bi$,
\[
p_\bi\equiv\tr\big[\P_{i_n}\cdots\P_{i_1}\R\P_{i_1}\cdots\P_{i_n}\big],
\]
and volume of multi-macrostate $\bi$,
\[
V_\bi\equiv\tr[\P_{i_n}\cdots\P_{i_1}\cdots\P_{i_n}].
\]

Now we prove $S(\R)\leq S_{O(\C_1,\dots,\C_n)}(\R)$ plus the equality condition.
Defining
\[
a_x^{(\bi)}\equiv\frac{\bra{x}\P_{i_1}\cdots\P_{i_n}\cdots\P_{i_1}\ket{x}}{V_\bi}
\]
for $V_\bi\neq 0$ and $a_x^{(\bi)}\equiv 0$ for $V_\bi= 0$, and then using the spectral decomposition of $\R$ we have
\[\label{eq:p_over_V}
\frac{p_\bi}{V_\bi}=\frac{\sum_x \rho_x \bra{x}\P_{i_1}\cdots\P_{i_n}\cdots\P_{i_1}\ket{x}}{V_\bi}=\sum_x \rho_x a_x^{(\bi)}.
\]
Using the cyclic property of trace, $V_\bi=\tr[\P_{i_1}\cdots\P_{i_n}\cdots\P_{i_1}]=\sum_x\bra{x}\P_{i_1}\cdots\P_{i_n}\cdots\P_{i_1}\ket{x}$, we derive
\[\label{eq:sum_ax}
\sum_xa_x^{(\bi)}=1.
\]
Using the fact that sets of projectors form a complete set, $\sum_{i_k} \P_{i_k}=\hat{I}$, we also have
\[\label{sum_V_ax}
\sum_\bi V_\bi a_x^{(\bi)}=\sum_\bi\bra{x}\P_{i_1}\cdots\P_{i_n}\cdots\P_{i_1}\ket{x}=\braket{x}{x}=1.
\]

Series of equalities and inequalities follow
\[
\begin{split}
&S_{O(\C_1,\dots,\C_n)}(\R)=-\sum_{\bi}p_\bi\ln \frac{p_\bi}{V_{\bi}}\\
&=-\sum_{\bi}V_{\bi}\frac{p_\bi}{V_{\bi}}\ln \frac{p_\bi}{V_{\bi}}\\
&=\sum_{\bi}V_{\bi}\left(-\sum_x \rho_x a_x^{(\bi)}\ln \sum_x \rho_x a_x^{(\bi)}\right)\\
&\geq \sum_{\bi}V_{\bi} \left(-\sum_x a_x^{(\bi)} \rho_x\ln \rho_x\right)\\
&=-\sum_x\left( \sum_{\bi}V_{\bi} a_x^{(\bi)}\right) \rho_x\ln \rho_x=S(\R).
\end{split}
\]
The third equality comes from Eq.~\eqref{eq:p_over_V}, and the last equality comes from Eq.~\eqref{sum_V_ax}. We have applied the Jensen's Theorem (Theorem~\ref{thm:Jensen}) on strictly concave function $f(x)=-x\ln x$ to derive the inequality. We have chosen $a_x\equiv a_x^{(\bi)}$ and $b_x=\rho_x$ for the Theorem. This is a valid choice because of $0\leq a_x^{(\bi)}\leq 1$ and Eq.~\eqref{eq:sum_ax}. This proves the first inequality.

According to the Jensen's Theorem, the inequality becomes equality if and only if
\[\label{eq:first_eq_condition_old}
\begin{split}
&(\forall \bi)(\forall x, \tilde{x}|\bra{x}\P_{i_1}\!\cdots\P_{i_n}\!\cdots\P_{i_1}\ket{x}\!\neq\! 0, \bra{\tilde{x}}\P_{i_1}\!\cdots\P_{i_n}\!\cdots\P_{i_1}\ket{\tilde{x}}\!\neq\! 0)\\
&(\rho_x=\rho_{\tilde{x}}).
\end{split}
\]
To explain, the inequality becomes equality when for a given multi-index $\bi$, all eigenvectors of the density matrix $\ket{x}$ such that $\bra{x}\P_{i_1}\cdots\P_{i_n}\cdots\P_{i_1}\ket{x}\neq 0$ have the same associated eigenvalue $\rho_x$ with them. In other words, we can associate this unique eigenvalue to the multi-index $\bi$ itself, $\rho_\bi \equiv\rho_x$, where $\rho_x$ is given by any representative $x$ such that $\bra{x}\P_{i_1}\cdots\P_{i_n}\cdots\P_{i_1}\ket{x}\neq 0$. For the inequality to become equality this must hold for every multi-index $\bi$. Thus we have a unique map which attaches some eigenvalue of the density matrix to each multi-index $\bi$. In addition, realizing that from the definition of norm follows $\bra{x}\P_{i_1}\cdots\P_{i_n}\cdots\P_{i_1}\ket{x}\neq 0$ if and only if $\P_{i_n}\cdots\P_{i_1}\ket{x}\neq 0$, we can write Eq.~\eqref{eq:first_eq_condition_old} as
\[\label{eq:first_eq_condition}
(\forall \bi)(\forall x, \tilde{x}|\ \P_{i_n}\!\cdots\P_{i_1}\ket{x}\!\neq\! 0,\P_{i_n}\!\cdots\P_{i_1}\ket{\tilde{x}}\!\neq\! 0)(\rho_x=\rho_{\tilde{x}}\equiv\rho_\bi).
\]
Defining set
\[
I^{(\bi)}=\{x|\rho_x=\rho_\bi\},
\]
using the above condition, and $\sum_x\ket{x}\bra{x}=\hat{I}$, we can write
\[\label{eq:finer_rho}
\begin{split}
\P_{i_n}\cdots\P_{i_1}&=\P_{i_n}\cdots\P_{i_1}\sum_x\ket{x}\bra{x}=\sum_x\P_{i_n}\cdots\P_{i_1}\ket{x}\bra{x}\\
&=\sum_{x\in I^{(\bi)}}\P_{i_n}\cdots\P_{i_1}\ket{x}\bra{x}=\P_{i_n}\cdots\P_{i_1}\sum_{x\in I^{(\bi)}}\ket{x}\bra{x}\\
&=\P_{i_n}\cdots\P_{i_1}\P_{\rho_\bi}
\end{split}
\]
The third equality holds because for every $x\notin I^{(\bi)}$, $\P_{i_n}\!\cdots\P_{i_1}\ket{x}= 0$, so these terms disappear in the sum. $\P_{\rho_\bi}$ denotes a projector associated with eigenvalue $\rho_\bi$ from the uniquely defined spectral decomposition of the density matrix, $\R=\sum_{\rho}\rho\P_{\rho}$. For every multi-index $\bi$ we have found a projector $\P_{\rho_\bi}\in\C_\R$ such that Eq.~\eqref{eq:finer_rho} holds, which by Def.~\ref{def:finer_set_coarse_graining} means that $\C_\R\hookrightarrow (\C_1,\dots,\C_n)$.

Now that we have shown implication $S(\R)=S_{O(\C_1,\dots,\C_n)}(\R)\Rightarrow \C_\R\hookrightarrow (\C_1,\dots,\C_n)$, we will make sure that the opposite implication also holds. By multiplying Eq.~\eqref{eq:finer_rho} by $\P_{\rho}$, where $\rho\neq \rho_\bi$, from the orthogonality of projectors we find
\[
\P_{i_n}\cdots\P_{i_1}\P_{\rho}=\P_{i_n}\cdots\P_{i_1}\P_{\rho_\bi}\P_{\rho}=0.
\]
Therefore, assuming Eq.~\eqref{eq:finer_rho} holds, we compute
\[
\begin{split}
p_\bi&=\tr\big[\P_{i_n}\cdots\P_{i_1}\sum_{\rho}\rho\P_{\rho}\P_{i_1}\cdots\P_{i_n}\big]\\
&=\rho_\bi\tr\big[\P_{i_n}\cdots\P_{i_1}\P_{\rho_\bi}\P_{i_1}\cdots\P_{i_n}\big]\\
&=\rho_\bi\tr\big[\P_{i_n}\cdots\P_{i_1}\cdots\P_{i_n}\big]=\rho_\bi V_\bi.
\end{split}
\]
Moreover, using Eq.~\eqref{eq:finer_rho} we have
\[
\begin{split}
\tr[\P_{\rho}]&=\sum_\bi\tr[\P_{i_n}\cdots\P_{i_1}\P_{\rho}\P_{i_1}\cdots\P_{i_n}]\\
&=\sum_{\bi\in I^{(\rho)}}\tr[\P_{i_n}\cdots\P_{i_1}\P_{\rho}\P_{i_1}\cdots\P_{i_n}]\\
&=\sum_{\bi\in I^{(\rho)}}\tr[\P_{i_n}\cdots\P_{i_1}\cdots\P_{i_n}]=\sum_{\bi\in I^{(\rho)}}V_\bi,
\end{split}
\]
where $I^{(\rho)}=\{\bi|\rho_\bi=\rho\}$. The second equality holds because $\P_{i_n}\cdots\P_{i_1}\P_{\rho}=0$ for $\bi\notin I^{(\rho)}$. Combining the above two equations we derive
\[
\begin{split}
S_{O(\C_1,\dots,\C_n)}(\R)&=-\sum_{\bi}\rho_\bi V_\bi\ln \frac{\rho_\bi V_\bi}{V_{\bi}}=-\sum_{\rho}\bigg(\sum_{\bi\in I^{(\rho)}} V_\bi \bigg)\rho \ln \rho\\
&=-\sum_{\rho}\tr[\P_{\rho}]\rho \ln \rho=S(\R).
\end{split}
\]
This concludes the proof of the equality conditions $S(\R)=S_{O(\C_1,\dots,\C_n)}(\R)$.

Now we prove $S_{O(\C_1,\dots,\C_n)}(\R)\leq \ln \dim \HS$ plus the equality condition.
\[
\begin{split}
&S_{O(\C_1,\dots,\C_n)}(\R)=\sum_{\bi:p_\bi\neq 0}p_\bi\ln \frac{V_{\bi}}{p_\bi}\leq \ln\left(\sum_{\bi:p_\bi\neq 0}p_\bi\frac{V_{\bi}}{p_\bi}\right)\\
&\leq \ln\left(\sum_{\bi}V_{\bi}\right)=\ln \tr{\hat{I}}=\ln \dim\; \!\HS.
\end{split}
\]
The first inequality comes from the Jensen's Theorem applied on strictly concave function $f(x)=\ln x$ when choosing $a_\bi\equiv p_\bi$ and $b_\bi\equiv\frac{V_{\bi}}{p_\bi}$ for the Theorem. $0\leq a_\bi\leq 1$ and $\sum_\bi a_\bi=1$ so this is a valid choice. The second inequality comes from $V_i\geq 0$ and the fact that logarithm is an increasing function. The second equality comes from $\sum_{i_k} \P_{i_k}=\hat{I}$ and the definition of $V_\bi$.

The first inequality becomes identity if and only if
\[
(\forall \bi, \boldsymbol{j}| p_\bi\neq 0, p_{\boldsymbol{j}}\neq 0)\left(\frac{V_\bi}{p_\bi}=\frac{V_{\boldsymbol{j}}}{p_{\boldsymbol{j}}}=c\right)
\]
where $c$ is some real constant. To determine this constant we express the condition as $V_\bi=c p_\bi$ and sum over all multi-indexes $\bi$ such that $p_\bi\neq 0$, which gives $c=\sum_{\bi:p_\bi\neq 0}V_\bi$. The first equality condition can be then written as
\[
(\forall p_\bi\neq 0)\left(p_\bi=\frac{V_\bi}{\sum_{\bi:p_\bi\neq 0}V_\bi}\right).
\]
Since logarithm is a strictly increasing function, the second inequality becomes equality if and only if for all $\bi$ such that $p_\bi=0$ also $V_\bi=0$. Assuming the second condition is satisfied, we can write $\sum_{\bi:p_\bi\neq 0}V_\bi=\sum_{\bi}V_\bi=\dim \HS$ for the first condition, which comes from $\sum_{i_k}\P_{i_k}=\hat{I}$ and the definition of $V_\bi$. Combining both equality conditions yields that $S_{O(\C_1,\dots,\C_n)}(\R)=\ln \dim\; \!\HS$ if and only if
\[
(\forall p_\bi)\left(p_\bi=\frac{V_\bi}{\dim \HS}\right),
\]
which completes the proof.
\end{proof}

\subsection{Proof of Corollary~\ref{thm:pure_state_max_entropy}}
\begin{proof}
We have $p_i=\frac{\tr \P_i}{\mathrm{dim}\HS}$ for both $\R$ and $\R_{\mathrm{id}}$. The statement therefore follows directly from Theorem~\ref{thm:bounded}.
\end{proof}

\subsection{Proof of Theorem~\ref{thm:extensivity}}
\begin{proof}
The statement follows from $\tr[\P_{i_1}\otimes\dots\otimes\P_{i_m}\R^{(1)}\otimes\cdots\otimes\R^{(m)}]=\tr[\P_{i_1}\R^{(1)}]\cdots\tr[\P_{i_m}\R^{(m)}]$, $\tr[\P_{i_1}\otimes\dots\otimes\P_{i_m}]=\tr[\P_{i_1}]\cdots\tr[\P_{i_m}]$, and from the properties of logarithm.
\end{proof}

\subsection{Proof of Theorem~\ref{thm:constant_entropies}}
\begin{proof}
Since for all $\P_i\in\C$, $[\P_i,\hat{H}]=0$ we have $p_i=\tr[\P_iU(t)\R_0 U(t)^\dag]=\tr[U(t)\P_i\R_0 U(t)^\dag]=\tr[\P_i\R_0]$ which proves the first part of the Theorem.
Hermitian operators commute if and only if projectors from their spectral decompositions commute~\cite{blank2008hilbert}. In other words, assuming $\hat{A}=\sum_ia_i\P_i$ and $[\hat{A},\hat{H}]=0$ implies $[\P_i,\hat{H}]=0$ for every $i$ which concludes the proof.
\end{proof}

\subsection{Proof of Theorem~\ref{second_law_thermo}}
\begin{proof}
We assume that
\[\label{eq:assumption_for_pi}
p_i\equiv\tr[\P_i\R_t] \geq p_i^{(max)}\equiv\frac{1}{1+\frac{\min_{k\neq i}\tr[\P_k]}{\tr[\P_i]}},
\]
where $p_i^{(max)}$ is the point where function
\[\label{eq:function_f}
f(p_i)=-p_i\ln\frac{p_i}{\tr[\P_i]}-(1-p_i)\ln\frac{1-p_i}{\min_{k\neq i}\tr[\P_k]}
\]
achieves its maximum. Then
\[
\begin{split}
&S_{O(\C)}(\R_{t})=-\sum_kp_k\ln\frac{p_k}{\tr[\P_k]}\\
&=-p_i\ln\frac{p_i}{\tr[\P_i]}-\sum_{k\neq i}p_k\ln p_k+\sum_{k\neq i}p_k\ln \tr[\P_k]\\
&=-p_i\ln\frac{p_i}{\tr[\P_i]}-\sum_{k\neq i}p_k\sum_{k\neq i}\bigg(\frac{p_k}{\sum_{k\neq i}p_k}\bigg)\ln \bigg(\frac{p_k}{\sum_{k\neq i}p_k}\bigg)\\
&-\sum_{k\neq i}p_k\ln\sum_{k\neq i}p_k+\sum_{k\neq i}p_k\ln \tr[\P_k]\\
&\geq-p_i\ln\frac{p_i}{\tr[\P_i]}+0-\sum_{k\neq i}p_k\ln\sum_{k\neq i}p_k+\sum_{k\neq i}p_k\ln \min_{k\neq i}\tr[\P_k]\\
&=-p_i\ln\frac{p_i}{\tr[\P_i]}-(1-p_i)\ln\frac{1-p_i}{\min_{k\neq i}\tr[\P_k]}\\
&\geq\ln\tr[\P_i]=S_{O(\C)}(\R_0)
\end{split}
\]
The first inequality holds because the second term after the third equal sign is positive (it is a Shannon entropy) and because logarithm in the fourth term is an increasing function. We have used $\sum_{k\neq i}p_k=1-p_i$ for the equality that follows. The second inequality holds because of assumption~\eqref{eq:assumption_for_pi} and because function $f$ from Eq.~\eqref{eq:function_f} is a decreasing function on interval $p_i^{(max)}\leq p_i\leq 1$.

All we have to do now is to find how small time $t$ must be such that the assumption~\eqref{eq:assumption_for_pi} holds. Since $\P_i\R_0\P_i=\R_0$, then $\tr[\P_i\R_0]=1$ and Eq.~\eqref{eq:assumption_for_pi} can be rewritten as
\[\label{eq:assumption_for_pi2}
\tr[\P_i(\R_0-\R_{t})]\leq \frac{1}{1+\frac{\tr[\P_i]}{\min_{k\neq i}\tr[\P_k]}}.
\]
Expanding the left hand side up to the second order in $t$ using $\R_t=U(t)\R_0 U(t)^\dag$, $U(t)=e^{-i\hat{H}t}$, and $\P_i\R_0\P_i=\R_0$, we find
\[
\tr[\P_i(\R_0-\R_{t})]=\tr\big[(\hat{I}-\P_i)\hat{H}\R_0\hat{H}\big]t^2+o(t^2),
\]
where $o(t^2)$ denotes scaling in the little-o notation, $\lim_{t\rightarrow 0}\frac{o(t^2)}{t^2}=0$. Inserting this expression into Eq.~\eqref{eq:assumption_for_pi2} and ignoring term $o(t^2)$ yields
\[
t\lessapprox \left(\tr\big[(\hat{I}-\P_i)\hat{H}\R_0\hat{H}\big]\left(1+\frac{\tr[\P_i]}{\min_{j\neq i}\tr[\P_j]}\right)\right)^{-\frac{1}{2}},
\]
which proves the Theorem.
\end{proof}

\subsection{Proof of Lemma~\ref{thm:joint_coarsegraining}}
\begin{proof}
We prove the uniqueness first. We assume that two joint coarse-grainings $\C_{1,2}^{(1)}=\{\P_k^{(1)}\}_k$ and  $\C_{1,2}^{(2)}=\{\P_l^{(2)}\}_l$ both satisfy Eq.~\eqref{def:joint_coarsegraining}. Then by definition $\C_{1,2}^{(1)}\hookrightarrow \C_{1,2}^{(2)}$ and $\C_{1,2}^{(2)}\hookrightarrow \C_{1,2}^{(1)}$, thus
\[
\begin{split}
\P_l^{(2)}=\sum_{k\in I_l}\P_k^{(1)}=\sum_{k\in I_l}\sum_{\tilde l\in I_k}\P_{\tilde l}^{(2)}.
\end{split}
\]
Both index sets $I_k$ and $I_l$ must contain a single element. If they did not, then there would be an index $\tilde l\neq l$ and a non-zero vector $\ket{\psi}\in \HS_{\tilde l}$ such that $0=\P_l^{(2)}\ket{\psi}=\sum_{k\in I_l}\sum_{\tilde l\in I_k}\P_{\tilde l}^{(2)}\ket{\psi}=\ket{\psi}$. Therefore for every $l$ there exists exactly one $k$ such that $\P_l^{(2)}=\P_k^{(1)}$ and vice versa. In other words, sets $\C_{1,2}^{(1)}$ and $\C_{1,2}^{(2)}$ are identical.

Now we prove the second part of the Theorem. Clearly, coarse-graining given by $\{\P_{i_1}\P_{i_2}\}_{{i_1},{i_2}}\!\setminus\!\{0\}$ is finer than both $\C_{1}$ and $\C_{2}$. All we need to prove that it is the roughest such coarse-graining. Let $\C=\{\hat{\tilde{P}}_k\}_k$ be such that $\C_{1}\hookrightarrow \C$ and $\C_{2}\hookrightarrow \C$. We choose $\P_{i_1}\P_{i_2}\neq 0$. Then
\[
\begin{split}
\P_{i_1}\P_{i_2}&=\sum_{k\in I_{i_1}}\hat{\tilde{P}}_k\sum_{l\in I_{i_2}}\hat{\tilde{P}}_l=\sum_{k\in I_{i_1},l\in I_{i_2}}\hat{\tilde{P}}_k\hat{\tilde{P}}_l\\
&=\sum_{k\in I_{i_1},l\in I_{i_2}}\delta_{kl}\hat{\tilde{P}}_k=\sum_{k\in I_{i_1}\cap I_{i_2}}\hat{\tilde{P}}_k
\end{split}
\]
which by definition means $\{\P_{i_1}\P_{i_2}\}_{{i_1},{i_2}}\!\setminus\!\{0\}\hookrightarrow \C$ and therefore $\C_{1,2}=\{\P_{i_1}\P_{i_2}\}_{{i_1},{i_2}}\!\setminus\!\{0\}$.
\end{proof}

\subsection{Proof of Theorem~\ref{thm:non-increase}}
\begin{proof}
We will denote $p_{i_1,\dots,i_n,i_{n+1}}\equiv p_{\bi,i_{n+1}}$, and $V_{i_1,\dots,i_n,i_{n+1}}\equiv V_{\bi,i_{n+1}}$. Other notation remains the same. Using trivial identities,
\begin{subequations}
\begin{align}
p_{\bi}&=\sum_{i_{n+1}}p_{\bi,i_{n+1}},\\
V_{\bi}&=\sum_{i_{n+1}}V_{\bi,i_{n+1}},
\end{align}
\end{subequations}
and Jensen's Theorem~\ref{thm:Jensen}, we derive,
\[
\begin{split}
&S_{O(\C_1,\dots,\C_n)}(\R)=-\sum_{\bi}p_\bi\ln \frac{p_\bi}{V_{\bi}}\\
&=-\sum_{\bi}\sum_{i_{n+1}}p_{\bi,i_{n+1}}\ln \frac{\sum_{i_{n+1}}p_{\bi,i_{n+1}}}{V_{\bi}}\\
&=-\sum_{\bi}V_{\bi}\bigg(\sum_{i_{n+1}}\frac{p_{\bi,i_{n+1}}}{V_{\bi,i_{n+1}}}\frac{V_{\bi,i_{n+1}}}{V_{\bi}}\bigg)\ln \bigg(\sum_{i_{n+1}}\frac{p_{\bi,i_{n+1}}}{V_{\bi,i_{n+1}}}\frac{V_{\bi,i_{n+1}}}{V_{\bi}}\bigg)\\
&\geq \sum_{\bi}V_{\bi}\sum_{i_{n+1}}\frac{V_{\bi,i_{n+1}}}{V_{\bi}}\bigg(-\frac{p_{\bi,i_{n+1}}}{V_{\bi,i_{n+1}}}\ln \frac{p_{\bi,i_{n+1}}}{V_{\bi,i_{n+1}}}\bigg)\\
&=-\sum_{\bi,i_{n+1}}p_{\bi,i_{n+1}}\ln \frac{p_{\bi,i_{n+1}}}{V_{\bi,i_{n+1}}}=S_{O(\C_1,\dots,\C_n,\C_{n+1})}(\R),
\end{split}
\]
where we have used $f(x)=-x\ln x$, $a_{i_{n+1}}=\frac{V_{\bi,i_{n+1}}}{V_{\bi}}$, and $b_{i_{n+1}}=\frac{p_{\bi,i_{n+1}}}{V_{\bi,i_{n+1}}}$ for the Jensen's Theorem.

The equality condition from the Jensen's inequality turn into an equation that is similar to Eq.~\eqref{eq:equality_monotonic}. After some simple algebra, one finds that  $S_{O(\C_1,\dots,\C_n)}(\R)=S_{O(\C_1,\dots,\C_n,\C_{n+1})}(\R)$ if and only if
\[
(\forall \bi)(\forall i_{n+1})\left(p_{\bi,i_{n+1}}=\frac{V_{\bi,i_{n+1}}}{V_{\bi}}p_{\bi}\right).
\]

Assuming that $p_{\bi}\neq 0$, and rewriting the above condition as $p(i_{n+1}|\bi)\equiv \frac{p_{\bi,i_{n+1}}}{p_{\bi}}=\frac{V_{\bi,i_{n+1}}}{V_{\bi}}$, the above equality says that the entropy will not decrease with additional coarse-graining $\C_{n+1}$ if the conditional probability of the outcome $i_{n+1}$ is given by the ratio of the volumes of macrostates.

The above equality condition is for example satisfied when the set of coarse-grainings $(\C_1,\dots,\C_n)$ projects onto a pure state, i.e., for all density matrices and every $\bi$ we can write, $\pro{\psi_{\bi}}{\psi_{\bi}}=\frac{\P_{i_n}\cdots\P_{i_1}\R\P_{i_1}\cdots\P_{i_n}}{p_{\bi}}$ (note that the left hand side does not depend on $\R$ anymore). Since this holds for every density matrix, it also holds for $\R_{\mathrm{id}}=\frac{1}{\dim \HS}\hat{I}$, which gives $\pro{\psi_{\bi}}{\psi_{\bi}}=\frac{\P_{i_n}\cdots\P_{i_1}\cdots\P_{i_n}}{V_{\bi}}$. Then
\[
\begin{split}
p_{\bi,i_{n+1}}&=\tr[\P_{i_{n+1}}\pro{\psi_{\bi}}{\psi_{\bi}}]p_{\bi}\\
&=\tr[\P_{i_{n+1}}\frac{\P_{i_n}\cdots\P_{i_1}\cdots\P_{i_n}}{V_{\bi}}]p_{\bi}=\frac{V_{\bi,i_{n+1}}}{V_{\bi}}p_{\bi}.
\end{split}
\]

Another example when the equality condition is satisfied is when $\C_{n+1}\hookrightarrow (\C_n,\dots,\C_1)$. By definition, for every multi-index $\bi$ there exists index $i_{n+1}^{(\bi)}$ such that
$\P_{i_1}\cdots\P_{i_n}\P_{i_{n+1}^{(\bi)}}=\P_{i_1}\cdots\P_{i_n}$. It follows that for every other index ${i_{n+1}}\neq i_{n+1}^{(\bi)}$, $\P_{i_1}\cdots\P_{i_n}\P_{i_{n+1}}=0$. Then for index $i_{n+1}^{(\bi)}$,
\[
\begin{split}
p_{\bi,i_{n+1}^{(\bi)}}&=\tr[\P_{i_n}\cdots\P_{i_1}\R\P_{i_1}\cdots\P_{i_n}\P_{i_{n+1}^{(\bi)}}]\\
&=\tr[\P_{i_n}\cdots\P_{i_1}\R\P_{i_1}\cdots\P_{i_n}]=\frac{V_{\bi,i_{n+1}^{(\bi)}}}{V_{\bi}}p_{\bi},
\end{split}
\]
because $\frac{V_{\bi,i_{n+1}^{(\bi)}}}{V_{\bi}}=1$, and for every other index ${i_{n+1}}\neq i_{n+1}^{(\bi)}$,
\[
\begin{split}
p_{\bi, i_{n+1}}&=\tr[\P_{i_n}\cdots\P_{i_1}\R\P_{i_1}\cdots\P_{i_n}\P_{i_{n+1}}]\\
&=0=\frac{V_{\bi, i_{n+1}}}{V_{\bi}}p_{\bi},
\end{split}
\]
because $\frac{V_{\bi,i_{n+1}}}{V_{\bi}}=0$.
\end{proof}

\section{Properties of Definition~\ref{def:finer_set_coarse_graining}}\label{app:finer_definition}
First we show that attachment $\bi\rightarrow j$ in the Def.~\ref{def:finer_set_coarse_graining} is unique for $\P_{i_n}\cdots\P_{i_1}\neq0$. For contradiction, we assume that there are two projectors $\P_j,\P_{\tilde j}\in \C$ such that $\P_{i_n}\cdots\P_{i_1}\P_j=\P_{i_n}\cdots\P_{i_1}$, $\P_{i_n}\cdots\P_{i_1}\P_{\tilde j}=\P_{i_n}\cdots\P_{i_1}$. Then multiplying the first equation by $\P_{\tilde j}$ and using orthogonality of the projectors we obtain
\[
0=\P_{i_n}\cdots\P_{i_1}\P_j\P_{\tilde j}=\P_{i_n}\cdots\P_{i_1}\P_{\tilde j}=\P_{i_n}\cdots\P_{i_1}\neq 0,
\]
which is a contradiction.

Assuming that $\C\hookrightarrow (\C_1,\dots,\C_n)$, i.e., for every $\bi$ exists $j$ such that $\P_{i_n}\cdots\P_{i_1}\P_j=\P_{i_n}\cdots\P_{i_1}$, then for the multi-index $(i_1,\dots,i_n,i_{n+1})$ we take the same $j$ that is obtained from the first $n$ indexes. Then
\[
\P_{i_{n+1}}\P_{i_n}\cdots\P_{i_1}\P_j=\P_{i_{n+1}}\P_{i_n}\cdots\P_{i_1},
\]
which by definition means $\C\hookrightarrow (\C_1,\dots,\C_n,\C_{n+1})$.

Finally, we are going to show that Definitions~\ref{def:finer_coarse_graining} and~\ref{def:finer_set_coarse_graining} coincide for $n=1$. Assuming that $\C\hookrightarrow \C_1$ by Def.~\ref{def:finer_set_coarse_graining}, for every $i_1$ exists $j$ such that $\P_{i_1}\P_j=\P_{i_1}$. Using that for every other $\P_{\tilde j}$ we have $\P_{i_1}\P_{\tilde j}=\P_{i_1}\P_j\P_{\tilde j}=0$, and $\sum_{{i_1}} \P_{{i_1}}=\hat{I}$, we have
\[\label{eq:P_j_by_i1}
\P_j=\sum_{{i_1}} \P_{{i_1}}\P_j=\sum_{{i_1}\in I^{(j)}} \P_{{i_1}}\P_j=\sum_{{i_1}\in I^{(j)}} \P_{{i_1}}
\]
where we have defined $I^{(j)}\equiv \{i_1|\P_{{i_1}}\P_j\neq 0\}$. The above equation means that $\C\hookrightarrow \C_1$ by Def.~\ref{def:finer_coarse_graining}.

Now let us consider the opposite implication, assuming that Eq.~\eqref{eq:P_j_by_i1} holds. For each ${\tilde i_1}$, we find $j$ such that $\P_{{\tilde i_1}}\P_j\neq 0$. Such $j$ must exist, because the set of projectors form a complete set. Multiplying Eq.~\eqref{eq:P_j_by_i1} with this $j$ by $\P_{{\tilde i_1}}$ and using orthogonality of the projectors, we find
\[
\P_{{\tilde i_1}}\P_j=\P_{{\tilde i_1}}\sum_{{i_1}\in I^{(j)}} \P_{{i_1}}=\sum_{{i_1}\in I^{(j)}}\delta_{i_1,{\tilde i_1}}\P_{{\tilde i_1}}=\P_{{\tilde i_1}},
\]
which means that $\C\hookrightarrow \C_1$ by Def.~\ref{def:finer_set_coarse_graining}.

\section{Bounds on the factorized Observational entropy}\label{app:bounds_on_FOE}

Here we are going to prove Eqs.~\eqref{eq:bounds_on_FOE}, \eqref{eq:FOE_of_diagonal}, \eqref{eq:FOE_of_thermal}, and \eqref{eq:correctionToThermal}, i.e., $S_{F}(\R)+O(\epsilon)\leq S(\R_d) \leq S(\R_{th})$, $S_{F}(\R_d)+O(\epsilon)= S(\R_d)$, $S_{F}(\R_{th})+O(\epsilon)= S(\R_{th})$, and the explicit form of $O(\epsilon)$ term in this last equation.

The following sequence of inequalities and identities holds.
\[
\begin{split}
S_{F}(\R)&\equiv S_{O{\displaystyle (}\C_{\hat{H}^{(1)}}\otimes \C_{\hat{H}^{(2)}}{\displaystyle )}}(\R)\\
&\leq S_{O{\displaystyle (}\C_{\hat{H}}{\displaystyle )}}(\R)+O(\epsilon)\\
&= S_{O{\displaystyle (}\C_{\hat{H}}{\displaystyle )}}(\R_d)+O(\epsilon)\\
&= S(\R_d)+O(\epsilon).
\end{split}
\]

The first inequality is a consequence of identities
\[\label{eq:P_E}
\begin{split}
\P_E&=\P_{E_-}\Bigr|_{E_-=E}+O(\epsilon).\\
\P_{E_-}&=\sum_{E_1,E_2}\delta_{E_-,E_1+E_2}\P_{E_1}\!\!\otimes\!\!\P_{E_2}
\end{split}
\]
where  $\hat{H}=\sum_EE\P_E$, $\hat{H}^{(1)}=\sum_{E_1}E_1\P_{E_1}$, $\hat{H}^{(2)}=\sum_{E_1}E_2\P_{E_2}$, and $\hat{H}_-=\hat{H}-\epsilon\hat{H}^{(\mathrm{int})}=\sum_{E_-}{E_-}\P_{E_-}$ denotes the full Hamiltonian without the interaction part.
By definition, we have $\C_{\hat{H}_-}\hookrightarrow\C_{\hat{H}^{(1)}}\otimes \C_{\hat{H}^{(2)}}$ (and $\C_{\hat{H}_-}=\C_{\hat{H}^{(1)}}\otimes \C_{\hat{H}^{(2)}}$ when $\hat{H}_-$ is non-degenerate), i.e., the factorized coarse-graining is finer than the coarse-graining by the full Hamiltonian without the interaction part. From Theorem~\ref{thm:monotonic} we have
\[
S_{O{\displaystyle (}\C_{\hat{H}^{(1)}}\otimes \C_{\hat{H}^{(2)}}{\displaystyle )}}(\R)\leq S_{O{\displaystyle (}\C_{\hat{H}_-}{\displaystyle )}}(\R),
\]
and by explicit computation\footnote{For simplicity, assuming that both $\hat{H}_{-}$ and $\hat{H}=\hat{H}_{-}+\epsilon\hat{H}^{(\mathrm{int})}$ are non-degenerate, and assuming structure
\begin{align}
E&=E_{-}+\epsilon E^{(1)},\\
p_E&=p_{E_{-}}+\epsilon p_E^{(1)},\\
\end{align}
we derive
\[\label{eq:expansionofentropy}
\begin{split}
&S_{O{\displaystyle (}\C_{\hat{H}}{\displaystyle )}}(\R)
=-\sum_Ep_E\ln p_E\\
&=-\sum_E(p_{E_{-}}+\epsilon p_E^{(1)})\ln (p_{E_{-}}+\epsilon p_E^{(1)})\\
&=-\sum_Ep_{E_{-}}\ln p_{E_{-}}-\epsilon\sum_Ep_E^{(1)}\ln p_{E_{-}}-\sum_Ep_{E_{-}}\frac{\epsilon p_E^{(1)}}{p_{E_{-}}}\\
&=S_{O{\displaystyle (}\C_{\hat{H}_-}{\displaystyle )}}(\R)-\epsilon\sum_Ep_E^{(1)}\ln p_{E_{-}},
\end{split}
\]
where we have used $\sum_Ep_E^{(1)}=0$. We have $p_{E_{-}}=\tr[\P_{E_{-}}\R]$, and using standard perturbation theory we derive
\[
p_E^{(1)}=2\sum_{\tilde{E}\neq E}\frac{\Re(\tr[\P_{E_{-}}\R\P_{\tilde{E}_{-}}\hat{H}^{(\mathrm{int})}])}{E_{-}-\tilde{E}_{-}}.
\] Correction amplitude $\sum_Ep_E^{(1)}\ln p_{E_{-}}$ can be quite large, especially when $p_{E_{-}}\approx 0$ while $p_E^{(1)}\neq0$, which happens for example when state $\R\approx \ket{\psi}\bra{\psi}$ is an eigenstate of Hamiltonian $\hat{H}_-$, i.e., $\ket{\psi}\approx \ket{E_1}\ket{E_2}$. In that case it can be easily checked that $S_{O{\displaystyle (}\C_{\hat{H}_-}{\displaystyle )}}(\R)\approx 0$, while $S_{O{\displaystyle (}\C_{\hat{H}}{\displaystyle )}}(\R)$ can be some non-zero and possibly large number. This problematic behavior of diverging amplitude points to cases when perturbative expansion of entropy is not valid, at least not in this form. We can still see, however, than even in this pathological case of $\ket{\psi}\approx \ket{E_1}\ket{E_2}$, inequality  $S_{F}(\R)\leq S(\R_d)$ holds.} using Eq.~\eqref{eq:P_E} we obtain
\[
S_{O{\displaystyle (}\C_{\hat{H}_-}{\displaystyle )}}(\R)= S_{O{\displaystyle (}\C_{\hat{H}}{\displaystyle )}}(\R)+O(\epsilon),
\]
which is valid up to pathological cases, when $\R$ is either eigenstate of $\hat{H}_{-}$ or eigenstate of $\hat{H}$. In the first case $S_{O{\displaystyle (}\C_{\hat{H}_-}{\displaystyle )}}(\R)=0$ and $S_{O{\displaystyle (}\C_{\hat{H}}{\displaystyle )}}(\R)$ is potentially large, while in the second $S_{O{\displaystyle (}\C_{\hat{H}}{\displaystyle )}}(\R)= 0$ and $S_{O{\displaystyle (}\C_{\hat{H}_-}{\displaystyle )}}(\R)$ is potentialy large.

The following identity $S_{O(\C_{\hat{H}})}(\R)= S_{O(\C_{\hat{H}})}(\R_d)$ holds because from the definition of $\R_d$, $p_E(\R)=\tr[\P_E\R]=\tr[\P_E\R_d]=p_E(\R_d)$.

The last identity $S_{O(\C_{\hat{H}})}(\R_d)= S(\R_d)$ comes from $\C_{\R_d}\hookrightarrow \C_{\hat{H}}$ and the equality condition in Theorem~\ref{thm:bounded}.

The second inequality from Eq.~\eqref{eq:bounds_on_FOE},
\[
S(\R_d)\leq S(\R_{th})
\]
is a simple consequence of the fact that $\R_{th}$ maximizes the von Neumann entropy with the constraint on energy $\overline{E}=\tr[\hat{H}\R_t]=\tr[\hat{H}\R_d]$~\cite{jaynes1957information2}.

Now we prove the equalities. Given definition of the diagonal density matrix, $\R_d=\sum_E \frac{p_E(\R_t)}{\tr[\P_E]}\P_E$, we define its \emph{minus} counterpart as $\R_{d_-}=\sum_{E_-} \frac{p_{E_-}(\R_t)}{\tr[\P_{E_-}]}\P_{E_-}$. Since by Eq.~\eqref{eq:P_E}, $\C_{\R_{d_-}}\hookrightarrow \C_{\hat{H}^{(1)}}\otimes \C_{\hat{H}^{(2)}}$, hence from equality condition in Theorem~\ref{thm:bounded} we have
\[\label{eq:FOEandReducedHequivalence}
S_{F}(\R_{d_-})\equiv S_{O{\displaystyle (}\C_{\hat{H}^{(1)}}\otimes \C_{\hat{H}^{(2)}}{\displaystyle )}}(\R_{d_-})=S(\R_{d_-}).
\]
Since $\R_{d_-}=\R_{d}+O(\epsilon)$, we have $S_{F}(\R_{d_-})=S_F(\R_{d})+O(\epsilon)$ and $S(\R_{d_-})=S(\R_{d})+O(\epsilon)$, which in combination with the above equation proves $S_{F}(\R_d)+O(\epsilon)= S(\R_d)$.

The second equality, $S_{F}(\R_{th})+O(\epsilon)= S(\R_{th})$, is a direct consequence of the previous equality, because $\R_{th}$ is a special case of $\R_{d}$ due to its form, $\R_{th}=\frac{1}{Z}\sum_E\exp(-\beta E)\P_E$. This inequality can be also obtained directly, by maximizing the Observational entropy $S_{F}(\R)$ with condition on the mean energy $\ov{E}=\tr[\R\hat{H}_-]+O(\epsilon)$, which gives $\R^{(\max)}=\frac{1}{Z}\sum_{E_1,E_2}\exp(-\beta (E_1+E_2))\P_{E_1}\!\otimes\!\P_{E_2}$, and  $S_{F}(\R^{(\max)})=S(\R_{th})+O(\epsilon)$.

Finally, we derive explicit form of correction term $O(\epsilon)$ in equation $S_{F}(\R_{th})+O(\epsilon)= S(\R_{th})$ (Eq.~\eqref{eq:FOE_of_thermal}) in terms physical quantities. We recall $\Rt=e^{-\beta \hat{H}}/Z$, $Z=\tr[e^{-\beta \hat{H}}]$, and define $\Rtt=e^{-\beta \hat{H}_-}/Z_-$, $Z_-=\tr[e^{-\beta \hat{H}_-}]$. Moreover, we denote mean of an operator $\hat{A}$ as $\mean{\hat{A}}_\R\equiv\tr[\hat{A}\R]$ for any density matrix $\R$.
We have
\[\label{eq:totalO}
\begin{split}
O(\epsilon)&=S(\Rt)-S_{F}(\Rt)\\
&=S(\Rt)-S(\Rtt)+S(\Rtt)-S_{F}(\Rt)\\
&=O_1(\epsilon)+O_2(\epsilon).
\end{split}
\]
We will study separately the terms $O_1(\epsilon)$ and $O_2(\epsilon)$.

First, we study term
\[
O_1(\epsilon)=S(\Rt)-S(\Rtt).
\]
Assuming that $\Ha_-$ and $\Hi$ commute (which means that we effectively study the classical corrections), we can write Taylor expansion
\[
\Rt=\Rtt+\epsilon\beta\Rtt(\mean{\Hi}_\Rtt-\Hi).
\]
If $\R_0$ and $\R^{(1)}$ in expansion $\R=\R_0+\epsilon \R^{(1)}$ commute, we can write $S(\R)=S(\R_0)-\epsilon\tr[\R^{(1)}\ln\R_0]$ (similar to Eq.~\eqref{eq:expansionofentropy}). Therefore, we have
\[\label{eq:O1final}
\begin{split}
O_1(\epsilon)&=-\epsilon\beta\tr[\Rtt(\mean{\Hi}_\Rtt-\Hi)\ln\Rtt]\\
&=-\epsilon\beta\tr[\Rtt(\mean{\Hi}_\Rtt-\Hi)(-\beta \Ha_-+\ln Z_-)]\\
&=-\epsilon\beta^2(\mean{\Ha_- \Hi}_{\Rtt}-\mean{\Ha_-}_{\Rtt}\mean{\Hi}_{\Rtt})\\
&=-\epsilon\beta^2\mean{\Ha_- \Hi}_C
\end{split}
\]
where we have defined covariance as $\mean{\Ha_- \Hi}_C = \mean{\Ha_- \Hi}_{\Rtt}-\mean{\Ha_-}_{\Rtt}\mean{\Hi}_{\Rtt}$.

Now we move onto term $O_2(\epsilon)$. For simplicity we assume that both $\Ha_-$ and $\Ha$ are non-degenerate (although it makes no difference if they are degenerate - it just makes for a complicated notation), and that $\Ha_-$ and $\Hi$ commute. Then we can write
\[
\begin{split}
&p_{E_-}(\Rt)=\bra{E_-}\Rt\ket{E_-}\\
&=\bra{E_-}\Rtt\ket{E_-}\\
&+\epsilon\beta\bra{E_-}\Rtt\ket{E_-}(\mean{\Hi}_\Rtt-\bra{E_-}\Hi\ket{E_-})\\
&=p_{E_-}(\Rtt)+\epsilon\beta p_{E_-}(\Rtt)(\mean{\Hi}_\Rtt-\bra{E_-}\Hi\ket{E_-})
\end{split}
\]
and using Eq.~\eqref{eq:expansionofentropy} we have
\[
\begin{split}
S_{F}(\Rt)&=S_{F}(\Rtt)\\
&-\epsilon\beta\sum_{E_-}p_{E_-}\ln p_{E_-} (\mean{\Hi}_\Rtt-\bra{E_-}\Hi\ket{E_-})
\end{split}
\]
where we have used a simplified notation $p_{E_-}\equiv p_{E_-}(\Rtt)$. Considering $S_{F}(\Rtt)=S(\Rtt)$ (Eq.~\eqref{eq:FOEandReducedHequivalence}), we derive
\[
O_2(\epsilon)=\epsilon\beta\sum_{E_-}p_{E_-}\ln p_{E_-} (\mean{\Hi}_\Rtt-\bra{E_-}\Hi\ket{E_-}).
\]
$\mean{\Hi}_\Rtt$ represents the canonical average of operator $\Hi$, and $\bra{E_-}\Hi\ket{E_-}$ represents the microcanonical average. To signify the dependence on temperature, energy respectively, we denote $\mean{\Hi}_\Rtt\equiv \mean{\Hi}_\beta$ and $\bra{E_-}\Hi\ket{E_-}\equiv\mean{\Hi}_{E_-}$. For the purposes of simplifying notation of the following derivation, we also write simply $E$ instead of $E_-$ (so $p_{E_-}$ turns into $p_{E}$, all averages are averages in reduced Hamiltonain $H_-$). We can turn the sum into an integral, and write
\[\label{eq:o2goingthere}
O_2(\epsilon)=\epsilon\beta\int \rho(E) p_E\ln p_E (\mean{\Hi}_\beta-\mean{\Hi}_E)\ dE,
\]
where $\rho(E)$ denotes the energy density of states, i.e., $\rho(E)dE$ denotes number of energy eigenstates with energy $\tilde{E}$ in interval $E\leq\tilde{E}<E+dE$. Since the energy density of states can be written using the microcanonical entropy $S(E)\equiv S_{\mathrm{micro}}(E)$ as $\rho(E)=e^{S(E)}$ (Eq.~\eqref{eq:microcanonical_entropy}; while ignoring the unimportant term $\Delta E$), we can expand exponent of function $p_E\rho(E)$ around its maximum $E_*$ (which turns out to be defined implicitly by $\beta=\partial_ES(E_*)$),
\[\label{eq:fE}
\begin{split}
&f(E-E_*)\equiv p_E\rho(E)=\frac{e^{-\beta E+S(E)}}{Z}\\
&=\frac{e^{-\beta E_*+S(E_*)+\frac{1}{2}\partial_E^2S|_{E=E_*}(E-E_*)^2+\frac{1}{6}\partial_E^3S|_{E=E_*}(E-E_*)^3+\cdots}}{Z}\\
&=\frac{e^{\frac{1}{2\sigma^2}(E-E_*)^2}}{\sqrt{2\pi\sigma^2}}\bigg(1+\frac{1}{6}\partial_E^3S|_{E=E_*}(E-E_*)^3+\cdots\bigg)\\
&\equiv g(E-E_*)\bigg(1+c(E-E_*)^3+\cdots\bigg),
\end{split}
\]
where $\sigma\equiv(\partial_E^2S|_{E=E_*})^{-1/2}$, and $c=\frac{1}{6}\partial_E^3S|_{E=E_*}$. The second term in the expansion of the exponent was zero, because we expanded around the maximum. We can derive explicit form of $\sigma$ and $c$ as
\begin{subequations}
\begin{align}
\sigma^{-2}&=\frac{\partial}{\partial E}\bigg(\frac{\partial S}{\partial E}\bigg)\!=\!\frac{\partial}{\partial E}\bigg(\frac{1}{T}\bigg)\!=\!-\frac{1}{T^2}\frac{\partial T}{\partial E}\!=\!-\frac{1}{T^2c_E},\\
c&=\frac{1}{6}\frac{\partial}{\partial E}\bigg(-\frac{1}{T^2c_E}\bigg),
\end{align}
\end{subequations}
where $T$ is the temperature, and $c_E$ is the specific heat.
Although we are not going to use this form for our final result, it helps us determine scaling with $N$ (number of particles) in the thermodynamic limit. $c_E\sim N$ is extensive, $T\sim 1$, and thermodynamic energy $E\sim N$ is also extensive. Therefore, we have $\sigma^2\sim N$ and $c\sim 1/N^2$. We do not have to consider higher-order corrections in Eq.~\eqref{eq:fE}, because their scaling lead to the subleading order in $N$ in the final result. $g(E-E_*)$ represents Gaussian function peaked around point $E_*$, which is normalized to 1, because the partition function is defined as $Z=\int e^{-\beta E}\rho(E)dE$. Further, using $Z=e^{-\beta E_*+S(E_*)}\sqrt{2\pi\sigma^2}$ we have
\[\label{eq:lnp}
\begin{split}
\ln p_E&=-\beta E-\ln Z\\
&=-\beta (E- E_*)-S(E_*)-\ln\sqrt{2\pi\sigma^2}
\end{split}
\]
Combining Eqs.\eqref{eq:fE} and~\eqref{eq:lnp} with Eq.~\eqref{eq:o2goingthere} we can write
\[\label{eq:o2middlepart}
\begin{split}
O_2&(\epsilon)=\epsilon\beta\int f(E-E_*)\ln p_E (\mean{\Hi}_\beta-\mean{\Hi}_E)dE\\
&=-\epsilon\beta(S(E_*)+\ln\sqrt{2\pi\sigma^2})\mean{\Hi}_\beta\int f(E-E_*) dE\\
&+\epsilon\beta(S(E_*)+\ln\sqrt{2\pi\sigma^2})\int f(E-E_*)\mean{\Hi}_E dE\\
&-\epsilon\beta^2\mean{\Hi}_\beta\!\!\int\!\! f(E-E_*)(E-E_*) dE\\
&+\epsilon\beta^2\!\!\int\!\! f(E-E_*)(E-E_*)\mean{\Hi}_E dE
\end{split}
\]
The first two terms cancel each other, since $\int f(E-E_*) dE=1$ and because the canonical average is the canonical mean of the microcanonical averages,
\[
\mean{\Hi}_\beta=\sum_Ep_E\mean{\Hi}_E=\int f(E-E_*)\mean{\Hi}_E dE.
\]
Using this equation, and expanding $\mean{\Hi}_E$ around point $E_*$, we can also compute
\[\label{eq:canonicalvsmicrocanonicalHi}
\begin{split}
\mean{&\Hi}_\beta=\int f(E-E_*)\mean{\Hi}_E dE\\
&=\int g(E-E_*)(1+c(E-E_*)^3)
\big(\mean{\Hi}|_{E=E_*}\\
&+\partial_E\mean{\Hi}|_{E=E_*}(E-E_*)\\
&+\frac{1}{2}\partial_E^2\mean{\Hi}|_{E=E_*}(E-E_*)^2\big) dE\\
&=\mean{\Hi}|_{E=E_*}\\
&+\frac{1}{2}\partial_E^2\mean{\Hi}|_{E=E_*}\int g(E-E_*)(E-E_*)^2 dE\\
&+\partial_E\mean{\Hi}|_{E=E_*}c\int g(E-E_*)(E-E_*)^4 dE\\
&=\mean{\Hi}|_{E=E_*}\\
&+\frac{1}{2}\partial_E^2\mean{\Hi}|_{E=E_*}\int f(E-E_*)(E-E_*)^2 dE\\
&+\partial_E\mean{\Hi}|_{E=E_*}c\int f(E-E_*)(E-E_*)^4 dE\\
&=\mean{\Hi}|_{E=E_*}+\frac{1}{2}\partial_E^2\mean{\Hi}|_{E=E_*}\mean{\Delta E^2}_\beta\\
&+\partial_E\mean{\Hi}|_{E=E_*}c\mean{\Delta E^4}_\beta,
\end{split}
\]
where have defined $\Delta E=E-E_*$. Terms with $g(E-E_*)(E-E_*)^{2k-1}$ in the integral had vanished because the integral was over an odd function. The first term in the last line scales as $\mean{\Hi}|_{E=E_*}\sim N$, while the other two as $\frac{1}{2}\partial_E^2\mean{\Hi}|_{E=E_*}\mean{\Delta E^2}_\beta\sim\partial_E\mean{\Hi}|_{E=E_*}c\mean{\Delta E^4}_\beta\sim 1$.
Similarly, we derive
\[
\int f(E-E_*)(E-E_*) dE=c\mean{\Delta E^4}_\beta\sim 1,
\]
and
\[
\begin{split}
\int &f(E-E_*)(E-E_*)\mean{\Hi}_E dE\\
&=\partial_E\mean{\Hi}|_{E=E_*}\mean{\Delta E^2}_\beta+\mean{\Hi}|_{E=E_*}c\mean{\Delta E^4}_\beta
\end{split}
\]
Inserting the above expressions into Eq.~\eqref{eq:o2middlepart} while considering just the leading terms ($\sim N$), we derive
\[\label{eq:O2almostdone}
\begin{split}
O_2(\epsilon)&=-\epsilon\beta^2\mean{\Hi}|_{E=E_*}c\mean{\Delta E^4}_\beta\\
&+\epsilon\beta^2\partial_E\mean{\Hi}|_{E=E_*}\mean{\Delta E^2}_\beta\\
&+\epsilon\beta^2\mean{\Hi}|_{E=E_*}c\mean{\Delta E^4}_\beta\\
&=\epsilon\beta^2\partial_E\mean{\Hi}|_{E=E_*}\mean{\Delta E^2}_\beta.
\end{split}
\]
As Eq.~\eqref{eq:canonicalvsmicrocanonicalHi} shows, canonical and microcanonical averages of energy are equal in the leading term, $\mean{\Hi}_\beta=\mean{\Hi}|_{E=E_*}+O(1)$, and the same statement can be derived in analogy for the energy itself, $\mean{E}_\beta\equiv\mean{\Ha_-}_\beta=\mean{\Ha_-}|_{E=E_*}+O(1)\equiv E_*+O(1)$. Thus considering just the leading terms we derive
\[\label{eq:littlederivationforderivative}
\begin{split}
&\partial_E\mean{\Hi}|_{E=E_*}\equiv\frac{\partial\mean{\Hi}}{\partial E}|_{E=E_*}=\frac{\partial\mean{\Hi}_{\beta}+O(1)}{\partial \mean{E}_{\beta}+O(1)}\\
&=\frac{\partial\beta}{\partial \mean{E}_{\beta}}\frac{\partial\mean{\Hi}_{\beta}}{\partial \beta}=-\frac{\mean{\Ha_-\Hi}_C}{\mean{\Delta E^2}_\beta},
\end{split}
\]
where we have used
\[
\frac{\partial\mean{E}_{\beta}}{\partial \beta}=\tr[\Ha_-\partial_\beta\Rtt]=-\mean{\Ha_-\Ha_-}_C=-\mean{\Delta E^2}_\beta,
\]
and
\[\frac{\partial\mean{\Hi}_{\beta}}{\partial \beta}=\tr[\Hi\partial_\beta\Rtt]=-\mean{\Ha_-\Hi}_C.
\]
Inserting Eq.~\eqref{eq:littlederivationforderivative} into Eq.~\eqref{eq:O2almostdone} we derive
\[\label{eq:O2final}
O_2(\epsilon)=-\epsilon\beta^2\mean{\Ha_-\Hi}_C.
\]

Combining Eqs.~\eqref{eq:totalO},~\eqref{eq:O1final}, and \eqref{eq:O2final}, we finally derive
\[\label{eq:finalOepsilon}
O(\epsilon)=-2\epsilon\beta^2\mean{\Ha_-\Hi}_C.
\]
Since $\hat{H}_-\equiv\hat{H}-\epsilon\hat{H}^{(\mathrm{int})}$, $\Rtt=\Rt+O(\epsilon)$, inserting these equations will lead only to the second-order corrections in $\epsilon$, so for the first-order correction we can as well write
\[
O(\epsilon)=-2\epsilon\beta^2\mean{\Ha\Hi}_C,
\]
which is Eq.~\eqref{eq:correctionToThermal}. This result scale as $\sim N$ in the thermodynamic limit, however, as sizes of regions grow bigger, then $\Hi$ grows smaller compared to $\Ha$, and therefore also their correlation $\mean{\Ha\Hi}_C$ gets smaller. $O(\epsilon)$ term therefore represents a finite-size effect.

On a lattice with local interactions, where each lattice site $i$ is described by its own local Hamiltonian $\Ha_i$ (such as in Hamiltonian written in Eq.~\eqref{FermionHam}), we can write $\Ha_-=\sum_{i\notin S}\Ha_i$, and approximate\footnote{In reality, however, the $\Hi$ can consist just of terms that enable interaction between the two regions, such as nearest-neighbor and next-nearest-neighbor hoppings and interactions between the two adjacent regions (Hamiltonian Eq.~\eqref{FermionHam}), and not of any sites themselves.} the interaction part of Hamiltonian as $\Hi=\sum_{j\in S}\Ha_j$, where $S$ denotes the ``surfaces'' of all regions. Then using Eq.~\eqref{eq:finalOepsilon} we can write
\[
O(\epsilon)=-2\epsilon\beta^2\sum_{i\notin S, j\in S}\mean{\Ha_i\Ha_j}_C,
\]
where $\mean{\Ha_i\Ha_j}_C$ denotes energy-energy correlation function. For a Hamiltonian with local interaction, this function fades with a growing distance between $i$ and $j$. This shows explicitly that $O(\epsilon)$ is really just a boundary term.

\section{Convergence of FOE to microcanonical entropy}
\label{sec-FOE_for_eigenstates}

In the first part of this Appendix we show that the factorized Observational entropy of energy eigenstates gives the microcanonical entropy for closed non-integrable systems. To do that we
follow a similar approach as was used in eigenstate thermalization hypothesis~\cite{deutsch1991quantum}. We point out that FOE will not give the thermodynamic entropy for every system, however. For example, as shown by our numerics in Sec. \ref{sec-simulation}, integrable Hamiltonians do not give the thermodynamic entropy when computing FOE for energy eigenstates.

In the second part of this Appendix we show that FOE of superposition of energy eigenstates with random phases is larger than the mean microcanonical entropy, and that FOE of a superposition of close energy eigenstates with random phases gives the microcanonical entropy.

Since state of the system with random phases represents a typical state in future, this says that the FOE converges to microcanonical entropy for such superpositions in the long-time limit.

\subsection{FOE of energy eigenstates}

Defining the Hamiltonian without the interaction part as $\hat{H}_-=\hat{H}-\epsilon\hat{H}^{(\mathrm{int})}=\sum_{E_-}{E_-}\P_{E_-}$ (which has been previously used in Appendix~\ref{app:bounds_on_FOE}), and for simplicity assuming that both Hamiltonians $\hat{H}$ and $\hat{H}_-$ are non-degenerate,\footnote{I.e., they do not have degenerate eigenvalues. This is expected to be roughly true for non-integrable systems, where eigenvalues are usually irrationally related, in other words, each eigenvalue of the Hamiltonian is related to some other by addition of an irrational number. Both $\hat{H}$ and $\hat{H}_-$ may have this property, since $\hat{H}_-$ consists of smaller non-integrable systems with the same property. The irrational relation of eigenvalues is due to the sufficient mixing from the interaction terms.} we have $\P_{E_-}=\P_{E_1}\otimes \P_{E_2}$ from Eq.~\eqref{eq:P_E} for some eigenvalues $E_1$, $E_2$, such that $E_1+E_2=E_-$. Then from the definition of the FOE, using that $\tr[\P_{E_-}]=1$ that holds for non-degenerate $\hat{H}_-$,
\[
\label{eq:SF=Shannon}
\begin{split}
S_{F}(\R)&\equiv S_{O{\displaystyle (}\C_{\hat{H}^{(1)}}\otimes \C_{\hat{H}^{(2)}}{\displaystyle )}}(\R)=S_{O(\C_{\hat{H}_-})}(\R)\\
&= - \sum_{E_-} p_{E_-} \ln p_{E_-},
\end{split}
\]
where we have defined
\[
p_{E_-} \equiv \braket{E_-|\R}{E_-}.
\]
For an eigenstate of energy $\R = \pro{E}{E}$ we have
\[
\label{eq:PE=E-Esq}
p_{E_-} = |\braket{E_-}{E}|^2.
\]

As we shall detail in a few paragraphs, if $p_{E_-}$ has a large enough value for a large enough set of $\ket{E_-}$ vectors, i.e., there a large set of vectors $\ket{E}$ that overlap with vector $\ket{E_-}$, then the FOE will approximate the microcanonical entropy for non-integrable systems.

To understand properties of eigenstates in non-integrable systems, one can consider an integrable Hamiltonian, for example a gas of non-interating particles, and add a perturbation, for example giving the particles a small hard core radius, which will make the Hamiltonian  non-integrable. Instead of trying to analyze properties of energy eigenstates with this perturbation, we replace the perturbation with a small random matrix where typical values of matrix elements are very small, but still much larger than the average separation between energy levels for energies close to the energy eigenstate that we are considering. There is a deep connection between non-integrable systems and random matrices that is an important area of research that started with the work of Wigner on this issue~\cite{wigner1955RandomMatrices}. It has been discussed in detail by many authors~\cite{berry1987bakerian,FeingoldPhysRevA.34.591,feingold1989semiclassical,DAlessio2016quantum}. In the present case, Hamiltonian $\hat{H}_-$ is integrable in the sense that as the size of the system goes to infinity, with a fixed box size, the system contains an infinite number of invariants. This is because as mentioned earlier in form of identity $\P_{E_-}=\P_{E_1}\otimes \P_{E_2}$, different boxes do not interact, we can write an arbitrary eigenstate $\ket{E_-}$  as the product of individual energy eigenstates in each box, $\ket{E_-} = \ket{E_1}\otimes\ket{E_2}$, where the $E_i$ denote the energy eigenstate of each separate box. We are adding to this Hamiltonian a perturbation $\hat{H}_r$ that will represents the interaction term $\epsilon \hat{H}^{(\mathrm{int})}$ that we have previously taken away, to produce the full Hamiltonian $\hat{H}$, which couples the different boxes together,
\[
\hat{H} = \hat{H}_- + \hat{H}_r
\]
where we are taking $\hat{H}_r$ to be a random matrix. To index the matrix, we choose an index to be monotonically related to the energy of a basis vector $\ket{E_-}$ (i.e., matrix ${H}_-$, representing operator $\hat{H}_-$ in a matrix form in its own eigenbasis, has increasing diagonal elements). In the basis of $\hat{H}_-$, we can write the matrix elements of $\hat{H}$, $\braket{E_-|\hat{H}}{E'_-}$ as
\[
H_{ij} = E_i \delta_{ij} + h_{ij}
\]
where $E_i \delta_{ij}$ corresponds to $\hat{H}_-$, and $h_{ij} = h_{ji}$, $\ov{h_{ij}h_{kl}} = \epsilon^2 \delta_{ik}\delta_{jl}$, corresponds  to the random matrix $\hat{H}_r$, Here the bar denotes an average over all possible random matrices in this ensemble. No ensemble average is taken for a particular Hamiltonian $\hat{H}$. A particular realization of the random matrix $h_{ij}$ corresponds to that choice of Hamiltonian $\hat{H}$. $h_{ij}$ is also banded with a width that is proportional to the temperature corresponding to the energy of the state under consideration~\cite{deutsch1991quantum}.

We now wish to diagonalize matrix $H_{ij}$ to determine its eigenvectors in the basis of matrix ${H}_-$. Denoting $c_i$ as an eigenvector of ${H}$, and its $j$'th element in basis of ${H}_-$ is defined as $(c_i)^{(j)}$, we define a matrix formed from these eigenvectors as $c_{ij}=(c_i)^{(j)}$ (i.e., this corresponds to the similarity transformation that diagonalizes $H$). We find that
\[
\label{eq:ccave=Lambda}
\ov{c_{ij}c_{kl}} = \Lambda_{ik} \delta_{ik}\delta_{jl},
\]
where $\Lambda_{ij}$ is a function that has been introduced in Ref.~\cite{deutsch1991quantum}, and which is well explained in Ref.~\cite{reimann2015eigenstate}. $\Lambda_{ij}$ can be well approximated as being of the form $\Lambda(i-j)$, and is a function that was shown to be well approximated by a Lorentzian~\cite{wigner1957BandedRandomMatrices,deutsch1991quantum}. If we think about the indices $i$ and $j$ as corresponding to energies $E$ and $E'$ so that the coefficients $c_{ij}$ can be described as $c_{E E'}$, then the Lorentzian has a width of energy  proportional to $\epsilon$~\cite{wigner1957BandedRandomMatrices,deutsch1991quantum}, but with a tail that dies off faster than an exponential~\cite{wigner1957BandedRandomMatrices,deutsch1991quantum}.

We therefore have the relationship between the eigenstates of $\hat{H}$ and $\hat{H}_-$,
\[
\label{eq:E=CEEE-}
\ket{E} = \sum_{E_-} c_{EE_-} \ket{E_-}.
\]
This implies that
\[
\label{eq:braketE-=c}
\braket{E_-}{E} =  c_{EE_-}
\]
and $c_{E,E_-}$ are random elements of eigenvectors as described above (in more detail, it is random in vectors $\ket{E}$, because $H$ depends on the random matrix, but not in $\ket{E_-}$ which come from diagonalization of $\hat{H}_-$).

We can then compute average values of quantities and their fluctuations. In the case of expectation values of observables, those fluctuation can be shown to be exceedingly small~\cite{deutsch1991quantum}.

In the present case, we know from Eqs.~\eqref{eq:PE=E-Esq},~\eqref{eq:ccave=Lambda}, and~\eqref{eq:braketE-=c} that the coefficients $c_{E,E_-}$ are related to $p_{E_-}$ by
\[
\label{eq:corr_in_cs}
\ov{p_{E_-}} = \ov{|c_{EE_-}|^2} = \Lambda_{E E_-}
\]

We will first incorrectly ignore fluctuations in the probabilities $p_{E_-}$ coming from the randomness of $\hat{H}_r$, and assume that $p_{E_-} = \ov{p_{E_-}}$. After understanding this simplified case, we will show how to treat this $p_{E_-}$ more accurately.

$p_{E_-}$ achieves the maximum at $E=E_-$, and is relatively large over a band of width $\epsilon$ around $E$. We will now argue that this choice of $p_{E_-}$ will give the thermodynamic entropy when substituted into Eq.~\eqref{eq:SF=Shannon}. For example, suppose we took $p_{E_-}$ to be uniform in the interval  ${E_- - \epsilon/2 < E < E_- + \epsilon/2}$. The density of states $\rho(E)$ is related in the usual way to the thermodynamic (microcanonical) entropy at energy $E$ as $\rho(E) = \exp(S_{\rm micro})$ (Eq.~\eqref{eq:microcanonical_entropy} and Ref.~\cite{reif2009fundamentals}). Then using $\sum_i p_i = 1$, we see that $p_i = 1/(\epsilon\exp(S_{\rm micro}))$. So the entropy becomes
\[
S_F = \mean{\ln (\epsilon \exp(S_{\rm micro}))} = \ln\epsilon + S_{\rm micro}
\]
Because $S_{\rm micro}$ is extensive, it grows in thermodynamic limit, while the $\ln\epsilon$ grows slowly or not at all and therefore does not contribute in the thermodynamic limit. The above equations shows that if Hamiltonian $\hat{H}$ has eigenvectors $\ket{E}$ that have a constant overlap $p_{E_-}$ with each $\ket{E_-}$, as long as eigenvalues $E$ are from within the distance $\epsilon$ from $E_-$, and this overlap is zero otherwise, then the FOE gives the microcanonical entropy $S_{\rm micro}$ up to a small correction term.

We will now assume that the overlap $p_{E_-}$ is not a top hat shape but we will still assume that there are no fluctuations in $p_{E_-}$, i.e., now we extend the above argument of a top hat shaped $\Lambda(E-E')$ to $\Lambda$ of any shape.

With the previously introduced approximation $\Lambda_{E E_-} \approx \Lambda(E-E')$ we define a function $\lambda(E-E')$ to rescale $\Lambda(E-E')$ as follows,
\[\label{eq:sub_lambda}
\Lambda(E-E') = \frac{1}{\cal N}\lambda(E-E')
\]
where $\int \lambda(x) dx = 1$. Using unitarity of the $c_{EE'}$ elements, we have that $\sum_{E'}\Lambda(E-E')\approx \sum_{E'} \Lambda_{EE'} = \sum_{E'} \ov{c_{EE'}^2} =  1$. Therefore, using approximation $\sum_{E'} \rightarrow \int dE' \rho(E')$, we obtain
\[
\begin{split}
1 & \approx \sum_{E'} \Lambda(E-E') = \int dE' \rho(E')\Lambda(E-E')\\
& \approx \int dE' \rho(E)\Lambda(E-E') \\
&  = \rho(E)\int dE' \frac{\lambda(E-E') }{{\cal N}} = \frac{\rho(E)}{\cal N}
\end{split}
\]
Where the second $\approx$ is due to our  assumption that the width of $\Lambda$ is stil $O(\epsilon)$, hence $\rho(E')$ can be considered to be approximately equal to a constant $\rho(E)$ on this small interval where $\Lambda$ is large. This gives the normalization ${\cal N} = \rho(E)$. Combining $p_{E_-} = \ov{p_{E_-}}$, Eqs.~\eqref{eq:corr_in_cs} and \eqref{eq:sub_lambda}, and plugging the result into the formula for the FOE, Eq.~\eqref{eq:SF=Shannon}, we obtain
\[
\label{eq:S_F=int_lambda_ln_lambda}
\begin{split}
S_F & =- \sum_{E_-} p_{E_-} \ln p_{E_-}=- \sum_{E_-} \Lambda_{E E_-} \ln \Lambda_{E E_-}\\
&\approx -\int dE'\rho(E') \frac{\lambda(E-E')}{\rho(E)}\ln\Big(\frac{\lambda(E-E')}{\rho(E)}\Big)\\
&\approx -\int de \lambda(e)\ln\lambda(e) + \ln\rho(E)
\end{split}
\]
Where we have used the substitution $e = E-E'$.

To investigate the effect of the width of $\lambda(e)$, which represents the energy spread of matrix elements connecting the energy eigenvectors $\ket{E_-}$ and $\ket{E}$, we introduce a function $\tilde{\lambda}$ which depends only on the shape but not the width of $\lambda$ as
\[
\lambda(e) = \frac{1}{\epsilon}\tilde{\lambda}\Big(\frac{e}{\epsilon}\Big)
\]
The argument of $\tilde{\lambda}$ is dimensionless. Eq.~\eqref{eq:S_F=int_lambda_ln_lambda} becomes
\[\label{eq:S_FaveIntegral}
\begin{split}
S_F &= -\int de \lambda(e)\ln\lambda(e) + \ln\rho(E)\\
&= \int dx \tilde{\lambda}(x) \ln \tilde{\lambda}(x) + \ln(\epsilon) + \ln\rho(E)
\end{split}
\]
The first term depends only on the shape of $\lambda$ and not its variance. The second term gives its dependence on the energy spread, and the last term is the microcanonical entropy. We note that Eq.~\eqref{eq:S_FaveIntegral} also represents FOE of a microcanonical state $\R^{\rm micro}_{E_0}=\frac{1}{\mathcal{N}}\sum_{|E-E_0|< \epsilon/2}\pro{E}{E}$, as pictured on Fig.~\ref{fig:SXEvsE}, where $\mathcal{N}=\tr\big[\sum_{|E-E_0|< \epsilon/2}\pro{E}{E}\big]$ is the normalization constant.

Now we understand how the entropy is obtained when we assume $p_{E_-} = \ov{p_{E_-}}$. But in fact, Eq.~\eqref{eq:corr_in_cs} gives us $p_{E_-}$, averaged over the random
matrix ensemble, and any particular realization will fluctuate giving
\[
\label{eq:pE_-=LambdaEta}
p_{E_-} = \Lambda_{E E_-} \eta_{E_-}^2,
\]
where  $\eta_{E_-}$ is random variable that fluctuates from eigenstate to eigenstate so that the value of it averaged over the ensemble of $\eta$'s, $\ov{\eta_{E_-}^2} = 1$. We will
call the distribution of values of $\eta$ to be $P(\eta)$. The form of this distribution is not important to our analysis as we will see. We can now calculate how this multiplicative term affects the entropy by averaging over possible realizations of the $\eta$'s. Rewriting Eq.~\eqref{eq:SF=Shannon} using Eq.~\eqref{eq:pE_-=LambdaEta}, we have
\[
\label{eq:SF=LambdaEtalog}
S_F = - \sum_{E_-} \Lambda_{E E_-} \eta_{E_-}^2 \ln(\Lambda_{E E_-} \eta_{E_-}^2).
\]
Because the distribution of $\eta_{E_-}$ is the same for all $E_-$, the average value will be also the same, which allows us to introduce a random variable $\eta$ with the exactly same probability distribution, and the same property $\ov{\eta^2} = 1$. Then we can write
\begin{subequations}\label{eq:eta_reintroduced}
\begin{align}
\ov{\eta_{E_-}^2}&=\ov{\eta^2}=1\\
- \ov{\eta_{E_-}^2 \ln(\eta_{E_-}^2)}&=- \ov{\eta^2 \ln(\eta^2)}
\end{align}
\end{subequations}
which holds for all $E_-$, i.e., the averages do not depend on specific $E_-$ anymore. Using the above, we compute the average of the FOE of an energy eigenstate (which we stress out now by adding in the dependence $(\ket{E})$) as
\[
\label{eq:SF=LambdaEtalog_average}
\begin{split}
\ov{S_F(\ket{E})} &\!=\! - \sum_{E_-} \Lambda_{E E_-} \ov{\eta_{E_-}^2} \ln\Lambda_{E E_-}
- \sum_{E_-} \Lambda_{E E_-} \ov{\eta_{E_-}^2 \ln\eta_{E_-}^2}\\
&\!= - \sum_{E_-} \Lambda_{E E_-} \ln\Lambda_{E E_-} - \ov{\eta^2 \ln\eta^2}\sum_{E_-} \Lambda_{E E_-}\\
&\!=- \sum_{E_-} \Lambda_{E E_-} \ln\Lambda_{E E_-} - \ov{\eta^2 \ln\eta^2}.\\
&\!=\ln\rho(E)+\int dx \tilde{\lambda}(x) \ln \tilde{\lambda}(x) + \ln(\epsilon) - \ov{\eta^2 \ln\eta^2}.
\end{split}
\]
For the third equality we have used $\sum_{E_-} \Lambda_{E E_-}=1$, and for the last equality we used combination of Eqs.~\eqref{eq:S_F=int_lambda_ln_lambda} and~\eqref{eq:S_FaveIntegral}. This gives the same entropy as calculated in Eq.~\eqref{eq:S_FaveIntegral} that ignored these fluctuations, save for an additional term of order 1.  The fluctuations in the entropy $\mathrm{Var}(S_F)$, assuming independent $\eta$'s, are straightforward to calculate and are negligibly small. Therefore, the FOE of energy eigenstate $\ket{E}$ is approximately equal to the mean value~\eqref{eq:SF=LambdaEtalog_average}, which is equal to microcanonical entropy, up to terms of order one, which become irrelevant in the thermodynamic limit. We note that Eq.~\eqref{eq:SF=LambdaEtalog_average} that represents FOE of an energy eigenstate state, as depicted on Fig.~\ref{fig:SXEvsE}, differs from the FOE of a microcanonical state, Eq.~\eqref{eq:S_FaveIntegral}, only by the last term, $- \ov{\eta^2 \ln(\eta^2)}$.

Therefore, FOE of an energy eigenstate gives the thermodynamic entropy, up to an additive constant. This argument is relying on the relationship between non-integrable systems and random matrix models. What we argued is that the interactions introduced by adding in $\epsilon \hat{H}^{(\mathrm{int})}$ to an integrable Hamiltonian causes an energy eigenstate $\ket{E}$ to have substantial non-zero overlap with the integralable states $\ket{E_-}$ for states within of order $\epsilon$ of its energy. In the integrable case, because of the infinite number of invariants, a lot of states have very small overlap, but without these invariants present, there is much more overlap. Because the thermodynamic entropy can be obtained
for a large variety of distribution, the details of the precise amount of overlap are irrelevant to the final answer.

\subsection{FOE of a superposition of energy eigenstates}\label{subsec:FOE_of_superposition}

Here we derive that FOE of a superposition of energy eigenstates with random phases give a value that is higher than the microcanonical entropy, and that the superposition of close energy eigenstates with random phases give the microcanonical entropy.

We consider an initial state
\[\label{eq:general_initial_pure}
\ket{\psi}=\sum_Ee^{i\phi_E}d_E\ket{E}
\]
where the phases are random, $d_E=\sqrt{p_E(\R)}$ are real positive or zero numbers such that $\sum_Ed_E^2=1$, and we calculate the FOE for this state.

Using the same notation as in Eq.~\eqref{eq:SF=Shannon},
we define
\[
\label{eq:def_p_E_psi}
p_{E_-} \equiv \bra{E_-}\R\ket{E_-}=|\braket{E_-}{\psi}|^2
\]
and we first we calculate its value when averaged over all phases $\phi_E$. Because this is equivalent mathematically to a random walk, we have
\[
\label{eq:ave_p_E_psi}
\ov{p_{E_-}} = \sum_{E}d_E^2|\braket{E}{E_-}|^2
\]
By the Central Limit Theorem, we know that the distribution of $\braket{E_-}{\psi}$ is a complex Gaussian. But first we will ignore these fluctuations as we did in the last section by setting $p_{E_-} = \ov{p_{E_-}}$. In other words, first we will assume that phases $\phi_E$ are such that $p_{E_-} = \ov{p_{E_-}}$. This will give us an approximation to $S_F(\R)$, that we denote $S_F^0(\R)$.

We also note that for a large number of $d_E$'s contributing to $\ket{\psi}$, Eq.~\eqref{eq:ave_p_E_psi} is self averaging, meaning that we can think of $|\braket{E}{E_-}|^2$ (which is highly fluctuating) as the mean value (where the mean is taken over random matrices $H_r$) which is given by $\Lambda_{EE_-}$ introduced in Eq.~\eqref{eq:corr_in_cs} plus a randomly fluctuating term so that
\[\label{eq:self_averaging}
\ov{p_{E_-}} =\sum_{E}d_E^2\Lambda_{EE_-} + O(n^{-1/2})
\]
where $n$ is number of elements $d_E$ that contribute substantially, and where we have used defining relation for $\Lambda_{EE_-}$, Eq.~\eqref{eq:corr_in_cs}.  Assuming that $n$ is sufficiently large, we can neglect the term. The above equation helped us to avoid using the highly fluctuating term $|\braket{E}{E_-}|^2$, by using the averaging over the random matrices. This ``smoothing out'' is a perfectly fine procedure as long as there are more than a few eigenstates in the superposition, Eq.~\eqref{eq:general_initial_pure}. Realizing that the correction term is small when using this approximation will allow us to neglect this correction term, which will soon give us precise estimates on the FOE.

Inserting Eq.~\eqref{eq:self_averaging} into Eq.~\eqref{eq:SF=Shannon} and using Jensen's Theorem (Theorem~\ref{thm:Jensen}) we obtain
\[\label{eq:S_F_superposition}
\begin{split}
S_F^0(\R)&\approx - \sum_{E_-} \sum_{E}d_E^2\Lambda_{EE_-} \ln \sum_{E}d_E^2\Lambda_{EE_-}\\
&\geq - \sum_{E}d_E^2\sum_{E_-}\Lambda_{EE_-} \ln \Lambda_{EE_-}\\
&=\sum_{E}d_E^2 \ov{S_F(\ket{E})}+\ov{\eta^2 \ln\eta^2}
\end{split}
\]
The inequality sign $\geq$ becomes equality when, for all non-zero $d_E$, $|\Lambda_{{E}{E_-}}|^2$ are equal, which is approximately true for close energy eigenstates $\ket{E}$ peaked around a given value of energy denoted $E_0$, i.e., when Eq.~\eqref{eq:general_initial_pure} denotes a PS state. The last equality comes from the third row in Eq.~\eqref{eq:SF=LambdaEtalog_average}, and $\eta_{E_-}$ is a coefficient describing the effect the random matrix (Gaussian random variable), defined in the  Eq.~\eqref{eq:pE_-=LambdaEta}.

Now we will turn to more general case when $p_{E_-} \neq \ov{p_{E_-}}$ and introduce a random complex variable $\zeta_{E_-}$, that will capture the statistical properties of $p_{E_-}$ written in Eq.~\eqref{eq:def_p_E_psi} coming from randomness of phases $\phi_E$. We define $\zeta_{E_-}$ by writing $p_{E_-} = \ov{p_{E_-}} |\zeta_{E_-}|^2$, where $\ov{|\zeta|^2} = 1$. In analogy to the previous section (Eq.~\eqref{eq:zeta_reintroduced}), we can also introduce the random variable $\zeta$ that has the same probability distribution, and for which
\begin{subequations}\label{eq:zeta_reintroduced}
\begin{align}
\ov{|\zeta_{E_-}|^2}&=\ov{\zeta^2}=1\\
-\ov{|\zeta_{E_-}|^2\ln|\zeta_{E_-}|^2}&=- \ov{|\zeta|^2\ln|\zeta|^2}
\end{align}
\end{subequations}
Using the definition of FOE, Eq.~\eqref{eq:SF=Shannon}, $p_{E_-} = \ov{p_{E_-}} |\zeta_{E_-}|^2$, the above equation, and the result for $p_{E_-}=\ov{p_{E_-}}$, Eq.~\eqref{eq:S_F_superposition}, we obtain
\[\label{eq:inequality_average_SF}
\begin{split}
\ov{S_F(\R)} &= S_F^0 (\R) - \ov{|\zeta|^2\ln|\zeta|^2}\\
&\gtrapprox\sum_{E}d_E^2 \ov{S_F(\ket{E})}-\ov{|\zeta|^2\ln|\zeta|^2}+\ov{\eta^2 \ln\eta^2}
\end{split}
\]
where $\gtrapprox$ becomes approximate equality $\approx$ for PS states peaked around energy $E_0$. Using $\sum_{E}d_E^2=1$, for such states we have
\[\label{eq:differences_between_eigen_and_super}
\begin{split}
\ov{S_F(\R_{E_0})} &\approx \sum_{E}d_E^2 \ov{S_F(\ket{E_0})}-\ov{|\zeta|^2\ln|\zeta|^2}+\ov{\eta^2 \ln\eta^2}\\
&=\ov{S_F(\ket{E_0})}-\ov{|\zeta|^2\ln|\zeta|^2}+\ov{\eta^2 \ln\eta^2}.
\end{split}
\]
We note that Eq.~\eqref{eq:differences_between_eigen_and_super} that represents FOE of a PS state with random amplitudes and phases, as depicted on Fig.~\ref{fig:SXEvsE}, differs from FOE of energy eigenstate, Eq.~\eqref{eq:SF=LambdaEtalog_average}, by $-\ov{|\zeta|^2\ln|\zeta|^2}$, and from the FOE of a microcanonical state, Eq.~\eqref{eq:S_FaveIntegral}, by $-\ov{|\zeta|^2\ln|\zeta|^2}+\ov{\eta^2 \ln\eta^2}$.

Now we calculate the relevant correction terms $\ov{|\zeta|^2\ln|\zeta|^2}$ and $\ov{\eta^2 \ln\eta^2}$. The Central limit Theorem gives probability distribution for $\zeta$ as
\[
P(\zeta) d^2\zeta = \frac{e^{-|\zeta|^2}}{\pi} d^2\zeta
\]
giving
\[\label{eq:zeta_mean}
\ov{|\zeta|^2\ln|\zeta|^2} \approx 0.422784336
\]
We can obtain a similar estimate for the second term: The distribution of the unitary matrix elements $c_{EE_-}$ should also be close to a Gaussian distribution, but these elements were assumed to be real (by time reversal invariance), in which case
\[\label{eq:eta_mean}
\ov{\eta^2\ln\eta^2} \approx 0.72963715
\]

Eq.~\eqref{eq:inequality_average_SF} then becomes
\[\label{eq:inequality_average_SF_numbers}
\ov{S_F(\R)} \gtrapprox \sum_{E}d_E^2 \ov{S_F(\ket{E})} + 0.3068528,
\]
and Eq.~\eqref{eq:differences_between_eigen_and_super} becomes
\[\label{eq:S_FapproxS_F+0.3}
\ov{S_F(\R_{E_0})} \approx \ov{S_F(\ket{E_0})} + 0.3068528.
\]

\begin{figure}[htp]
\centering
\includegraphics[width=1\hsize]{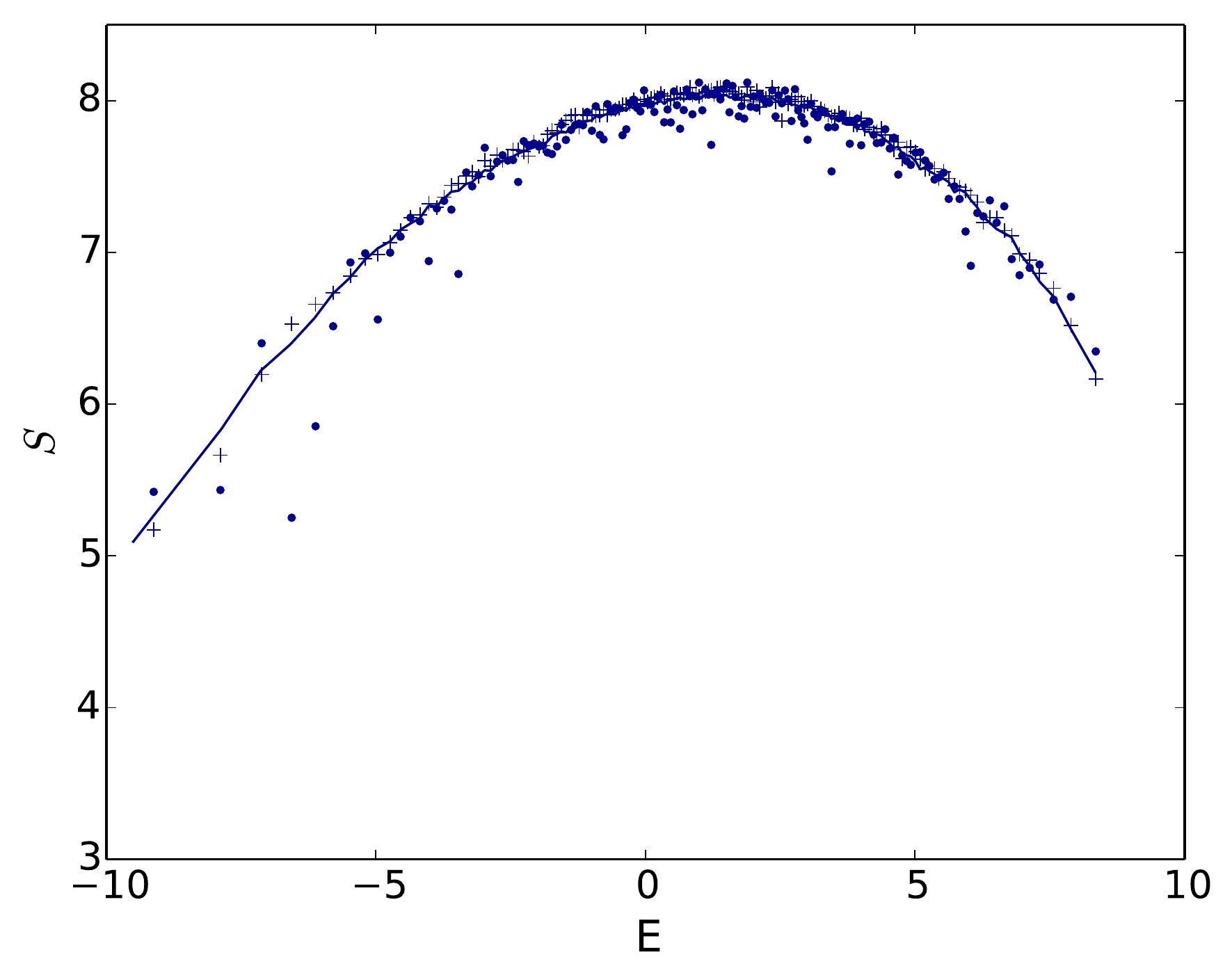}
\caption
{FOE of a microcanonical state $S_F(\R^{\rm micro}_{E})$ (line), adjusted FOE of a random PS state $S_F(\R_{E})+\ov{\eta^2 \ln\eta^2}-\ov{|\zeta|^2\ln|\zeta|^2}$ (crosses), and adjusted FOE of energy eigenstates $S_F(\ket{E})+\ov{\eta^2 \ln\eta^2}$ (dots). Order $1$ corrections $\ov{\eta^2 \ln\eta^2}$ and $\ov{|\zeta|^2\ln|\zeta|^2}$ have been calculated from the theory, and are given by Eqs.~\eqref{eq:zeta_mean} and \eqref{eq:eta_mean}. According to the theory, Eqs.~\eqref{eq:differences_in_FOE_not_numbers} and \eqref{eq:differences_in_FOE}, the above functions should be approximately equal. The plot shows that the curves nicely overlap, demonstrating that the observed differences between FOE's of different states match precisely our predictions of order 1 corrections. The parameters of the model are the same as in Fig.~\ref{fig:SXEvsE}(a).}
\label{fig:SXEvsEsubtract}
\end{figure}

We can also combine the results for all FOE of an energy eigenstate $\ket{E_0}$, Eq.~\eqref{eq:S_FaveIntegral}, FOE of a random PS state $\R_{E_0}$, Eq.~\eqref{eq:differences_between_eigen_and_super}, and FOE of microcanonical state $\R^{\rm micro}_{E_0}$, and see
\[\label{eq:differences_in_FOE_not_numbers}
\begin{split}
S_F(\R^{\rm micro}_{E_0})&\approx S_F(\R_{E_0})+\ov{\eta^2 \ln\eta^2}-\ov{|\zeta|^2\ln|\zeta|^2}\\
&\approx S_F(\ket{E_0})+\ov{\eta^2 \ln\eta^2},
\end{split}
\]
which after putting in numbers gives
\[\label{eq:differences_in_FOE}
\begin{split}
S_F(\R^{\rm micro}_{E_0})&\approx S_F(\R_{E_0})+0.3068528\\
&\approx S_F(\ket{E_0})+0.72963715.
\end{split}
\]
This explains offsets in FOE of different states seen in Fig.~\ref{fig:SXEvsE}. This figure can be adjusted to take into account these offsets, which we plotted in Fig.~\ref{fig:SXEvsEsubtract}. All curves nicely overlap, which confirms our analytical reasoning.

Now we will turn to more detailed discussion about the long-time limit. Let us assume now that we have such a PS state $\R_{E_0}$ with random phases. Neglecting the terms\footnote{We encountered four of them in total, three of them from Eq.~\eqref{eq:SF=LambdaEtalog_average}, and one of them from Eq.~\eqref{eq:inequality_average_SF}.} of order $1$ that become irrelevant in thermodynamic limit of large systems, according to Eq.~\eqref{eq:SF=LambdaEtalog_average}, elements on the right hand side approximate the microcanonical entropy,
\[\label{eq:S_F_approx_micro}
S_F(\R)\approx \ov{S_F(\R_{E_0})} \approx \ov{S_F(\ket{E_0})}\approx \ln\rho(E_0)\approx S_{\rm micro}(E_0).
\]
which is the microcanonical entropy at energy $E_0$.

We have therefore shown that for a random superposition of energy eigenstates, the FOE gives value that is larger than the averaged value of microcanonical entropies (Eq.~\eqref{eq:inequality_average_SF_numbers}), and for a random superposition of close energy eigenstates, this value is equal to the microcanonical entropy (Eq.~\eqref{eq:S_F_approx_micro}), up to terms of order 1 that become irrelevant in the thermodynamic limit. Since superposition of close energy eigenstates with random phases is a typical state at some late point in future, we can conclude that the factorized Observational entropy converges to the microcanonical entropy in the long-time limit for initial microcanonical states. Mathematically, we can write
\[\label{eq:convergence_eigenstate}
S_F(\R_t)\overset{t\rightarrow \infty}{\leadsto} S_{\rm micro}(E),
\]
for initial microcanonical states $\R_0$.

Now let us take a look at general pure states of form Eq.~\eqref{eq:general_initial_pure} that are not a superposition of close energy eigenstates, but rather a superposition of many energy eigenstates. For such many energy eigenstates with random phases, the second term in the last row of Eq.~\eqref{eq:self_averaging} will be really quite precisely zero. We can again start over above Eq.~\eqref{eq:S_F_superposition}, but now we switch the role of variables on which the Jensen's Theorem (Theorem~\ref{thm:Jensen}) is applied. In addition to inequality~\eqref{eq:S_F_superposition}, the following inequality is also true:
\[\label{eq:S_F_superposition_alternative}
\begin{split}
S_F^0(\R)&\approx - \sum_{E_-} \sum_{E}d_E^2\Lambda_{EE_-} \ln \sum_{E}d_E^2\Lambda_{EE_-}\\
&\geq - \sum_{E_-} \Lambda_{EE_-}\sum_{E}d_E^2 \ln d_E^2\\
&=-\sum_{E}d_E^2\ln d_E^2=S(\R_d)
\end{split}
\]
Where we have used $\sum_{E'} \Lambda_{EE'} =  1$, and the last equality is the consequence of the definition of the diagonal density matrix, Eq.~\eqref{eq:diagonal_density_matrix}, that we applied on the initial pure state of form~\eqref{eq:general_initial_pure}, while considering non-degenerate Hamiltonian $\hat{H}=\sum_E E\pro{E}{E}$. Considering the fluctuations $p_{E_-} = \ov{p_{E_-}} |\zeta_{E_-}|^2$, and the result from Eq.~\eqref{eq:inequality_average_SF}, we derive bound
\[
\ov{S_F(\R)} = S_F^0 (\R) - \ov{|\zeta|^2\ln|\zeta|^2}\gtrapprox S(\R_d) - \ov{|\zeta|^2\ln|\zeta|^2},
\]
where $\ov{|\zeta|^2\ln|\zeta|^2}\approx 0.422784336$.
Combining the above equation with the upper bound, Eq.~\eqref{eq:bounds_on_FOE}, we have
\[\label{eq:greater_S_F_than_diagonal}
S(\R_d)- 0.422784336\lessapprox S_{F}(\R)\leq S(\R_d)+O(\epsilon)
\]
where the left hand side approximation $\lessapprox$ depends on how well the second term in Eq.~\eqref{eq:self_averaging} can be approximated to be zero, i.e., it depends how many energy eigenstates with non-negligible $d_E$ are considered in the superposition, Eq.~\eqref{eq:general_initial_pure}, so that random phases can effectively average this term to zero. More such states are considered (i.e., bigger the superposition), the better the inequality $\lessapprox$. The validity of the right hand side (meaning that $O(\epsilon)$ is small) depends on the interaction strength between partitions of the Hilbert space and the fact that $\R$ spans across many energy eigenstates. Smaller the interaction strength (but importantly, non-zero, so the thermalization can take place), and the larger superpositions considered, better the inequality.

Ignoring the order 1 corrections that will become irrelevant in the thermodynamic limit, we can write
\[\label{eq:convergence_spanning}
S_F(\R_t)\overset{t\rightarrow \infty}{\leadsto} S(\R_d),
\]
for initial states $\R_0$ that span across many different energy eigenstates. (Although we have proved it only for pure states, it is relatively easy to generalize to mixed states.)

The two inequalities, Eqs.~\eqref{eq:inequality_average_SF_numbers} and~\eqref{eq:greater_S_F_than_diagonal}, can be also combined into a single inequality that considers both superposition of close energy eigenstates and superpositions of many energy eigenstates as
\[
S_{F}(\R)\gtrapprox \max\Big\{\sum_{E}p_E(\R)S_F(\ket{E})+ 0.307,\ \!S(\R_d)- 0.423\Big\}
\]
for general states with random phases, where $p_E(\R)=\tr[\P_E\R]$. Since such states correspond to states of the system at some long time in future, we can also combine Eqs.~\eqref{eq:convergence_eigenstate},~\eqref{eq:greater_S_F_than_diagonal}, and~\eqref{eq:convergence_spanning}, while considering approximation given by Eq.~\eqref{eq:S_F_approx_micro} and ignoring order $1$ corrections, and write
\[
S_F(\R_t)\overset{t\rightarrow \infty}{\leadsto}\max\Big\{\sum_{E}p_E(\R_0)S_{\rm micro}(E),\ \!S(\R_d)\Big\}.
\]
The above equation says, that for general states and in closed non-integrable systems, the FOE converges to either mean value of corresponding microcanonical entropies, or to the von Neumann entropy of the diagonal state, whichever is bigger, up to order 1 corrections that become irrelevant in the thermodynamic limit.

\section{Correspondence of $S_{xE}$ and FOE for very small $\epsilon$}
\label{sec-S_xE_FOE_correspondence}
In this section we show that Observational entropy $S_{xE}$ and FOE $S_F$ gives the same result, when the coarse-grained position projectors match the partitions of the Hilbert space for the FOE (also meaning that the number of coarse-grained position equals the number of partitions, $p=m$), and when we consider the interaction strength between different partitions to be zero, i.e., $\epsilon=0$, or to be so small that the differences between between energy eigenvalues of Hamiltonian with such zero interaction $\hat{H}_-$ and the full Hamiltonian $\hat{H}$ is much smaller than the typical energy difference between the eigenvalues of $\hat{H}$. This assumption then assures that each energy eigenstate of $\hat{H}_-$ have almost zero overlap with all energy eigenstates of $\hat{H}$ but one, which corresponds to the same energy eigenstate with a slight modification due to $\epsilon$. For simplicity we also assume that $\hat{H}_-$ is non-degenerate.

We start by considering the density matrix to be in a pure state, $\R = \pro{\psi}{\psi}$. We write
\[\label{eq:TrPEPxrhoPx}
p_{\vec x E} \equiv \tr[\P_E \P_{\vec x}^{(\delta)} \pro{\psi}{\psi} \P_{\vec x}^{(\delta)}\P_E]=  |\bra{E}\P_{\vec x}^{(\delta)}\ket{\psi}|^2.
\]
Similarly, we can write
\[\label{eq:TrPEPx}
V_{\vec x E} \equiv \tr[\P_E\P_{\vec x}^{(\delta)}\P_E] = \bra{E}\P_{\vec x}^{(\delta)}\ket{E}.
\]

We would like to simplify these two expressions in the limit of large system sizes in order to evaluate $S_{xE}$.

We can write the Hamiltonian $\hat{H}$ by dividing it up as in Eq.~\eqref{eq:H=H1+H2+Hint}. Let us employ the same strategy as in the last section and set $\epsilon=0$ and call that Hamiltonian $\hat{H}_-$. In this case, $\hat{H}_-$ is block diagonal in the $\vec{x}$ basis. The projectors $\P_{\vec x}^{(\delta)}$ are diagonal in eigenbasis of $\hat{H}_-$ with zero diagonal elements everywhere except for when $\vec{\tilde{x}} \in C_{\vec{x}}$ where the diagonal value is unity. Therefore $[\hat{H}_-,\P_{\vec x}^{(\delta)}]=0$ for all $C_{\vec{x}}$
and we can simultaneously diagonalize all of these operators.

$\bra{E_-}$ can be chosen to simultaneously be an eigenstate of all of the $P_{\vec x}$'s (choosing such $\bra{E_-}$ is possible because the common eigenbasis of $\hat{H}_-$, $\P_{\vec x}^{(\delta)}$ exists). Note that because of the diagonal form of these operators in this special basis, that $P_{\vec x}\ket{E_-}=0$ for all $\vec x$ but one, say, $\vec x'$. For that projector, $P_{\vec x'} \ket{E_-} = \ket{E_-}$ (a projector can only have eigenvalues $0$ or $1$). In other words, for each energy eigenvector $\ket{E_-}$ there is exactly one vector $\vec x'$ such that $P_{\vec x'} \ket{E_-} = \ket{E_-}$. This defines a function $x'=\vec x(E_-)$, and we can write $P_{\vec x(E_-)} \ket{E_-} = \ket{E_-}$.

In Eq.~\eqref{eq:TrPEPxrhoPx} we will approximate $\bra{E}$ by $\bra{E_-}$. We can do that, because of our assumption that $\epsilon$ is small. We take vector $\ket{\psi} = \sum_{E_-'}\braket{E_-'}{\psi}\ket{E_-'}$ written in the eigenbasis of $\hat{H}_-$. Then we can write
\[
\begin{split}
p_{\vec x E} &\approx  |\bra{E_-}\P_{\vec x}^{(\delta)}\ket{\psi}|^2=|\sum_{E_-'}\braket{E_-'}{\psi}\delta_{\vec x,\vec x(E_-')}\delta_{E_-,E_-'}|^2\\
&=|\braket{E_-}{\psi}\delta_{\vec x,\vec x(E_-)}|^2.
\end{split}
\]

With the same approximation $\bra{E}\approx\bra{E_-}$ we can write Eq.~\eqref{eq:TrPEPx} as
\[
V_{\vec x E} \approx \bra{E_-}\P_{\vec x}^{(\delta)}\ket{E_-}=\delta_{\vec x,\vec x(E_-')}.
\]

Then assuming we do not sum over elements such that $V_{\vec x E}=0$, we can write
\[\label{eq:Sxe_equals_S_F}
\begin{split}
S_{xE}&\equiv\sum_{\vec x,E}p_{\vec x E}\ln\frac{p_{\vec x E}}{V_{\vec x E}}\\
&\approx \sum_{\vec x,E_-,\vec x=\vec x(E_-)}|\braket{E_-}{\psi}\delta_{\vec x,\vec x(E_-)}|^2\ln\frac{|\braket{E_-}{\psi}\delta_{\vec x,\vec x(E_-)}|^2}{\delta_{\vec x,\vec x(E_-')}}\\
&=\sum_{E_-}|\braket{E_-}{\psi}|^2\ln|\braket{E_-}{\psi}|^2=S_{O(\C_{\hat{H}_-})}(\R)\\
&=S_{O{\displaystyle (}\C_{\hat{H}^{(1)}}\otimes \cdots\otimes \C_{\hat{H}^{(m)}}{\displaystyle )}}(\R)\equiv S_{F}(\R),
\end{split}
\]
where the previous from the last equality is due to the non-degeneracy of Hamiltonian without interaction $\hat{H_-}$, and assuming that partitions $\C_{\hat{H}^{(1)}}\otimes \cdots\otimes \C_{\hat{H}^{(m)}}$ copy the positional coarse-graining.

This analysis can be generalized to mixed states $\R$, by obtaining
\[
p_{\vec x E} \approx \bra{E_-} \P_{\vec x}^{(\delta)} \R \P_{\vec x}^{(\delta)}\ket{E_-}=  \bra{E_-} \R \ket{E_-}\delta_{\vec x,\vec x(E_-')},
\]
and calculating the same string of equalities as in Eq.~\eqref{eq:Sxe_equals_S_F}.

The above argument can be understood more intuitively by giving an example. Suppose we divide a one dimensional lattice system with $9$ sites $1,2,\dots,9$, and 4 particles into boxes of size $\delta = 3$. Here we are assuming that particle number is conserved. Then coarse graining in position, $C_{\vec{x}}$, separates basis states into different groupings. For example, the first particle could be in the box $\{1-3\}$, the next two could be in box $\{4-6\}$, and the final one could be in the box $\{7-9\}$. We can represent this coarse graining by the ``signature" $[1, 2, 1]$, when the particles are indistinguishable, which represents the number of particles in each box. The set of coarse grainings projectors $\C_1$ are isomorphic
to the set of allowed signatures, $[4,0,0]$, $[3,1,0]$, etc. When a projector $P_{\vec x}$ acts on a wavefunction, it is projecting out the components of the wavefunction with $P_{\vec x}$'s signature.

Now we consider eigenstates of $\hat{H}_-$. Because different boxes do not interact, we can write an arbitrary eigenstate $\ket{E_-}$  as the product of individual energy eigenstates in each box, $\ket{E_-} = \ket{E_1}\otimes\ket{E_2}\otimes\ket{E_3}$, where the $E_i$ denote the energy eigenstate of each separate box. But each of these eigenstates has a fixed particle number. Therefore to each total eigenstate, $\ket{E_-}$ we can associate a unique signature, for example $[1,2,1]$, meaning that $\ket{E_1}$ is a one particle eigenstate, $\ket{E_2}$ has two particles, etc. When we apply a projector to $\ket{E_-}$, $P_{\vec x}\ket{E_-}$, we will get zero unless the signature of $P_{\vec x}$ and $\ket{E_-}$
are the same. Therefore $P_{\vec x} \ket{E_-}$ will be zero unless ${\vec x}  = {\vec x(E_-)}$. And by orthogonality, $\braket{E_-'}{E_-}= 0$ unless $E_-'=E_-$, so we conclude that
$\bra{E_-'} P_{\vec x} \ket{E_-} = 0$ unless ${\vec x} = \vec x ({E_-})=\vec x ({E_-'})$, and $E_-'=E_-$, to which we arrived at above, by more general means.

\section{Convergence of $S_{xE}$ to microcanonical entropy}
\label{sec-S_xE_for_convergence}

In this first part of this section we show that the Observational entropy $S_{xE}$ is equal to the microcanonical entropy for energy eigenstates. In the second part, we show that $S_{xE}$ of a superposition of close energy eigenstates with random phases also gives the thermodynamical entropy, implying that the $S_{xE}$ converges to microcanonical entropy for initial microcanonical states.

\subsection{$S_{xE}$ of energy eigenstates}

We assume that the Hamiltonian is non-degenerate, which is the case for non-integrable systems, and for simplicity we assume that the Hamiltonian is real, i.e., all of its eigenvectors can be chosen to be real, which means that the Hamiltonian has spinless time-reversal symmetry.

We will also refer to the nonzero $x$ projector corresponding to $E_-$ as $x(E_-)$.
In contrast to the previous section, where we considered $\epsilon$ to be small, here we consider $S_{xE}$ with any finite $\epsilon$.  We will concentrate on the
entropy of an energy eigenstate $\ket{\psi} = \ket{E'}$, in which case Eq.~\eqref{eq:TrPEPxrhoPx}
becomes
\[
p_{\vec{x}E} =  |\bra{E}\P_{\vec x}^{(\delta)}\ket{E'}|^2.
\]
Using Eq.~\eqref{eq:E=CEEE-} this becomes
\[
\begin{split}
&\bra{E} \P_{\vec x}^{(\delta)}\ket{E'} = \sum_{E_-,E'_-,\vec{x}=\vec{x}(E_-)} c_{E'E'_-}c_{E E_-} \braket{E_-}{E'_-}\\
&=\sum_{E_-,\vec{x}=\vec{x}(E_-)} c_{E'E'_-}c_{E E_-}
\end{split}
\]
Computing the average over the $c$'s similar to what was done in Appendix~\ref{sec-FOE_for_eigenstates}
\[
\ov{p_{\vec{x}E}}=\ov{|\bra{E} \P_{\vec x}^{(\delta)}\ket{E'}|^2} =\!\!\!\!\!
\sum_{\substack{E_-,E'_-,\\ \vec{x}(E_-)=\vec{x}(E'_-)= \vec{x}}}\!\!\!\!\! \ov{c_{EE_-}c_{E'E_-} c_{EE'_-}c_{E'E'_-}}
\]
We take the $c$'s to be Gaussian random variables. They are also real from the definition, because the eigenvectors are assumed to be real. They are not completely independent
because they are all orthogonal,
\[
\sum_{E_-} c_{EE_-}c_{E'E_-} = \delta_{EE'}
\]
This leads to extra four point correlations which are derived in appendix \ref{app:4PntCorr}, giving
\[
\begin{split}
\label{eq:aveCCCC=}
&\ov{c_{EE_-}c_{E'E_-} c_{EE'_-}c_{E'E'_-}} =\\
&-\frac{\Lambda^2_{E'E_-}\Lambda^2_{EE-}}{\sum_{E''_-}\Lambda^2_{EE''_-}\Lambda^2_{E'E''_-}}
+\delta_{E_-E'_-}\ov{c^2_{E'E_-}c^2_{EE_-}}
\end{split}
\]
Here as given in Eq.~\eqref{eq:ccave=Lambda}, $\ov{c^2_{EE-}} = \Lambda_{EE-}$. Using Wick's
Theorem,
\[
\label{eq:aveCsqCsq}
\ov{c^2_{E'E_-}c^2_{EE_-}} = \Lambda^2_{E E_-}\Lambda^2_{E'E_-} + 4 \delta_{EE'} (\Lambda^2_{E'E_-})^2
\]
Thus $p_{\vec{x}E}$ is the sum of three terms and the magnitude of all of them can be estimated. Denote the number of energy levels contributing to $\Lambda$ as $N$, and the number of $\vec{x}$ signatures that are being summed over as $N_{\vec{x}}$.  Thus the first term in Eq.~\eqref{eq:aveCCCC=} is of order $\Lambda^2 \frac{N_{\vec{x}}^2}{N^2}$, and the first in Eq.~\eqref{eq:aveCsqCsq} is of order $\frac{N_{\vec{x}}}{N^2}$ and the second is of order $\frac{1}{N^2}$. If the size of a box is large, we have that $\frac{N_x}{N} \ll 1$. Thus the surviving term in this limit is
\[
\label{eq:pxE=sumLambdaLambda}
p_{\vec{x}E} = \sum_{E_-,\vec{x}=\vec{x}(E_-)} \Lambda^2_{EE_-}\Lambda^2_{E'E_-} \equiv \Omega_{E E';\vec{x}},
\]
where we have defined $\Omega_{E E';\vec{x}}$ by the above equation, in order to use it further.

Similarly
\[
\label{eq:VxE=sumLambda}
\begin{split}
&V_{\vec{x}E} = \tr[\P_{\vec x}^{(\delta)}\P_E] = \braket{E|\P_{\vec x}^{(\delta)}}{E} =\\
&\sum_{E_-,E'_-,\vec{x}=\vec{x}(E_-)} c_{EE_-}c_{EE'_-}\braket{E_-}{E'_-} = \sum_{E_-,\vec{x}=\vec{x}(E_-)} \Lambda^2_{E E_-}
\end{split}
\]
Because $\sum_{E_-} \Lambda^2_{EE_-} = 1$, Eq.~\eqref{eq:VxE=sumLambda} can be rewritten
using Eq.~\eqref{eq:pxE=sumLambdaLambda} as
\[
V_{xE} = \sum_{E'} \Omega_{E E';x}
\]
and therefore
\[
S_{xE}(\ket{E'}) = -\sum_{x,E} \Omega_{E E';x} \ln\frac{\Omega_{E E';x}}{\sum_{E'}{\Omega_{E E';x}}}
\]

We are assuming that $\Lambda^2_{EE_-}$ only is sizable for $|E-E_-| \ll E$, and can be written as
$\Lambda^2_{EE_-} = \Lambda^2(E-E_-) $.  Because  $\sum_{x,E} p_{\vec{x}E}  = \sum_{x,E} \Omega_{EE';x} = 1$,
Therefore $\Omega$ from Eq.~\eqref{eq:pxE=sumLambdaLambda}
is of the form
\[\label{eq:norm_Omega}
\sum_{\vec{x},E} \Omega_{E E';\vec{x}} = \sum_x\int \Omega_{EE';x} \rho(E) dE = 1,
\]
where $\rho(E)$ denotes the density of states. Because, as we explained in Appendix~\ref{sec-FOE_for_eigenstates}, $\Lambda$ is approximated by a Lorenztian and is therefore highly peaked, and according to Eq.~\eqref{eq:pxE=sumLambdaLambda} $\Omega$ is a convolution of two $\Lambda$ and therefore is also highly peaked, then we can write this function as
\[
\Omega_{E E';\vec{x}} =\frac{1}{\mathcal{N}} \lambda_x(E-E'),
\]
where $\mathcal{N}$ is a normalization constant and $\lambda_x(E-E')$ denotes a highly peaked distribution with normalization $\int de \lambda_x(e) = 1$. From Eq.~\eqref{eq:norm_Omega}, we obtain the normalization constant
\[
\mathcal{N}=N_x\rho(E').
\]
Rewriting the entropy in terms of $\lambda$,
\[
\begin{split}
S_{xE}(\ket{E'}) &= -\sum_{x,E} \frac{\lambda_x(E-E')}{N_x\rho(E')} \ln\frac{\lambda_x(E-E')}{\rho(E')}\\
&=-\sum_x\int dE \frac{\lambda_x(E-E')}{N_x} \ln\frac{\lambda_x(E-E')}{\rho(E')}.
\end{split}
\]
$\rho(E')$ disappeared from the denominator because of approximation $\sum_E\rightarrow \int_E dE$. Substituting $e \equiv E-E'$
\[
S_{xE}(\ket{E'}) =\ln\rho(E')-\sum_x\frac{1}{N_x}\int de \lambda_x(e) \ln\lambda_x(e) .
\]
The second term on the right hand side is of order 1 (proportional to $\ln \Delta E$ in comparison with definition of microcanonical entropy, Eq.~\eqref{eq:microcanonical_entropy}) unless $\lambda_x$ has a pathological form (i.e., for example $\lambda_x$ being a long-tailed function). The first term is the microcanonical entropy. This shows that $S_{xE}$ of an energy eigenstate gives the microcanonical entropy, up to a constant of order 1 that becomes irrelevant in the thermodynamic limit.

\subsection{$S_{xE}$ of a superposition of energy eigenstates}

In analogy with the second part of Appendix~\ref{sec-FOE_for_eigenstates}, we  consider an initial state $\R=\pro{\psi}{\psi}$, $\ket{\psi}=\sum_{E'}e^{i\phi_{E'}}c_{E'}\ket{{E'}}$, with random phases $\phi_{E'}$. Then, in analogy of Eq.~\eqref{eq:self_averaging}, considering a simplified argument (without ``smoothing out'' by function $\Lambda_{EE_-}$, without random variable $\eta$ coming from averaging over random matrices, and without $\zeta$ coming from averaging over phases), and ignoring order 1 corrections, we have
\[
\begin{split}
p_{\vec{x}E} &=  |\bra{E}\P_{\vec x}^{(\delta)}\ket{\psi}|^2=\sum_{E'}c_{E'}^2|\bra{E}\P_{\vec x}^{(\delta)}\ket{E'}|^2\\
&+\sum_{E'\neq E''}c_{E'}c_{E''}e^{i (\phi_{E'}- \phi_{E''})}\bra{E}\P_{\vec x}^{(\delta)}\ket{E'}\bra{E''}\P_{\vec x}^{(\delta)}\ket{E},
\end{split}
\]
where the second term is small due to the randomness of $\phi_{E'}$ and $\phi_{E''}$ and we will neglect it.

In analogy of Eq.~\eqref{eq:S_F_superposition},
\[\label{eq:S_F_superposition_sxe}
S_{xE}(\R)\gtrapprox \sum_{E}d_E^2S_{xE}(\ket{E}).
\]
where the inequality becomes approximate equality when for all non-zero $d_E$, eigenvectors $\ket{E}$ are close to each other and peaked around some eigenvalue $E_0$, i.e., for PS states with random phases. For such state we have
\[
S_{xE}(\R)\approx \sum_{E}\!d_E^2S_{xE}(\ket{E_0})=S_{xE}(\ket{E_0})\sum_{E}\!d_E^2=S_{xE}(\ket{E_0}).
\]
Since random phases indicate the state of the system some time in future, we can conclude that for initial PS states, $S_{xE}$ converges to the microcanonical entropy.

The above arguments ignore order 1 corrections, because we ignored fluctuations in eigenvectors by averaging of the $c$'s. However we can follow the same logic {\em mutatis mutandis} of subsection~\ref{subsec:FOE_of_superposition} which leads to analogous corrections and a similar relation to Eq.~\eqref{eq:differences_in_FOE}.

\section{Four point correlations of random eigenvectors}
\label{app:4PntCorr}

Consider two normalized N dimensional vectors $\vec{x}$ and $\vec{y}$ thar are orthogonal $\vec{x}\cdot\vec{y} = 0$. Aside from that constraint, the two vectors were drawn from Gaussian distributions so that every component is independent and $\mean{x_i^2} = \sigma^2_i$ and $\mean{y_i^2} = \omega^2_i$. We would like to compute $\mean{x_n y_n x_m y_m}$ for $n\ne m$.

The probability of any particular $\vec{x}$ and $\vec{y}$ is
\[
P(\vec{x},\vec{y}) \propto \exp\Big(-\sum_i \frac{x_i^2}{2\sigma^2_i} +\frac{y_i^2}{2\omega^2_i}\Big)\delta\Big(\sum_i x_i y_i\Big)
\]
And therefore
\[
\mean{x_n y_n x_m y_m} = \int\prod_i (dx_i dy_i) P(\vec{x},\vec{y}) x_n y_n x_m y_m
\]
This can be written in the form
\[
\label{eq:meanxyxy=partial}
\mean{x_n y_n x_m y_m} = \frac{\partial^2 \ln Z}{\partial \epsilon_m\epsilon_m}\bigg\rvert_{\epsilon_m=\epsilon_n = 0}
\]
where
\[
\begin{split}
Z(\epsilon_m,\epsilon_n) &=\int\prod_j (dx_j dy_j) \exp\Big(-\sum_i \frac{x_i^2}{2\sigma^2_i} +\frac{y_i^2}{2\omega^2_i} \\
&+ \epsilon_m \delta_{im} +\epsilon_n\delta_{in}\Big) \delta(\vec{x}\cdot\vec{y})
\end{split}
\]
We evaluate this by using the Fourier representation of the $\delta$ function and ignoring
irrelevant const prefactors:
\[
\begin{split}
Z(\epsilon_m,\epsilon_n) &\propto\int_{-i\infty}^{i\infty} d\lambda \int\prod_j dx_j dy_j \exp\Big(-\sum_i \frac{x_i^2}{2\sigma^2_i} +\frac{y_i^2}{2\omega^2_i} \\
&+ \epsilon_m \delta_{im} +\epsilon_n\delta_{in} + \lambda x_i y_i\Big)
\end{split}
\]
Now the integral pairs $dx_j dy_j$ can be integrated separately and we can use that
\[
\int dx dy \exp{-\frac{1}{2}\Big(\frac{x^2}{\sigma^2}+\frac{y^2}{\omega^2} + 2\lambda x y\Big)} = 2\pi\Big(\frac{1}{\sigma^2\omega^2} -\lambda^2\Big)^{-1/2}
\]
to write
\[
\begin{split}
&Z(\epsilon_m,\epsilon_n) \propto\int_{-i\infty}^{i\infty} d\lambda \exp\bigg(-\frac{1}{2}\sum_{i\ne n,m} \ln\Big(\frac{1}{\sigma_i^2\omega_i^2} -\lambda^2\Big)\\
&-\frac{1}{2}\ln\Big(\frac{1}{\sigma_m^2\omega_m^2} -\lambda^2-2\lambda\epsilon_m\Big)-\frac{1}{2}\ln\Big(\frac{1}{\sigma_n^2\omega_n^2} -\lambda^2-2\lambda\epsilon_n\Big)\bigg)
\end{split}
\]
Because we are differentiating with respect to $\epsilon$, the terms of order $\epsilon^2$ have been dropped. And we can further make use of this by Taylor expanding the two final logarithms, giving
\[
\begin{split}
Z(\epsilon_m,\epsilon_n) &\propto\int_{-i\infty}^{i\infty} d\lambda \exp\bigg(-\frac{1}{2}\sum_{i\ne n,m} \ln\Big(\frac{1}{\sigma_i^2\omega_i^2} -\lambda^2\Big)\\
&-\frac{\lambda\epsilon_m\sigma_m^2\omega_m^2}{1-\lambda^2\sigma_m^2\omega_m^2}-\frac{\lambda\epsilon_n\sigma_n^2\omega_n^2}{1-\lambda^2\sigma_n^2\omega_n^2}\bigg)
\end{split}
\]
If we consider the terms $\ln(\frac{1}{\sigma_i^2\omega_i^2} -\lambda^2)$, we can write this as $\ln(1 - \sigma_i^2\omega_i^2\lambda^2)$, plus an unimportant additive constant. This in turn can be expanded to second order in $\lambda$. To that order
\[
-\sum_{i\ne n,m} \ln\big(1 - \sigma_i^2\omega_i^2\lambda^2\big) =  - \frac{\lambda^2}{2} \sum_{i\ne n,m}  \sigma_i^2\omega_i^2
\]
When integrating of $\lambda$, this gives a Gaussian with a variance of $1/\sum_{i\ne n,m}  \sigma_i^2\omega_i^2$. Using units where the maximum of $\sigma$ and $\omega$ is unity, and their distribution has a width of $N$, we can then say that this variance is $O(1/N)$. This means that in the integrand, $\lambda\gg 1/\sqrt(N)$ will give a negligible contribution to the integral and we can ignore all such contributions. Therefore the terms $\lambda^2\sigma_m^2\omega_m^2$ and$\lambda^2\sigma_n^2\omega_n^2$ also give
negligible contributions and we can write:
\[
\begin{split}
Z(\epsilon_m,\epsilon_n) &\propto\int_{-i\infty}^{i\infty} d\lambda \exp\bigg(\frac{\lambda^2}{2}\sum_{i} \sigma_i^2\omega_i^2\\
&+\lambda\big(\epsilon_m\sigma_m^2\omega_m^2+\epsilon_n\sigma_n^2\omega_n^2\big)\bigg)
\end{split}
\]
We can now integrate over $\lambda$ obtaining
\[
Z(\epsilon_m,\epsilon_n) \propto
e^{-\frac{(\epsilon_m\sigma_m^2\omega_m^2+\epsilon_n\sigma_n^2\omega_n^2)^2}{2\sum_i\sigma_i^2\omega_i^2}}
\]
Using \eqref{eq:meanxyxy=partial}, we differentiate with respect to the $\epsilon$'s to obtain
\[
\mean{x_n y_n x_m y_m} =
-\frac{\sigma_m^2\omega_m^2\sigma_n^2\omega_n^2}{\sum_i^N\sigma_i^2\omega_i^2}
\]
for $n\ne m$.

\begin{figure}[t]
\begin{center}
\includegraphics[width=1\hsize]{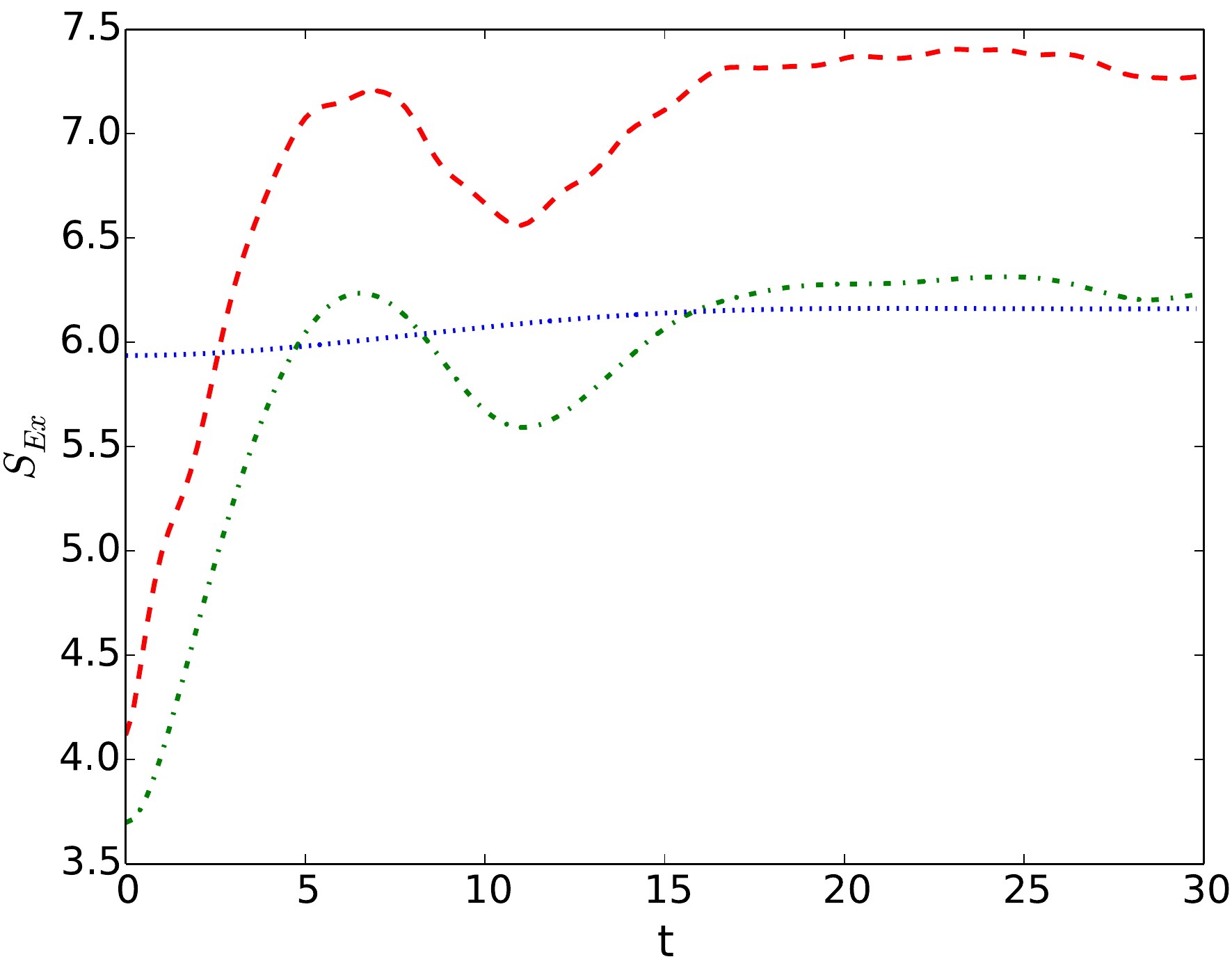}
\caption
{
Observational entropy of measuring energy with resolution $\Delta$ and position with resolution $\delta$ for a system of length $L=16$. The system starts contained within the first 8 sites with hard wall boundary conditions, in energy eigenstate number 11 of Hamiltonian $\hat{H}^{(1-8)}$. At $t=0$ the right wall is expanded so that
$L=16$ and the system evolves.  We study the integrable system, defined by parameters $t=V=1$, and $t'=V'=0.0$. The resolution in position is 4 sites corresponding to $\delta=4$. The resolution in energy is ${\Delta=\frac{E_{\max}-E_{\min}}{M}}$, where $E_{\max}$ and $E_{\min}$ are maximum and minimum eigenvalues of the Hamiltonian, and $M$ is the number of energy bins. The three lines correspond to different resolutions in measuring energy: M=1 (red dashed), M=8 (green half-dashed), M=64 (blue dotted). M=1 represents an inability to measure energy, and the resulting entropy is then observational entropy coarse-grained only in position.
}
\label{fig:SEx}
\end{center}
\end{figure}

\section{$S_{Ex}$, Observational entropy of measuring energy and then position}
\label{sec-S_Ex}

We already mentioned in the text, that the reverse order of two projections, Eqs.~\eqref{eq:S_Ex_entropy_projectors}, i.e., first measuring the energy, and then measuring the coarse-grained position, leads to an entropy that is independent of time, and therefore does not seem to have a good interpretation of entropy that has desirable properties for closed systems out of equilibrium. Here we explore this even further, by considering the same order of operations, but we also coarse-graining in energy, which introduces a non-trivial time-dependence. We will argue that neither this choice leads to a meaningful non-equilibrium entropy.

We consider two coarse-grained sets of projectors in position and energy
\begin{subequations}\label{eq:S_xE_entropy_projectors_coarse-grained}
\begin{align}
\C_{{\hat{X}}^{(\delta)}}&=\{\P_{\vec x}^{(\delta)}\}_{\vec x},\quad \P_{\vec{x}}^{(\delta)}=\sum_{\vec{\tilde{x}} \in C_{\vec{x}}} \pro{\vec{\tilde{x}}}{\vec{\tilde{x}}},\\
\C_{{\hat{H}}^{(\Delta)}}&=\{\P_E^{(\Delta)}\}_E,\quad \P_E^{(\Delta)}=\!\!\!\!\!\sum_{\tilde{E}\in [E,E+\Delta]} \!\!\!\!\!\pro{\tilde{E}}{\tilde{E}},
\end{align}
\end{subequations}
where as before in Eq.~\eqref{eq:x_projectors}, $\vec{x}=(x^{(1)},\dots,x^{(N)})$ is a vector denoting positions of $N$ particles, and its
elements take values of any $x_1,\dots,x_p$. $C_{\vec{x}}$ denotes a hypercube that starts at vector
$\vec{x}$ and is of width $\delta$. $\Delta$ denotes the width of coarse-graining in energy.

Now we will study the Observational entropy
\[
S_{Ex}(\R)\equiv
S_{O(\C_{{\hat{H}}^{(\Delta)}},\C_{{\hat{X}}^{(\delta)}})}(\R).
\]
Evolution of this entropy in the one-dimensional fermionic chain as a function of time is plotted in Fig.~\ref{fig:SEx}, for different values of $\Delta$. The resolution in measuring position is fixed to $4$ sites. The resolution in measuring energy was varied. The energy bins span the entire energy spectrum from the lowest to highest eigenvalue, $E_{\min}$ and $E_{\max}$ respectively. We define $\Delta=\frac{E_{\max}-E_{\min}}{M}$, where $M$ is the number of energy bings. One can see that as this number increases, the dynamics become smoother and vary less rapidly. In the limit where each bin contains only one energy level ($M\rightarrow \infty$), it is easily seen that there is no time dependence.

Theoretically, when a fine grain projection over the energy is applied, a time independent quantity is obtained because such a projection makes any state
stationary. Coarsening the energy projection makes the resultant quantity time dependent, but in a way that depends strongly on the amount of energy coarse graining, rather than the underlying dynamics. Therefore the dynamics of this kind of entropy depend on the choice of energy bin size, and do not reflect the
underlying microscopic dynamics of the system.

\bibliographystyle{apsrev}
\bibliography{observational_entropy}

\begin{thebibliography}{88}
\expandafter\ifx\csname natexlab\endcsname\relax\def\natexlab#1{#1}\fi
\expandafter\ifx\csname bibnamefont\endcsname\relax
  \def\bibnamefont#1{#1}\fi
\expandafter\ifx\csname bibfnamefont\endcsname\relax
  \def\bibfnamefont#1{#1}\fi
\expandafter\ifx\csname citenamefont\endcsname\relax
  \def\citenamefont#1{#1}\fi
\expandafter\ifx\csname url\endcsname\relax
  \def\url#1{\texttt{#1}}\fi
\expandafter\ifx\csname urlprefix\endcsname\relax\def\urlprefix{URL }\fi
\providecommand{\bibinfo}[2]{#2}
\providecommand{\eprint}[2][]{\url{#2}}

\bibitem[{\citenamefont{Jaynes}(1965)}]{jaynes1965gibbs}
\bibinfo{author}{\bibfnamefont{E.~T.} \bibnamefont{Jaynes}},
  \bibinfo{journal}{\href{http://aapt.scitation.org/doi/10.1119/1.1971557}{Am.
  J. Phys}} \textbf{\bibinfo{volume}{33}}, \bibinfo{pages}{391}
  (\bibinfo{year}{1965}).

\bibitem[{\citenamefont{Sevick et~al.}(2008)\citenamefont{Sevick, Prabhakar,
  Williams, and Searles}}]{sevick2008fluctuation}
\bibinfo{author}{\bibfnamefont{E.~M.} \bibnamefont{Sevick}},
  \bibinfo{author}{\bibfnamefont{R.}~\bibnamefont{Prabhakar}},
  \bibinfo{author}{\bibfnamefont{S.~R.} \bibnamefont{Williams}},
  \bibnamefont{and} \bibinfo{author}{\bibfnamefont{D.~J.}
  \bibnamefont{Searles}}, \bibinfo{journal}{Annu. Rev. Phys. Chem.}
  \textbf{\bibinfo{volume}{59}}, \bibinfo{pages}{603} (\bibinfo{year}{2008}).

\bibitem[{\citenamefont{\AA{}berg}(2018)}]{aberg2016fully}
\bibinfo{author}{\bibfnamefont{J.}~\bibnamefont{\AA{}berg}},
  \bibinfo{journal}{\href{https://link.aps.org/doi/10.1103/PhysRevX.8.011019}{Phys.
  Rev. X}} \textbf{\bibinfo{volume}{8}}, \bibinfo{pages}{011019}
  (\bibinfo{year}{2018}).

\bibitem[{\citenamefont{Gharibyan and Tegmark}(2014)}]{gharibyan2014sharpening}
\bibinfo{author}{\bibfnamefont{H.}~\bibnamefont{Gharibyan}} \bibnamefont{and}
  \bibinfo{author}{\bibfnamefont{M.}~\bibnamefont{Tegmark}},
  \bibinfo{journal}{\href{https://journals.aps.org/pre/pdf/10.1103/PhysRevE.90.032125}{Phys.
  Rev. E}} \textbf{\bibinfo{volume}{90}}, \bibinfo{pages}{032125}
  (\bibinfo{year}{2014}).

\bibitem[{\citenamefont{Cramer et~al.}(2008)\citenamefont{Cramer, Dawson,
  Eisert, and Osborne}}]{cramer2008exact}
\bibinfo{author}{\bibfnamefont{M.}~\bibnamefont{Cramer}},
  \bibinfo{author}{\bibfnamefont{C.~M.} \bibnamefont{Dawson}},
  \bibinfo{author}{\bibfnamefont{J.}~\bibnamefont{Eisert}}, \bibnamefont{and}
  \bibinfo{author}{\bibfnamefont{T.~J.} \bibnamefont{Osborne}},
  \bibinfo{journal}{\href{https://journals.aps.org/prl/abstract/10.1103/PhysRevLett.100.030602}{Phys.
  Rev. Lett}} \textbf{\bibinfo{volume}{100}}, \bibinfo{pages}{030602}
  (\bibinfo{year}{2008}).

\bibitem[{\citenamefont{Reimann}(2008)}]{reimann2008foundation}
\bibinfo{author}{\bibfnamefont{P.}~\bibnamefont{Reimann}},
  \bibinfo{journal}{\href{https://journals.aps.org/prl/abstract/10.1103/PhysRevLett.101.190403}{Phys.
  Rev. Lett}} \textbf{\bibinfo{volume}{101}}, \bibinfo{pages}{190403}
  (\bibinfo{year}{2008}).

\bibitem[{\citenamefont{Linden et~al.}(2009)\citenamefont{Linden, Popescu,
  Short, and Winter}}]{linden2009quantum}
\bibinfo{author}{\bibfnamefont{N.}~\bibnamefont{Linden}},
  \bibinfo{author}{\bibfnamefont{S.}~\bibnamefont{Popescu}},
  \bibinfo{author}{\bibfnamefont{A.~J.} \bibnamefont{Short}}, \bibnamefont{and}
  \bibinfo{author}{\bibfnamefont{A.}~\bibnamefont{Winter}},
  \bibinfo{journal}{\href{https://journals.aps.org/pre/abstract/10.1103/PhysRevE.79.061103}{Phys.
  Rev. E}} \textbf{\bibinfo{volume}{79}}, \bibinfo{pages}{061103}
  (\bibinfo{year}{2009}).

\bibitem[{\citenamefont{Cramer}(2012)}]{cramer2012thermalization}
\bibinfo{author}{\bibfnamefont{M.}~\bibnamefont{Cramer}},
  \bibinfo{journal}{\href{http://iopscience.iop.org/article/10.1088/1367-2630/14/5/053051/meta}{New
  J. Phys.}} \textbf{\bibinfo{volume}{14}}, \bibinfo{pages}{053051}
  (\bibinfo{year}{2012}).

\bibitem[{\citenamefont{Short and Farrelly}(2012)}]{short2012quantum}
\bibinfo{author}{\bibfnamefont{A.~J.} \bibnamefont{Short}} \bibnamefont{and}
  \bibinfo{author}{\bibfnamefont{T.~C.} \bibnamefont{Farrelly}},
  \bibinfo{journal}{\href{http://iopscience.iop.org/article/10.1088/1367-2630/14/1/013063/meta}{New
  J. Phys.}} \textbf{\bibinfo{volume}{14}}, \bibinfo{pages}{013063}
  (\bibinfo{year}{2012}).

\bibitem[{\citenamefont{Reimann and Kastner}(2012)}]{reimann2012equilibration}
\bibinfo{author}{\bibfnamefont{P.}~\bibnamefont{Reimann}} \bibnamefont{and}
  \bibinfo{author}{\bibfnamefont{M.}~\bibnamefont{Kastner}},
  \bibinfo{journal}{\href{http://iopscience.iop.org/article/10.1088/1367-2630/14/4/043020/meta}{New
  J. Phys.}} \textbf{\bibinfo{volume}{14}}, \bibinfo{pages}{043020}
  (\bibinfo{year}{2012}).

\bibitem[{\citenamefont{Brand{\~a}o et~al.}(2012)\citenamefont{Brand{\~a}o,
  {\'C}wikli{\'n}ski, Horodecki, Horodecki, Korbicz, and
  Mozrzymas}}]{brandao2012convergence}
\bibinfo{author}{\bibfnamefont{F.~G. S.~L.} \bibnamefont{Brand{\~a}o}},
  \bibinfo{author}{\bibfnamefont{P.}~\bibnamefont{{\'C}wikli{\'n}ski}},
  \bibinfo{author}{\bibfnamefont{M.}~\bibnamefont{Horodecki}},
  \bibinfo{author}{\bibfnamefont{P.}~\bibnamefont{Horodecki}},
  \bibinfo{author}{\bibfnamefont{J.~K.} \bibnamefont{Korbicz}},
  \bibnamefont{and}
  \bibinfo{author}{\bibfnamefont{M.}~\bibnamefont{Mozrzymas}},
  \bibinfo{journal}{\href{https://journals.aps.org/pre/abstract/10.1103/PhysRevE.86.031101}{Phys.
  Rev. E}} \textbf{\bibinfo{volume}{86}}, \bibinfo{pages}{031101}
  (\bibinfo{year}{2012}).

\bibitem[{\citenamefont{{\v{Z}}nidari{\v{c}}
  et~al.}(2012)}]{vznidarivc2012subsystem}
\bibinfo{author}{\bibfnamefont{M.}~\bibnamefont{{\v{Z}}nidari{\v{c}}}}
  \bibnamefont{et~al.},
  \bibinfo{journal}{\href{http://iopscience.iop.org/article/10.1088/1751-8113/45/12/125204/meta}{J.
  Phys. A}} \textbf{\bibinfo{volume}{45}}, \bibinfo{pages}{125204}
  (\bibinfo{year}{2012}).

\bibitem[{\citenamefont{Masanes et~al.}(2013)\citenamefont{Masanes, Roncaglia,
  and Ac\'{i}n}}]{masanes2013complexity}
\bibinfo{author}{\bibfnamefont{L.}~\bibnamefont{Masanes}},
  \bibinfo{author}{\bibfnamefont{A.~J.} \bibnamefont{Roncaglia}},
  \bibnamefont{and} \bibinfo{author}{\bibfnamefont{A.}~\bibnamefont{Ac\'{i}n}},
  \bibinfo{journal}{\href{https://journals.aps.org/pre/abstract/10.1103/PhysRevE.87.032137}{Phys.
  Rev. E}} \textbf{\bibinfo{volume}{87}}, \bibinfo{pages}{032137}
  (\bibinfo{year}{2013}).

\bibitem[{\citenamefont{Garc\'{\i}a-Pintos
  et~al.}(2017)\citenamefont{Garc\'{\i}a-Pintos, Linden, Malabarba, Short, and
  Winter}}]{garcia2015equilibration}
\bibinfo{author}{\bibfnamefont{L.~P.} \bibnamefont{Garc\'{\i}a-Pintos}},
  \bibinfo{author}{\bibfnamefont{N.}~\bibnamefont{Linden}},
  \bibinfo{author}{\bibfnamefont{A.~S.~L.} \bibnamefont{Malabarba}},
  \bibinfo{author}{\bibfnamefont{A.~J.} \bibnamefont{Short}}, \bibnamefont{and}
  \bibinfo{author}{\bibfnamefont{A.}~\bibnamefont{Winter}},
  \bibinfo{journal}{\href{https://link.aps.org/doi/10.1103/PhysRevX.7.031027}{Phys.
  Rev. X}} \textbf{\bibinfo{volume}{7}}, \bibinfo{pages}{031027}
  (\bibinfo{year}{2017}).

\bibitem[{\citenamefont{Gluza et~al.}(2016)\citenamefont{Gluza, Krumnow,
  Friesdorf, Gogolin, and Eisert}}]{gluza2016equilibration}
\bibinfo{author}{\bibfnamefont{M.}~\bibnamefont{Gluza}},
  \bibinfo{author}{\bibfnamefont{C.}~\bibnamefont{Krumnow}},
  \bibinfo{author}{\bibfnamefont{M.}~\bibnamefont{Friesdorf}},
  \bibinfo{author}{\bibfnamefont{C.}~\bibnamefont{Gogolin}}, \bibnamefont{and}
  \bibinfo{author}{\bibfnamefont{J.}~\bibnamefont{Eisert}},
  \bibinfo{journal}{\href{https://journals.aps.org/prl/abstract/10.1103/PhysRevLett.117.190602}{Phys.
  Rev. Lett}} \textbf{\bibinfo{volume}{117}}, \bibinfo{pages}{190602}
  (\bibinfo{year}{2016}).

\bibitem[{\citenamefont{Eisert et~al.}(2015)\citenamefont{Eisert, Friesdorf,
  and Gogolin}}]{eisert2015quantum}
\bibinfo{author}{\bibfnamefont{J.}~\bibnamefont{Eisert}},
  \bibinfo{author}{\bibfnamefont{M.}~\bibnamefont{Friesdorf}},
  \bibnamefont{and} \bibinfo{author}{\bibfnamefont{C.}~\bibnamefont{Gogolin}},
  \bibinfo{journal}{\href{http://www.nature.com/nphys/journal/v11/n2/full/nphys3215.html}{Nat.
  Phys}} \textbf{\bibinfo{volume}{11}}, \bibinfo{pages}{124}
  (\bibinfo{year}{2015}).

\bibitem[{\citenamefont{Gogolin and Eisert}(2016)}]{gogolin2016equilibration}
\bibinfo{author}{\bibfnamefont{C.}~\bibnamefont{Gogolin}} \bibnamefont{and}
  \bibinfo{author}{\bibfnamefont{J.}~\bibnamefont{Eisert}},
  \bibinfo{journal}{\href{http://iopscience.iop.org/article/10.1088/0034-4885/79/5/056001/meta}{Rep.
  Prog. Phys}} \textbf{\bibinfo{volume}{79}}, \bibinfo{pages}{056001}
  (\bibinfo{year}{2016}).

\bibitem[{\citenamefont{Reimann}(2016)}]{reimann2016typical}
\bibinfo{author}{\bibfnamefont{P.}~\bibnamefont{Reimann}},
  \bibinfo{journal}{\href{https://www.nature.com/articles/ncomms10821}{Nat.
  Commun}} \textbf{\bibinfo{volume}{7}}, \bibinfo{pages}{10821}
  (\bibinfo{year}{2016}).

\bibitem[{\citenamefont{von Neumann}(2010)}]{von2010proof}
\bibinfo{author}{\bibfnamefont{J.}~\bibnamefont{von Neumann}},
  \bibinfo{journal}{\href{http://dx.doi.org/10.1140/epjh/e2010-00008-5}{Eur.
  Phys. J. H}} \textbf{\bibinfo{volume}{35}}, \bibinfo{pages}{201}
  (\bibinfo{year}{2010}).

\bibitem[{\citenamefont{Tolman}(1938)}]{tolman1938principles}
\bibinfo{author}{\bibfnamefont{R.~C.} \bibnamefont{Tolman}},
  \emph{\bibinfo{title}{The principles of statistical mechanics}}
  (\bibinfo{publisher}{Courier Corporation}, \bibinfo{year}{1938}).

\bibitem[{\citenamefont{Ter~Haar and Band}(1954)}]{ter1954elements}
\bibinfo{author}{\bibfnamefont{D.}~\bibnamefont{Ter~Haar}} \bibnamefont{and}
  \bibinfo{author}{\bibfnamefont{W.}~\bibnamefont{Band}},
  \bibinfo{journal}{\href{http://aapt.scitation.org/doi/10.1119/1.1933869}{Am.
  J. Phys}} \textbf{\bibinfo{volume}{22}}, \bibinfo{pages}{641}
  (\bibinfo{year}{1954}).

\bibitem[{\citenamefont{Jaynes}(1957)}]{jaynes1957information2}
\bibinfo{author}{\bibfnamefont{E.~T.} \bibnamefont{Jaynes}},
  \bibinfo{journal}{\href{https://journals.aps.org/pr/pdf/10.1103/PhysRev.108.171}{Phys.
  Rev}} \textbf{\bibinfo{volume}{108}}, \bibinfo{pages}{171}
  (\bibinfo{year}{1957}).

\bibitem[{\citenamefont{Polkovnikov}(2011)}]{polkovnikov2011microscopic}
\bibinfo{author}{\bibfnamefont{A.}~\bibnamefont{Polkovnikov}},
  \bibinfo{journal}{\href{http://www.sciencedirect.com/science/article/pii/S0003491610001557}{Ann.
  Phys.}} \textbf{\bibinfo{volume}{326}}, \bibinfo{pages}{486}
  (\bibinfo{year}{2011}).

\bibitem[{\citenamefont{Ingarden and Urbanik}(1962)}]{ingarden1962quantum}
\bibinfo{author}{\bibfnamefont{R.~S.} \bibnamefont{Ingarden}} \bibnamefont{and}
  \bibinfo{author}{\bibfnamefont{K.}~\bibnamefont{Urbanik}},
  \bibinfo{journal}{Acta Physica Polonica} \textbf{\bibinfo{volume}{21}},
  \bibinfo{pages}{281} (\bibinfo{year}{1962}).

\bibitem[{\citenamefont{Grabowski and
  Staszewski}(1977)}]{grabowski1977continuity}
\bibinfo{author}{\bibfnamefont{M.}~\bibnamefont{Grabowski}} \bibnamefont{and}
  \bibinfo{author}{\bibfnamefont{P.}~\bibnamefont{Staszewski}},
  \bibinfo{journal}{\href{http://www.sciencedirect.com/science/article/pii/0034487777900659}{Rep.
  Math. Phys.}} \textbf{\bibinfo{volume}{11}}, \bibinfo{pages}{233}
  (\bibinfo{year}{1977}).

\bibitem[{\citenamefont{Anz{\`a} and Vedral}(2017)}]{anza2017information}
\bibinfo{author}{\bibfnamefont{F.}~\bibnamefont{Anz{\`a}}} \bibnamefont{and}
  \bibinfo{author}{\bibfnamefont{V.}~\bibnamefont{Vedral}},
  \bibinfo{journal}{\href{https://www.nature.com/articles/srep44066}{Sci. Rep}}
  \textbf{\bibinfo{volume}{7}}, \bibinfo{pages}{44066} (\bibinfo{year}{2017}).

\bibitem[{\citenamefont{Anz{\`a}}(2018)}]{anza2018new}
\bibinfo{author}{\bibfnamefont{F.}~\bibnamefont{Anz{\`a}}},
  \bibinfo{journal}{\href{https://www.mdpi.com/1099-4300/20/10/744}{Entropy}}
  \textbf{\bibinfo{volume}{20}}, \bibinfo{pages}{744} (\bibinfo{year}{2018}).

\bibitem[{\citenamefont{Anz{\`a} et~al.}(2018)\citenamefont{Anz{\`a}, Gogolin,
  and Huber}}]{anza2018eigenstate}
\bibinfo{author}{\bibfnamefont{F.}~\bibnamefont{Anz{\`a}}},
  \bibinfo{author}{\bibfnamefont{C.}~\bibnamefont{Gogolin}}, \bibnamefont{and}
  \bibinfo{author}{\bibfnamefont{M.}~\bibnamefont{Huber}},
  \bibinfo{journal}{\href{https://link.aps.org/doi/10.1103/PhysRevLett.120.150603}{Phys.
  Rev. Lett.}} \textbf{\bibinfo{volume}{120}}, \bibinfo{pages}{150603}
  (\bibinfo{year}{2018}).

\bibitem[{\citenamefont{Deutsch}(2010)}]{deutsch2010thermodynamic}
\bibinfo{author}{\bibfnamefont{J.~M.} \bibnamefont{Deutsch}},
  \bibinfo{journal}{\href{http://iopscience.iop.org/article/10.1088/1367-2630/12/7/075021/meta}{New
  J. Phys.}} \textbf{\bibinfo{volume}{12}}, \bibinfo{pages}{075021}
  (\bibinfo{year}{2010}).

\bibitem[{\citenamefont{Deutsch et~al.}(2013)\citenamefont{Deutsch, Li, and
  Sharma}}]{deutsch2013microscopic}
\bibinfo{author}{\bibfnamefont{J.~M.} \bibnamefont{Deutsch}},
  \bibinfo{author}{\bibfnamefont{H.}~\bibnamefont{Li}}, \bibnamefont{and}
  \bibinfo{author}{\bibfnamefont{A.}~\bibnamefont{Sharma}},
  \bibinfo{journal}{\href{https://journals.aps.org/pre/abstract/10.1103/PhysRevE.87.042135}{Phys.
  Rev. E}} \textbf{\bibinfo{volume}{87}}, \bibinfo{pages}{042135}
  (\bibinfo{year}{2013}).

\bibitem[{\citenamefont{Santos et~al.}(2012)\citenamefont{Santos, Polkovnikov,
  and Rigol}}]{santos2012weak}
\bibinfo{author}{\bibfnamefont{L.~F.} \bibnamefont{Santos}},
  \bibinfo{author}{\bibfnamefont{A.}~\bibnamefont{Polkovnikov}},
  \bibnamefont{and} \bibinfo{author}{\bibfnamefont{M.}~\bibnamefont{Rigol}},
  \bibinfo{journal}{\href{https://journals.aps.org/pre/abstract/10.1103/PhysRevE.86.010102}{Phys.
  Rev. E}} \textbf{\bibinfo{volume}{86}}, \bibinfo{pages}{010102}
  (\bibinfo{year}{2012}).

\bibitem[{\citenamefont{Sagawa and Ueda}(2012)}]{sagawa2012nonequilibrium}
\bibinfo{author}{\bibfnamefont{T.}~\bibnamefont{Sagawa}} \bibnamefont{and}
  \bibinfo{author}{\bibfnamefont{M.}~\bibnamefont{Ueda}},
  \bibinfo{journal}{\href{https://journals.aps.org/pre/abstract/10.1103/PhysRevE.85.021104}{Phys.
  Rev. E}} \textbf{\bibinfo{volume}{85}}, \bibinfo{pages}{021104}
  (\bibinfo{year}{2012}).

\bibitem[{\citenamefont{Konig et~al.}(2009)\citenamefont{Konig, Renner, and
  Schaffner}}]{konig2009operational}
\bibinfo{author}{\bibfnamefont{R.}~\bibnamefont{Konig}},
  \bibinfo{author}{\bibfnamefont{R.}~\bibnamefont{Renner}}, \bibnamefont{and}
  \bibinfo{author}{\bibfnamefont{C.}~\bibnamefont{Schaffner}},
  \bibinfo{journal}{\href{http://ieeexplore.ieee.org/abstract/document/5208530/}{IEEE
  Trans. Inf. Theory}} \textbf{\bibinfo{volume}{55}}, \bibinfo{pages}{4337}
  (\bibinfo{year}{2009}).

\bibitem[{\citenamefont{Modi et~al.}(2010)\citenamefont{Modi, Paterek, Son,
  Vedral, and Williamson}}]{modi2010unified}
\bibinfo{author}{\bibfnamefont{K.}~\bibnamefont{Modi}},
  \bibinfo{author}{\bibfnamefont{T.}~\bibnamefont{Paterek}},
  \bibinfo{author}{\bibfnamefont{W.}~\bibnamefont{Son}},
  \bibinfo{author}{\bibfnamefont{V.}~\bibnamefont{Vedral}}, \bibnamefont{and}
  \bibinfo{author}{\bibfnamefont{M.}~\bibnamefont{Williamson}},
  \bibinfo{journal}{\href{https://journals.aps.org/prl/abstract/10.1103/PhysRevLett.104.080501}{Phys.
  Rev. Lett.}} \textbf{\bibinfo{volume}{104}}, \bibinfo{pages}{080501}
  (\bibinfo{year}{2010}).

\bibitem[{\citenamefont{Deffner and Lutz}(2010)}]{deffner2010generalized}
\bibinfo{author}{\bibfnamefont{S.}~\bibnamefont{Deffner}} \bibnamefont{and}
  \bibinfo{author}{\bibfnamefont{E.}~\bibnamefont{Lutz}},
  \bibinfo{journal}{\href{https://journals.aps.org/prl/abstract/10.1103/PhysRevLett.105.170402}{Phys.
  Rev. Lett.}} \textbf{\bibinfo{volume}{105}}, \bibinfo{pages}{170402}
  (\bibinfo{year}{2010}).

\bibitem[{\citenamefont{Del~Rio et~al.}(2011)\citenamefont{Del~Rio, {\AA}berg,
  Renner, Dahlsten, and Vedral}}]{del2011thermodynamic}
\bibinfo{author}{\bibfnamefont{L.}~\bibnamefont{Del~Rio}},
  \bibinfo{author}{\bibfnamefont{J.}~\bibnamefont{{\AA}berg}},
  \bibinfo{author}{\bibfnamefont{R.}~\bibnamefont{Renner}},
  \bibinfo{author}{\bibfnamefont{O.}~\bibnamefont{Dahlsten}}, \bibnamefont{and}
  \bibinfo{author}{\bibfnamefont{V.}~\bibnamefont{Vedral}},
  \bibinfo{journal}{\href{https://www.nature.com/articles/nature10123}{Nature}}
  \textbf{\bibinfo{volume}{474}}, \bibinfo{pages}{61} (\bibinfo{year}{2011}).

\bibitem[{\citenamefont{\ifmmode~\check{S}\else \v{S}\fi{}afr\'anek
  et~al.}(2019)\citenamefont{\ifmmode~\check{S}\else \v{S}\fi{}afr\'anek,
  Deutsch, and Aguirre}}]{SafranekDeutschAguirreObservationalEntropyLetter}
\bibinfo{author}{\bibfnamefont{D.}~\bibnamefont{\ifmmode~\check{S}\else
  \v{S}\fi{}afr\'anek}}, \bibinfo{author}{\bibfnamefont{J.~M.}
  \bibnamefont{Deutsch}}, \bibnamefont{and}
  \bibinfo{author}{\bibfnamefont{A.}~\bibnamefont{Aguirre}},
  \bibinfo{journal}{\href{https://link.aps.org/doi/10.1103/PhysRevA.99.010101}{Phys.
  Rev. A}} \textbf{\bibinfo{volume}{99}}, \bibinfo{pages}{010101}
  (\bibinfo{year}{2019}).

\bibitem[{\citenamefont{Griffiths}(2017)}]{griffiths2014consistent}
\bibinfo{author}{\bibfnamefont{R.~B.} \bibnamefont{Griffiths}}, in
  \emph{\bibinfo{booktitle}{The Stanford Encyclopedia of Philosophy}}, edited
  by \bibinfo{editor}{\bibfnamefont{E.~N.} \bibnamefont{Zalta}}
  (\bibinfo{publisher}{Metaphysics Research Lab, Stanford University},
  \bibinfo{year}{2017}), \bibinfo{edition}{spring 2017} ed.

\bibitem[{\citenamefont{Von~Neumann}(1955)}]{von1955mathematical}
\bibinfo{author}{\bibfnamefont{J.}~\bibnamefont{Von~Neumann}},
  \emph{\bibinfo{title}{Mathematical foundations of quantum mechanics}},
  \bibinfo{number}{2}
  (\bibinfo{publisher}{\href{http://press.princeton.edu/titles/2113.html}{Princeton
  university press}}, \bibinfo{year}{1955}).

\bibitem[{\citenamefont{Lebowitz}(1999)}]{lebowitz1999microscopic}
\bibinfo{author}{\bibfnamefont{J.~L.} \bibnamefont{Lebowitz}},
  \bibinfo{journal}{\href{https://www.sciencedirect.com/science/article/pii/S0378437198005147}{Physica
  A}} \textbf{\bibinfo{volume}{263}}, \bibinfo{pages}{516}
  (\bibinfo{year}{1999}).

\bibitem[{\citenamefont{Wehrl}(1978)}]{wehrl1978general}
\bibinfo{author}{\bibfnamefont{A.}~\bibnamefont{Wehrl}},
  \bibinfo{journal}{\href{https://journals.aps.org/rmp/abstract/10.1103/RevModPhys.50.221}{Rev.
  Mod. Phys}} \textbf{\bibinfo{volume}{50}}, \bibinfo{pages}{221}
  (\bibinfo{year}{1978}).

\bibitem[{\citenamefont{Callens et~al.}(2004)\citenamefont{Callens, De~Roeck,
  Jacobs, Maes, and Neto{\v{c}}n{\'y}}}]{callens2004quantum}
\bibinfo{author}{\bibfnamefont{I.}~\bibnamefont{Callens}},
  \bibinfo{author}{\bibfnamefont{W.}~\bibnamefont{De~Roeck}},
  \bibinfo{author}{\bibfnamefont{T.}~\bibnamefont{Jacobs}},
  \bibinfo{author}{\bibfnamefont{C.}~\bibnamefont{Maes}}, \bibnamefont{and}
  \bibinfo{author}{\bibfnamefont{K.}~\bibnamefont{Neto{\v{c}}n{\'y}}},
  \bibinfo{journal}{\href{https://www.sciencedirect.com/science/article/pii/S0167278903003774}{Physica
  D}} \textbf{\bibinfo{volume}{187}}, \bibinfo{pages}{383}
  (\bibinfo{year}{2004}).

\bibitem[{\citenamefont{Alonso-Serrano and Visser}(2017)}]{alonso2017coarse}
\bibinfo{author}{\bibfnamefont{A.}~\bibnamefont{Alonso-Serrano}}
  \bibnamefont{and} \bibinfo{author}{\bibfnamefont{M.}~\bibnamefont{Visser}},
  \bibinfo{journal}{\href{http://www.mdpi.com/1099-4300/19/5/207/htm}{Entropy}}
  \textbf{\bibinfo{volume}{19}}, \bibinfo{pages}{207} (\bibinfo{year}{2017}).

\bibitem[{\citenamefont{Blank et~al.}(2008)\citenamefont{Blank, Exner, and
  Havlicek}}]{blank2008hilbert}
\bibinfo{author}{\bibfnamefont{J.}~\bibnamefont{Blank}},
  \bibinfo{author}{\bibfnamefont{P.}~\bibnamefont{Exner}}, \bibnamefont{and}
  \bibinfo{author}{\bibfnamefont{M.}~\bibnamefont{Havlicek}},
  \emph{\bibinfo{title}{Hilbert space operators in quantum physics}}
  (\bibinfo{publisher}{Springer Science \& Business Media},
  \bibinfo{year}{2008}).

\bibitem[{\citenamefont{Goldstein
  et~al.}(2010{\natexlab{a}})\citenamefont{Goldstein, Lebowitz, Tumulka, and
  Zangh{\`\i}}}]{goldstein2010long}
\bibinfo{author}{\bibfnamefont{S.}~\bibnamefont{Goldstein}},
  \bibinfo{author}{\bibfnamefont{J.~L.} \bibnamefont{Lebowitz}},
  \bibinfo{author}{\bibfnamefont{R.}~\bibnamefont{Tumulka}}, \bibnamefont{and}
  \bibinfo{author}{\bibfnamefont{N.}~\bibnamefont{Zangh{\`\i}}},
  \bibinfo{journal}{\href{https://link.springer.com/article/10.1140/epjh/e2010-00007-7}{Eur.
  Phys. J. H}} \textbf{\bibinfo{volume}{35}}, \bibinfo{pages}{173}
  (\bibinfo{year}{2010}{\natexlab{a}}).

\bibitem[{\citenamefont{Klein}(1952)}]{klein1952ergodic}
\bibinfo{author}{\bibfnamefont{M.~J.} \bibnamefont{Klein}},
  \bibinfo{journal}{\href{https://journals.aps.org/pr/abstract/10.1103/PhysRev.87.111}{Phys.
  Rev}} \textbf{\bibinfo{volume}{87}}, \bibinfo{pages}{111}
  (\bibinfo{year}{1952}).

\bibitem[{\citenamefont{Bocchieri and Loinger}(1958)}]{bocchieri1958ergodic}
\bibinfo{author}{\bibfnamefont{P.}~\bibnamefont{Bocchieri}} \bibnamefont{and}
  \bibinfo{author}{\bibfnamefont{A.}~\bibnamefont{Loinger}},
  \bibinfo{journal}{\href{https://journals.aps.org/pr/abstract/10.1103/PhysRev.111.668}{Phys.
  Rev}} \textbf{\bibinfo{volume}{111}}, \bibinfo{pages}{668}
  (\bibinfo{year}{1958}).

\bibitem[{\citenamefont{Goldstein
  et~al.}(2010{\natexlab{b}})\citenamefont{Goldstein, Lebowitz, Mastrodonato,
  Tumulka, and Zangh{\`\i}}}]{goldstein2010normal}
\bibinfo{author}{\bibfnamefont{S.}~\bibnamefont{Goldstein}},
  \bibinfo{author}{\bibfnamefont{J.~L.} \bibnamefont{Lebowitz}},
  \bibinfo{author}{\bibfnamefont{C.}~\bibnamefont{Mastrodonato}},
  \bibinfo{author}{\bibfnamefont{R.}~\bibnamefont{Tumulka}}, \bibnamefont{and}
  \bibinfo{author}{\bibfnamefont{N.}~\bibnamefont{Zangh{\`\i}}}, in
  \emph{\bibinfo{booktitle}{\href{http://rspa.royalsocietypublishing.org/content/466/2123/3203.short}{Proceedings
  of the Royal Society of London A: Mathematical, Physical and Engineering
  Sciences}}} (\bibinfo{organization}{The Royal Society},
  \bibinfo{year}{2010}{\natexlab{b}}), vol. \bibinfo{volume}{466}, pp.
  \bibinfo{pages}{3203--3224}.

\bibitem[{\citenamefont{Brody et~al.}(2007)\citenamefont{Brody, Hook, and
  Hughston}}]{brody2007unitarity}
\bibinfo{author}{\bibfnamefont{D.~C.} \bibnamefont{Brody}},
  \bibinfo{author}{\bibfnamefont{D.~W.} \bibnamefont{Hook}}, \bibnamefont{and}
  \bibinfo{author}{\bibfnamefont{L.~P.} \bibnamefont{Hughston}},
  \bibinfo{journal}{\href{http://iopscience.iop.org/article/10.1088/1751-8113/40/26/F01/meta}{J.
  Phys. A}} \textbf{\bibinfo{volume}{40}}, \bibinfo{pages}{F503}
  (\bibinfo{year}{2007}).

\bibitem[{\citenamefont{Asadi et~al.}(2015)\citenamefont{Asadi, Bakhshinezhad,
  and Rezakhani}}]{asadi2015quantum}
\bibinfo{author}{\bibfnamefont{P.}~\bibnamefont{Asadi}},
  \bibinfo{author}{\bibfnamefont{F.}~\bibnamefont{Bakhshinezhad}},
  \bibnamefont{and} \bibinfo{author}{\bibfnamefont{A.~T.}
  \bibnamefont{Rezakhani}},
  \bibinfo{journal}{\href{http://iopscience.iop.org/article/10.1088/1751-8113/49/5/055301/meta}{J.
  Phys. A}} \textbf{\bibinfo{volume}{49}}, \bibinfo{pages}{055301}
  (\bibinfo{year}{2015}).

\bibitem[{\citenamefont{Zhang et~al.}(2016)\citenamefont{Zhang, Quan, and
  Wu}}]{zhang2016ergodicity}
\bibinfo{author}{\bibfnamefont{D.}~\bibnamefont{Zhang}},
  \bibinfo{author}{\bibfnamefont{H.~T.} \bibnamefont{Quan}}, \bibnamefont{and}
  \bibinfo{author}{\bibfnamefont{B.}~\bibnamefont{Wu}},
  \bibinfo{journal}{\href{https://journals.aps.org/pre/abstract/10.1103/PhysRevE.94.022150}{Phys.
  Rev. E}} \textbf{\bibinfo{volume}{94}}, \bibinfo{pages}{022150}
  (\bibinfo{year}{2016}).

\bibitem[{\citenamefont{Dymarsky et~al.}(2018)\citenamefont{Dymarsky, Lashkari,
  and Liu}}]{dymarsky2018subsystem}
\bibinfo{author}{\bibfnamefont{A.}~\bibnamefont{Dymarsky}},
  \bibinfo{author}{\bibfnamefont{N.}~\bibnamefont{Lashkari}}, \bibnamefont{and}
  \bibinfo{author}{\bibfnamefont{H.}~\bibnamefont{Liu}},
  \bibinfo{journal}{\href{https://journals.aps.org/pre/abstract/10.1103/PhysRevE.97.012140}{Phys.
  Rev. E}} \textbf{\bibinfo{volume}{97}}, \bibinfo{pages}{012140}
  (\bibinfo{year}{2018}).

\bibitem[{\citenamefont{Garrison and Grover}(2018)}]{garrison2015does}
\bibinfo{author}{\bibfnamefont{J.~R.} \bibnamefont{Garrison}} \bibnamefont{and}
  \bibinfo{author}{\bibfnamefont{T.}~\bibnamefont{Grover}},
  \bibinfo{journal}{\href{https://link.aps.org/doi/10.1103/PhysRevX.8.021026}{Phys.
  Rev. X}} \textbf{\bibinfo{volume}{8}}, \bibinfo{pages}{021026}
  (\bibinfo{year}{2018}).

\bibitem[{\citenamefont{Bocchieri and Loinger}(1957)}]{bocchieri1957quantum}
\bibinfo{author}{\bibfnamefont{P.}~\bibnamefont{Bocchieri}} \bibnamefont{and}
  \bibinfo{author}{\bibfnamefont{A.}~\bibnamefont{Loinger}},
  \bibinfo{journal}{\href{https://journals.aps.org/pr/abstract/10.1103/PhysRev.107.337}{Phys.
  Rev}} \textbf{\bibinfo{volume}{107}}, \bibinfo{pages}{337}
  (\bibinfo{year}{1957}).

\bibitem[{\citenamefont{Bekenstein}(1981)}]{bekenstein1981universal}
\bibinfo{author}{\bibfnamefont{J.~D.} \bibnamefont{Bekenstein}},
  \bibinfo{journal}{\href{https://journals.aps.org/prd/abstract/10.1103/PhysRevD.23.287}{Phys.
  Rev. D}} \textbf{\bibinfo{volume}{23}}, \bibinfo{pages}{287}
  (\bibinfo{year}{1981}).

\bibitem[{\citenamefont{Penrose}(1979)}]{OPenrose1979FoundStatMech}
\bibinfo{author}{\bibfnamefont{O.}~\bibnamefont{Penrose}},
  \bibinfo{journal}{\href{http://stacks.iop.org/0034-4885/42/i=12/a=002}{Rep.
  Prog. Phys}} \textbf{\bibinfo{volume}{42}}, \bibinfo{pages}{1937}
  (\bibinfo{year}{1979}).

\bibitem[{\citenamefont{Girvin et~al.}(2009)\citenamefont{Girvin, Devoret, and
  Schoelkopf}}]{girvin2009circuit}
\bibinfo{author}{\bibfnamefont{S.}~\bibnamefont{Girvin}},
  \bibinfo{author}{\bibfnamefont{M.}~\bibnamefont{Devoret}}, \bibnamefont{and}
  \bibinfo{author}{\bibfnamefont{R.}~\bibnamefont{Schoelkopf}},
  \bibinfo{journal}{\href{http://iopscience.iop.org/article/10.1088/0031-8949/2009/T137/014012/meta}{Physica
  Scripta}} \textbf{\bibinfo{volume}{2009}}, \bibinfo{pages}{014012}
  (\bibinfo{year}{2009}).

\bibitem[{\citenamefont{Ruelle}(1999)}]{ruelle1999statistical}
\bibinfo{author}{\bibfnamefont{D.}~\bibnamefont{Ruelle}},
  \emph{\bibinfo{title}{Statistical Mechanics: Rigorous Results}}
  (\bibinfo{publisher}{\href{https://books.google.com/books?id=mNtWfcj\_xhwC}{World
  Scientific}}, \bibinfo{year}{1999}), ISBN \bibinfo{isbn}{9789810238629}.

\bibitem[{\citenamefont{Santos and Rigol}(2010)}]{santos2010onset}
\bibinfo{author}{\bibfnamefont{L.~F.} \bibnamefont{Santos}} \bibnamefont{and}
  \bibinfo{author}{\bibfnamefont{M.}~\bibnamefont{Rigol}},
  \bibinfo{journal}{\href{https://journals.aps.org/pre/abstract/10.1103/PhysRevE.81.036206}{Phys.
  Rev. E}} \textbf{\bibinfo{volume}{81}}, \bibinfo{pages}{036206}
  (\bibinfo{year}{2010}).

\bibitem[{\citenamefont{Reif}(2009)}]{reif2009fundamentals}
\bibinfo{author}{\bibfnamefont{F.}~\bibnamefont{Reif}},
  \emph{\bibinfo{title}{Fundamentals of statistical and thermal physics}}
  (\bibinfo{publisher}{Waveland Press}, \bibinfo{year}{2009}).

\bibitem[{\citenamefont{Lostaglio et~al.}(2017)\citenamefont{Lostaglio,
  Jennings, and Rudolph}}]{lostaglio2015thermodynamic}
\bibinfo{author}{\bibfnamefont{M.}~\bibnamefont{Lostaglio}},
  \bibinfo{author}{\bibfnamefont{D.}~\bibnamefont{Jennings}}, \bibnamefont{and}
  \bibinfo{author}{\bibfnamefont{T.}~\bibnamefont{Rudolph}},
  \bibinfo{journal}{\href{http://stacks.iop.org/1367-2630/19/i=4/a=043008}{New
  J. Phys.}} \textbf{\bibinfo{volume}{19}}, \bibinfo{pages}{043008}
  (\bibinfo{year}{2017}).

\bibitem[{\citenamefont{Halpern et~al.}(2016)\citenamefont{Halpern, Faist,
  Oppenheim, and Winter}}]{halpern2016microcanonical}
\bibinfo{author}{\bibfnamefont{N.~Y.} \bibnamefont{Halpern}},
  \bibinfo{author}{\bibfnamefont{P.}~\bibnamefont{Faist}},
  \bibinfo{author}{\bibfnamefont{J.}~\bibnamefont{Oppenheim}},
  \bibnamefont{and} \bibinfo{author}{\bibfnamefont{A.}~\bibnamefont{Winter}},
  \bibinfo{journal}{\href{https://www.nature.com/articles/ncomms12051}{Nat.
  Commun}} \textbf{\bibinfo{volume}{7}}, \bibinfo{pages}{12051}
  (\bibinfo{year}{2016}).

\bibitem[{\citenamefont{Guryanova et~al.}(2016)\citenamefont{Guryanova,
  Popescu, Short, Silva, and Skrzypczyk}}]{guryanova2016thermodynamics}
\bibinfo{author}{\bibfnamefont{Y.}~\bibnamefont{Guryanova}},
  \bibinfo{author}{\bibfnamefont{S.}~\bibnamefont{Popescu}},
  \bibinfo{author}{\bibfnamefont{A.~J.} \bibnamefont{Short}},
  \bibinfo{author}{\bibfnamefont{R.}~\bibnamefont{Silva}}, \bibnamefont{and}
  \bibinfo{author}{\bibfnamefont{P.}~\bibnamefont{Skrzypczyk}},
  \bibinfo{journal}{\href{https://www.nature.com/articles/ncomms12049}{Nat.
  Commun}} \textbf{\bibinfo{volume}{7}}, \bibinfo{pages}{12049}
  (\bibinfo{year}{2016}).

\bibitem[{\citenamefont{Bethe}(1931)}]{bethe1931theorie}
\bibinfo{author}{\bibfnamefont{H.}~\bibnamefont{Bethe}},
  \bibinfo{journal}{Zeitschrift f{\"u}r Physik} \textbf{\bibinfo{volume}{71}},
  \bibinfo{pages}{205} (\bibinfo{year}{1931}).

\bibitem[{\citenamefont{Karabach et~al.}(1997)\citenamefont{Karabach,
  M{\"u}ller, Gould, Tobochnik et~al.}}]{karabach1997introduction}
\bibinfo{author}{\bibfnamefont{M.}~\bibnamefont{Karabach}},
  \bibinfo{author}{\bibfnamefont{G.}~\bibnamefont{M{\"u}ller}},
  \bibinfo{author}{\bibfnamefont{H.}~\bibnamefont{Gould}},
  \bibinfo{author}{\bibfnamefont{J.}~\bibnamefont{Tobochnik}},
  \bibnamefont{et~al.},
  \bibinfo{journal}{\href{http://digitalcommons.uri.edu/phys_facpubs/45/}{Computers
  in Physics}} \textbf{\bibinfo{volume}{11}}, \bibinfo{pages}{36}
  (\bibinfo{year}{1997}).

\bibitem[{\citenamefont{Mehta}(2004)}]{mehta2004random}
\bibinfo{author}{\bibfnamefont{M.~L.} \bibnamefont{Mehta}},
  \emph{\bibinfo{title}{Random matrices}}, vol. \bibinfo{volume}{142}
  (\bibinfo{publisher}{\href{https://www.elsevier.com/books/random-matrices/lal-mehta/978-0-12-088409-4}{Academic
  press}}, \bibinfo{year}{2004}).

\bibitem[{\citenamefont{Wigner}(1951)}]{wigner1951class}
\bibinfo{author}{\bibfnamefont{E.~P.} \bibnamefont{Wigner}},
  \bibinfo{journal}{\href{http://www.jstor.org/stable/1969342}{Ann. Math}}
  \textbf{\bibinfo{volume}{53}}, \bibinfo{pages}{36} (\bibinfo{year}{1951}).

\bibitem[{\citenamefont{Rigol et~al.}(2008)\citenamefont{Rigol, Dunjko, and
  Olshanii}}]{rigol2008thermalization}
\bibinfo{author}{\bibfnamefont{M.}~\bibnamefont{Rigol}},
  \bibinfo{author}{\bibfnamefont{V.}~\bibnamefont{Dunjko}}, \bibnamefont{and}
  \bibinfo{author}{\bibfnamefont{M.}~\bibnamefont{Olshanii}},
  \bibinfo{journal}{\href{http://www.nature.com/nature/journal/v452/n7189/full/nature06838.html}{Nature}}
  \textbf{\bibinfo{volume}{452}}, \bibinfo{pages}{854} (\bibinfo{year}{2008}).

\bibitem[{\citenamefont{Alba}(2015)}]{alba2015eigenstate}
\bibinfo{author}{\bibfnamefont{V.}~\bibnamefont{Alba}},
  \bibinfo{journal}{\href{https://journals.aps.org/prb/abstract/10.1103/PhysRevB.91.155123}{Phys.
  Rev. B}} \textbf{\bibinfo{volume}{91}}, \bibinfo{pages}{155123}
  (\bibinfo{year}{2015}).

\bibitem[{\citenamefont{Beugeling et~al.}(2014)\citenamefont{Beugeling,
  Moessner, and Haque}}]{beugeling2014finite}
\bibinfo{author}{\bibfnamefont{W.}~\bibnamefont{Beugeling}},
  \bibinfo{author}{\bibfnamefont{R.}~\bibnamefont{Moessner}}, \bibnamefont{and}
  \bibinfo{author}{\bibfnamefont{M.}~\bibnamefont{Haque}},
  \bibinfo{journal}{\href{https://journals.aps.org/pre/abstract/10.1103/PhysRevE.89.042112}{Phys.
  Rev. E}} \textbf{\bibinfo{volume}{89}}, \bibinfo{pages}{042112}
  (\bibinfo{year}{2014}).

\bibitem[{\citenamefont{Deutsch}(1991)}]{deutsch1991quantum}
\bibinfo{author}{\bibfnamefont{J.~M.} \bibnamefont{Deutsch}},
  \bibinfo{journal}{\href{https://journals.aps.org/pra/abstract/10.1103/PhysRevA.43.2046}{Phys.
  Rev. A}} \textbf{\bibinfo{volume}{43}}, \bibinfo{pages}{2046}
  (\bibinfo{year}{1991}).

\bibitem[{\citenamefont{Srednicki}(1994)}]{srednicki1994chaos}
\bibinfo{author}{\bibfnamefont{M.}~\bibnamefont{Srednicki}},
  \bibinfo{journal}{\href{https://journals.aps.org/pre/abstract/10.1103/PhysRevE.50.888}{Phys.
  Rev. E}} \textbf{\bibinfo{volume}{50}}, \bibinfo{pages}{888}
  (\bibinfo{year}{1994}).

\bibitem[{\citenamefont{Kinoshita et~al.}(2006)\citenamefont{Kinoshita, Wenger,
  and Weiss}}]{kinoshita2006quantum}
\bibinfo{author}{\bibfnamefont{T.}~\bibnamefont{Kinoshita}},
  \bibinfo{author}{\bibfnamefont{T.}~\bibnamefont{Wenger}}, \bibnamefont{and}
  \bibinfo{author}{\bibfnamefont{D.~S.} \bibnamefont{Weiss}},
  \bibinfo{journal}{\href{https://www.nature.com/nature/journal/v440/n7086/abs/nature04693.html}{Nature}}
  \textbf{\bibinfo{volume}{440}}, \bibinfo{pages}{900} (\bibinfo{year}{2006}).

\bibitem[{\citenamefont{Hofferberth et~al.}(2007)\citenamefont{Hofferberth,
  Lesanovsky, Fischer, Schumm, and Schmiedmayer}}]{hofferberth2007non}
\bibinfo{author}{\bibfnamefont{S.}~\bibnamefont{Hofferberth}},
  \bibinfo{author}{\bibfnamefont{I.}~\bibnamefont{Lesanovsky}},
  \bibinfo{author}{\bibfnamefont{B.}~\bibnamefont{Fischer}},
  \bibinfo{author}{\bibfnamefont{T.}~\bibnamefont{Schumm}}, \bibnamefont{and}
  \bibinfo{author}{\bibfnamefont{J.}~\bibnamefont{Schmiedmayer}},
  \bibinfo{journal}{\href{http://www.nature.com/nature/journal/v449/n7160/abs/nature06149.html}{Nature}}
  \textbf{\bibinfo{volume}{449}}, \bibinfo{pages}{324} (\bibinfo{year}{2007}).

\bibitem[{\citenamefont{Trotzky et~al.}(2012)\citenamefont{Trotzky, Chen,
  Flesch, McCulloch, Schollw{\"o}ck, Eisert, and Bloch}}]{trotzky2012probing}
\bibinfo{author}{\bibfnamefont{S.}~\bibnamefont{Trotzky}},
  \bibinfo{author}{\bibfnamefont{Y.-A.} \bibnamefont{Chen}},
  \bibinfo{author}{\bibfnamefont{A.}~\bibnamefont{Flesch}},
  \bibinfo{author}{\bibfnamefont{I.~P.} \bibnamefont{McCulloch}},
  \bibinfo{author}{\bibfnamefont{U.}~\bibnamefont{Schollw{\"o}ck}},
  \bibinfo{author}{\bibfnamefont{J.}~\bibnamefont{Eisert}}, \bibnamefont{and}
  \bibinfo{author}{\bibfnamefont{I.}~\bibnamefont{Bloch}},
  \bibinfo{journal}{\href{http://www.nature.com/nphys/journal/v8/n4/abs/nphys2232.html}{Nat.
  Phys}} \textbf{\bibinfo{volume}{8}}, \bibinfo{pages}{325}
  (\bibinfo{year}{2012}).

\bibitem[{\citenamefont{Polkovnikov et~al.}(2011)\citenamefont{Polkovnikov,
  Sengupta, Silva, and Vengalattore}}]{polkovnikov2011nonequilibrium}
\bibinfo{author}{\bibfnamefont{A.}~\bibnamefont{Polkovnikov}},
  \bibinfo{author}{\bibfnamefont{K.}~\bibnamefont{Sengupta}},
  \bibinfo{author}{\bibfnamefont{A.}~\bibnamefont{Silva}}, \bibnamefont{and}
  \bibinfo{author}{\bibfnamefont{M.}~\bibnamefont{Vengalattore}},
  \bibinfo{journal}{\href{https://link.aps.org/doi/10.1103/RevModPhys.83.863}{Rev.
  Mod. Phys.}} \textbf{\bibinfo{volume}{83}}, \bibinfo{pages}{863}
  (\bibinfo{year}{2011}).

\bibitem[{\citenamefont{Kaufman et~al.}(2016)\citenamefont{Kaufman, Tai, Lukin,
  Rispoli, Schittko, Preiss, and Greiner}}]{kaufman2016quantum}
\bibinfo{author}{\bibfnamefont{A.~M.} \bibnamefont{Kaufman}},
  \bibinfo{author}{\bibfnamefont{M.~E.} \bibnamefont{Tai}},
  \bibinfo{author}{\bibfnamefont{A.}~\bibnamefont{Lukin}},
  \bibinfo{author}{\bibfnamefont{M.}~\bibnamefont{Rispoli}},
  \bibinfo{author}{\bibfnamefont{R.}~\bibnamefont{Schittko}},
  \bibinfo{author}{\bibfnamefont{P.~M.} \bibnamefont{Preiss}},
  \bibnamefont{and} \bibinfo{author}{\bibfnamefont{M.}~\bibnamefont{Greiner}},
  \bibinfo{journal}{\href{http://science.sciencemag.org/content/353/6301/794}{Science}}
  \textbf{\bibinfo{volume}{353}}, \bibinfo{pages}{794} (\bibinfo{year}{2016}).

\bibitem[{\citenamefont{Levin et~al.}(2012)\citenamefont{Levin, Fetter, and
  Stamper-Kurn}}]{levin2012ultracold}
\bibinfo{author}{\bibfnamefont{K.}~\bibnamefont{Levin}},
  \bibinfo{author}{\bibfnamefont{A.}~\bibnamefont{Fetter}}, \bibnamefont{and}
  \bibinfo{author}{\bibfnamefont{D.}~\bibnamefont{Stamper-Kurn}},
  \emph{\bibinfo{title}{Ultracold Bosonic and Fermionic Gases}},
  vol.~\bibinfo{volume}{5}
  (\bibinfo{publisher}{\href{https://doi.org/10.1016/B978-0-444-53857-4.00030-1}{Elsevier}},
  \bibinfo{year}{2012}).

\bibitem[{\citenamefont{Villa and De~Chiara}(2018)}]{Villa2018cavityassisted}
\bibinfo{author}{\bibfnamefont{L.}~\bibnamefont{Villa}} \bibnamefont{and}
  \bibinfo{author}{\bibfnamefont{G.}~\bibnamefont{De~Chiara}},
  \bibinfo{journal}{\href{https://doi.org/10.22331/q-2018-01-04-42}{Quantum}}
  \textbf{\bibinfo{volume}{2}}, \bibinfo{pages}{42} (\bibinfo{year}{2018}),
  ISSN \bibinfo{issn}{2521-327X}.

\bibitem[{\citenamefont{Schreiber et~al.}(2015)\citenamefont{Schreiber,
  Hodgman, Bordia, L{\"u}schen, Fischer, Vosk, Altman, Schneider, and
  Bloch}}]{schreiber2015observation}
\bibinfo{author}{\bibfnamefont{M.}~\bibnamefont{Schreiber}},
  \bibinfo{author}{\bibfnamefont{S.~S.} \bibnamefont{Hodgman}},
  \bibinfo{author}{\bibfnamefont{P.}~\bibnamefont{Bordia}},
  \bibinfo{author}{\bibfnamefont{H.~P.} \bibnamefont{L{\"u}schen}},
  \bibinfo{author}{\bibfnamefont{M.~H.} \bibnamefont{Fischer}},
  \bibinfo{author}{\bibfnamefont{R.}~\bibnamefont{Vosk}},
  \bibinfo{author}{\bibfnamefont{E.}~\bibnamefont{Altman}},
  \bibinfo{author}{\bibfnamefont{U.}~\bibnamefont{Schneider}},
  \bibnamefont{and} \bibinfo{author}{\bibfnamefont{I.}~\bibnamefont{Bloch}},
  \bibinfo{journal}{\href{http://science.sciencemag.org/content/349/6250/842}{Science}}
  \textbf{\bibinfo{volume}{349}}, \bibinfo{pages}{842} (\bibinfo{year}{2015}).

\bibitem[{\citenamefont{Ikeda et~al.}(2015)\citenamefont{Ikeda, Sakumichi,
  Polkovnikov, and Ueda}}]{ikeda2015second}
\bibinfo{author}{\bibfnamefont{T.~N.} \bibnamefont{Ikeda}},
  \bibinfo{author}{\bibfnamefont{N.}~\bibnamefont{Sakumichi}},
  \bibinfo{author}{\bibfnamefont{A.}~\bibnamefont{Polkovnikov}},
  \bibnamefont{and} \bibinfo{author}{\bibfnamefont{M.}~\bibnamefont{Ueda}},
  \bibinfo{journal}{\href{http://www.sciencedirect.com/science/article/pii/S0003491615000068}{Ann.
  Phys.}} \textbf{\bibinfo{volume}{354}}, \bibinfo{pages}{338}
  (\bibinfo{year}{2015}).

\bibitem[{\citenamefont{Wigner}(1955)}]{wigner1955RandomMatrices}
\bibinfo{author}{\bibfnamefont{E.~P.} \bibnamefont{Wigner}},
  \bibinfo{journal}{Ann. Math} \textbf{\bibinfo{volume}{62}},
  \bibinfo{pages}{548} (\bibinfo{year}{1955}).

\bibitem[{\citenamefont{Berry}(1987)}]{berry1987bakerian}
\bibinfo{author}{\bibfnamefont{M.~V.} \bibnamefont{Berry}}, in
  \emph{\bibinfo{booktitle}{Proceedings of the Royal Society of London A:
  Mathematical, Physical and Engineering Sciences}} (\bibinfo{organization}{The
  Royal Society}, \bibinfo{year}{1987}), vol. \bibinfo{volume}{413}, pp.
  \bibinfo{pages}{183--198}.

\bibitem[{\citenamefont{Feingold and Peres}(1986)}]{FeingoldPhysRevA.34.591}
\bibinfo{author}{\bibfnamefont{M.}~\bibnamefont{Feingold}} \bibnamefont{and}
  \bibinfo{author}{\bibfnamefont{A.}~\bibnamefont{Peres}},
  \bibinfo{journal}{\href{https://link.aps.org/doi/10.1103/PhysRevA.34.591}{Phys.
  Rev. A}} \textbf{\bibinfo{volume}{34}}, \bibinfo{pages}{591}
  (\bibinfo{year}{1986}).

\bibitem[{\citenamefont{Feingold et~al.}(1989)\citenamefont{Feingold, Leitner,
  and Piro}}]{feingold1989semiclassical}
\bibinfo{author}{\bibfnamefont{M.}~\bibnamefont{Feingold}},
  \bibinfo{author}{\bibfnamefont{D.~M.} \bibnamefont{Leitner}},
  \bibnamefont{and} \bibinfo{author}{\bibfnamefont{O.}~\bibnamefont{Piro}},
  \bibinfo{journal}{\href{https://journals.aps.org/pra/abstract/10.1103/PhysRevA.39.6507}{Phys.
  Rev. A}} \textbf{\bibinfo{volume}{39}}, \bibinfo{pages}{6507}
  (\bibinfo{year}{1989}).

\bibitem[{\citenamefont{D'Alessio et~al.}(2016)\citenamefont{D'Alessio, Kafri,
  Polkovnikov, and Rigol}}]{DAlessio2016quantum}
\bibinfo{author}{\bibfnamefont{L.}~\bibnamefont{D'Alessio}},
  \bibinfo{author}{\bibfnamefont{Y.}~\bibnamefont{Kafri}},
  \bibinfo{author}{\bibfnamefont{A.}~\bibnamefont{Polkovnikov}},
  \bibnamefont{and} \bibinfo{author}{\bibfnamefont{M.}~\bibnamefont{Rigol}},
  \bibinfo{journal}{\href{http://www.tandfonline.com/doi/full/10.1080/00018732.2016.1198134}{Adv.
  Phys.}} \textbf{\bibinfo{volume}{65}}, \bibinfo{pages}{239}
  (\bibinfo{year}{2016}).

\bibitem[{\citenamefont{Reimann}(2015)}]{reimann2015eigenstate}
\bibinfo{author}{\bibfnamefont{P.}~\bibnamefont{Reimann}},
  \bibinfo{journal}{\href{http://iopscience.iop.org/article/10.1088/1367-2630/17/5/055025/meta}{New
  J. Phys.}} \textbf{\bibinfo{volume}{17}}, \bibinfo{pages}{055025}
  (\bibinfo{year}{2015}).

\bibitem[{\citenamefont{Wigner}(1957)}]{wigner1957BandedRandomMatrices}
\bibinfo{author}{\bibfnamefont{E.~P.} \bibnamefont{Wigner}},
  \bibinfo{journal}{Ann. Math} \textbf{\bibinfo{volume}{65}},
  \bibinfo{pages}{203} (\bibinfo{year}{1957}).

\end{thebibliography}

\end{document}